\documentclass{aa}  

\usepackage{graphicx}
\usepackage{txfonts}
\newcommand{\teff}{$T_{\rm eff}$}

\newcommand{\msun}{${\rm M}_\odot$}
\newcommand{\lsun}{${\rm L}_\odot$}
\newcommand{\rsun}{${\rm R}_\odot$}

\newcommand{\msunyr}{${\rm M}_\odot {\rm yr}^{-1}$}

\def\arcsec{$''$}

\def\kms{km~s$^{-1}$}

\def\teff{$T_{\rm eff}$}
\def\logg{$\log g$}

\def\vsini{$v\sin i$}
\def\prot{${\rm P}_{rot}$}
\def\vrad{${\rm V}_{r}$}
\def\vmic{${\rm V}_{mic}$}
\def\ha{${\rm H}\alpha$}
\def\hb{${\rm H}\beta$}
\def\hc{${\rm H}\gamma$}
\def\hd{${\rm H}\delta$}
\def\he{${\rm H}\epsilon$}

\def\hei{${\rm HeI}$}
\def\heiir{${\rm HeI}$}

\def\brg{${\rm Br}\gamma$}
\def\brd{${\rm Br}\delta$}
\def\pab{${\rm Pa}\beta$}
\def\pac{${\rm Pa}\gamma$}
\def\pad{${\rm Pa}\delta$}

\def\heiir{HeI}

\def\macc{$\dot{M}_{\rm acc}$}
\def\lacc{$L_{\rm acc}$}

\def\rstar{${\rm R}_\star$}
\def\mstar{${\rm M}_\star$}
\def\lstar{${\rm L}_\star$}
\def\prot{${\rm P}_{rot}$}

\def\phirot{$\Phi_{rot}$}

\def\rcor{$r_{cor}$}

\begin{document} 



   \title{Stable accretion and episodic outflows in the young transition disk system GM Aurigae. }

   \subtitle{A semester-long optical and near-infrared spectrophotometric monitoring campaign\thanks{Based on observations obtained at the Canada-France-Hawaii Telescope (CFHT), at the Observatoire de Haute-Provence (OHP), at the European Organisation for Astronomical Research in the Southern Hemisphere (ESO), and at the Las Cumbres Observatory global telescope network (LCOGT).}\fnmsep\thanks{Tables containing the u'g'r'i' LCOGT photometric measurements are only available in electronic form
at the CDS via anonymous ftp to cdsarc.cds.unistra.fr (130.79.128.5)
or via https://cdsarc.cds.unistra.fr/cgi-bin/qcat?J/A+A/ }}


 \titlerunning{GMAur}
 \authorrunning{J. Bouvier, A. Sousa, K. Pouilly, et al.}

   \author{J. Bouvier  \inst{1}
          \and
          A. Sousa \inst{1}
          \and
          K. Pouilly \inst{2}
          \and
          J.M. Almenara \inst{1}
          \and
          J.-F. Donati \inst{3}
          \and 
          S. Alencar \inst{4}
          \and 
          A. Frasca \inst{5}
          \and
          K. Grankin \inst{6}
          \and
          A. Carmona \inst{1}
          \and
          G. Pantolmos \inst{1}
          \and
          B. Zaire \inst{4}
          \and
          X. Bonfils \inst{1}
          \and 
          A. Bayo \inst{7, 8}
          \and 
          L.M. Rebull \inst{9}
          \and
          J. Alonso-Santiago \inst{5}
          \and
          J. F. Gameiro\inst{10, 11}
          \and
          N. J. Cook\inst{12}
          \and
          E. Artigau\inst{12}
          \and
          the Spirou Legacy Survey (SLS) consortium
       }

   \institute{Univ. Grenoble Alpes, CNRS, IPAG, 38000 Grenoble, France 
         \and
        Department of Physics and Astronomy, Uppsala University, Box 516, SE-75120, Sweden
        \and
         Univ. de Toulouse, CNRS, IRAP, 14 avenue Belin, 31400 Toulouse, France
         \and
         Departamento de Fisica -- ICEx -- UFMG, Av. Antonio Carlos 6627, 30270-901 Belo Horizonte, MG, Brazil
         \and
         INAF – Osservatorio Astrofisico di Catania, via S. Sofia 78, 95123 Catania, Italy
         \and
         Crimean Astrophysical Observatory, Nauchny, 298409, Republic of Crimea
         \and
         Instituto de F\'{\i}sica y Astronom\'{\i}a, Facultad de Ciencias, Universidad de Valpara\'{\i}so, Chile
         \and
          European Southern Observatory, Karl-Schwarzschild-Strasse 2, 85748 Garching bei M\"unchen, Germany
          \and
          Infrared Science Archive (IRSA), IPAC, 1200 E. California Blvd., California Institute of Technology, Pasadena, CA 91125, USA
          \and
          Instituto de Astrof\'{i}sica e Ci\^{e}ncias do Espa\c{c}o,  Universidade do Porto, CAUP, Rua das Estrelas, 4150-762 Porto, Portugal
          \and
          Departamento de F\'{\i}sica e Astronomia, Faculdade de Ci\^encias, Universidade do Porto, rua do Campo Alegre 687, 4169-007 Porto. Portugal
          \and
          Research on Exoplanets, Universit\'e de Montr\'eal, D\'epartement de Physique, Montr\'eal, QC H3C 3J7, Canada
        }

   \date{Received 2 November 2022; accepted 5 January 2023}

  \abstract
{Young stellar systems actively accrete from their circumstellar disk and simultaneously launch outflows. The physical link between accretion and ejection processes remains to be fully understood.}  
{We investigate the structure and dynamics of magnetospheric accretion and associated outflows on a scale smaller than 0.1 au around the young transitional disk system GM Aur.}
{We devised a coordinated observing campaign to monitor the variability of the system on timescales ranging from days to months, including partly simultaneous high-resolution optical and near-infrared spectroscopy, multiwavelength photometry, and low-resolution near-infrared spectroscopy, over a total duration of six months, covering 30 rotational cycles. We analyzed the photometric and line profile variability to characterize the accretion and ejection processes.}
{The optical and near-infrared light curves indicate that the luminosity of the system is modulated by surface spots at the stellar rotation period of 6.04 $\pm$ 0.15 days. Part of the Balmer, Paschen, and Brackett hydrogen line profiles as well as the HeI 5876 \AA\ and HeI 10830 \AA\ line profiles are modulated on the same period. The Pa$\beta$ line flux correlates with the photometric excess in the u' band, which suggests that most of the line emission originates from the accretion process. High-velocity redshifted absorptions reaching below the continuum periodically appear in the near-infrared line profiles at the rotational phase in which the veiling and line fluxes are the largest. These are signatures of a stable accretion funnel flow and associated accretion shock at the stellar surface. This large-scale magnetospheric accretion structure appears fairly stable over at least 15 and possibly up to 30 rotational periods. In contrast, outflow signatures randomly appear as blueshifted absorption components in the Balmer and HeI 10830 \AA\ line profiles. They are not rotationally modulated and disappear on a timescale of a few days. The coexistence of a stable, large-scale accretion pattern and episodic outflows supports magnetospheric ejections as the main process occurring at the star-disk interface. 
} 
{Long-term monitoring of the variability of the GM Aur transitional disk system provides clues to the accretion and ejection structure and dynamics close to the star. Stable magnetospheric accretion and episodic outflows appear to be physically linked on a scale of a few stellar radii in this system. } 


   \maketitle

%

\section{Introduction}

Accretion and ejection processes are at the origin of most of the peculiar properties of young stellar systems. The structure and dynamics of the accretion flows within the disk and from the inner disk to the star, as well as the properties of the multiple outflows arising from the disk, from the star-disk interface, and from the stellar surface, remain to be fully deciphered, however. Low-mass pre-main-sequence stars, the so-called T Tauri stars (TTS), accrete from their circumstellar disks for a few million years, while contemporaneous planet formation impacts the disk structure and evolution. In the inner regions of the system, the disk is disrupted by the strong stellar magnetosphere that channels the accretion flow toward the star along magnetic field lines \citep[see, e.g., the review by][]{Hartmann16}. Thus, accretion funnel flows develop that connect the inner disk to the stellar surface, where the material is accreted at nearly free-fall velocity and is eventually halted in a strong accretion shock. Simultaneously, outflows are produced at the star-disk interface close to the magnetospheric truncation radius through the inflation and reconnection of magnetic field lines that are twisted by differential rotation \citep[e.g.,][]{Zanni13}. Ultimately, the release of gravitational energy delivered by the accretion process may trigger accretion-powered stellar winds \citep{Matt05}. The torque balance between accretion and ejection processes is a central issue for understanding the spin evolution of young stars \citep[e.g.,][]{Pantolmos20, Ireland21}

The star-disk interaction takes place on a distance of a few stellar radii \citep[e.g.,][]{Bessolaz08}, that is, on a scale of about 0.1 au or smaller. MHD models developed by several groups predict the structure and dynamics of the magnetospheric accretion region and associated outflows \citep[see, e.g., the review by][]{Romanova15}. Observationally, two main directions have been explored so far to investigate the properties of this region. On one hand, monitoring the spectroscopic and photometric variability of the system over a few rotational periods, that is, typically over a few weeks, allows identifying the signature of funnel flows, hot spots, and outflows, and relating them to the strength and topology of the surface magnetic field that is measured from spectropolarimetry \citep[e.g.,][]{Pouilly20, Pouilly21, Bouvier20b, Donati19, Donati20b, Alencar18}. On the other hand, a direct approach attempts to spatially resolve the star-disk interaction region on a scale of a few milliarcsecond on the sky, using long-baseline near-infrared interferometry \citep[e.g.,][]{Eisner14, Gravity20, Bouvier20a}.  Both approaches have been successful in mapping the inner region of accreting systems and have provided strong support to the magnetospheric accretion scenario and its MHD modeling. Following previous studies, we report here the results of  a new observing campaign devoted to the young stellar system GM Aur.  

GM Aur (RA = 04h55, Dec = +30\degr21, V = 12.1 mag)  is a solar-type pre-main-sequence star located in the Taurus-Auriga molecular cloud at a distance of 157.9 $\pm$ 1.2 pc \citep{Gaia21}. This classical T Tauri star (cTTS) has a spectral type K6 \citep{Herczeg14} and is surrounded by a circumstellar disk from which it actively accretes material at a rate of 0.6-2.0$\cdot$10$^{-8}$ \msunyr\ \citep{Robinson19}. 
 Based on its spectral energy distribution, which exhibits a small near-infrared excess compared to a significant mid-infrared one, the system has long been suspected to be in a transitional stage, that is, that it is surrounded by a disk whose inner regions are relatively devoid of matter \citep{Strom89}. High-resolution ALMA images of the circumstellar disk indeed reveal that it is highly structured. The large-scale disk, inclined at $\sim$53\degr\ on the line of sight, features a large inner dust cavity extending over $\sim$35-40 au and a succession of annular gaps and dusty rings on a wider scale up to 200 au \citep{Macias18, Huang20}. Much closer to the central star, long-baseline VLTI/GRAVITY interferometric observations unveil a compact dusty disk, whose inner edge was recently reported to be located at $r_{in}$ = 0.013$^{+0.015}_{-0.008}$ au from the central star \citep{Bohn22} and that extends over at least a few 0.1 au \citep{Akeson05}  and possibly up to 6.6 au \citep{Varga18,  Woitke19}. The gaseous component of the inner disk has been detected from CO 4.7 micron emission down to 0.5 $\pm$ 0.2 au \citep{salyk09}. The inclination and position angle of the major axis of the inner dusty  disk ($i$=68\degr$^{+16}_{-28}$, PA=37\degr$^{+31}_{-22}$) are found to be consistent with those of the outer disk, which suggests that the inner and outer disks are aligned \citep{Bohn22}.

In an attempt to decipher the physical processes at work at the heart of the system, GM Aur has been the subject of several multiwavelength monitoring campaigns. The long-term light curve presented by \cite{Grankin07} over the period 1986-1995 exhibits relatively low-level variability, with a V-band magnitude ranging from 11.74 to 12.35 mag. Photometric variations are modulated by surface spots at the stellar rotation period of 6.0-6.1 days \citep{Percy10, Artemenko12}. \cite{Ingleby15} reported variability over the full wavelength range from the  far-UV to the near-infrared, which they attributed in part to an accretion rate that varies by about a factor of 2 to 3 on a timescale of months, and for another part to dust inhomogeneities that are located in the inner disk close to the truncation radius.  Variations in the mass accretion rate of similar amplitude have also been reported on a shorter timescale of about a week by \cite{Robinson19}, and a connection between mass loss and mass accretion has been further suggested by \cite{Espaillat19}. \cite{McGinnis20} presented the results of a high-resolution optical spectroscopic monitoring campaign performed on a timescale of a week that illustrated the variability of the \ha, \hb, and HeI emission line profiles of the system. From the measured radial velocity variations of the \hei\ 5876 \AA\ line profile, whose narrow component traces the accretion shock, they deduced that GM Aur accretes material from its circumstellar disk through an inclined magnetosphere, whose axis is tilted by about 13\degr\ relative to the stellar rotational axis. GM Aur indeed harbors a strong surface magnetic field, with a mean value of 2.2 kG \citep{Johns-Krull07, Symington05}. Finally,  from a recent multiwavelength X-UV-optical campaign, \cite{Espaillat21} reported evidence for a transverse density stratification within the accretion shock at the base of the magnetic funnel flow. 

We report here the results of a new coordinated monitoring campaign on GM Aur that combines high-resolution optical spectroscopy and near-infrared spectropolarimetry, multiwavelength optical and near-infrared photometry, and long-term low-resolution near-infrared spectroscopy. Part of the observations have been obtained simultaneously over a timescale of a few weeks, while the total duration of the campaign amounted to six months. The goal of the campaign was to investigate the physical processes that cause variability in GM Aur on a scale of a few stellar radii, and in particular, to constrain the structure and dynamics of the magnetospheric accretion flow from the inner disk to the star. We devised a long-term campaign in order to be able to probe various timescales, from days to months, and obtain a sufficiently long temporal baseline to investigate the relation between accretion and ejection processes on small spatial scales from the stellar surface to the inner disk regions. 

The campaign whose results are reported here took place in the framework of a larger project led by the ODYSSEUS team\footnote{https://sites.bu.edu/odysseus/} \citep[see][]{Espaillat22}, which uses the Hubble UV Legacy Library of Young Stars as Essential Standards program \citep[ULLYSES\footnote{https://ullyses.stsci.edu/},][]{Roman-Duval20}, on HST Director’s Discretionary time, to monitor a sample of T Tauri stars in the UV domain, which includes GM Aur. Additional follow-up observations were acquired for this project at ESO in the framework of the PENELLOPE Large Program\footnote{https://sites.google.com/view/cfmanara/penellope}  \citep{Manara21}. 

In Section~\ref{sec:obs} we describe the observational techniques we implemented to perform the campaign. In Section~\ref{sec:results} we derive the properties of the system and analyze its photometric and spectroscopic variability over timescales from days to months, including veiling measurements and emission line profiles. We infer the global structure of the magnetospheric accretion flow from the observed variability and characterize associated outflows. In Section~\ref{sec:discussion} we discuss the dynamics of the accretion and ejection structure and show that short-lived episodic outflows coexist with a stable magnetospheric accretion pattern. In Section~\ref{sec:conclusion} we conclude on the ability of multiwavelength, multi-technique coordinated observational campaigns to unveil the physical processes at work in young stellar systems at the sub-au scale.

\section{Observations}
\label{sec:obs}

\begin{figure*}
   \centering
      \includegraphics[width=0.95\hsize]{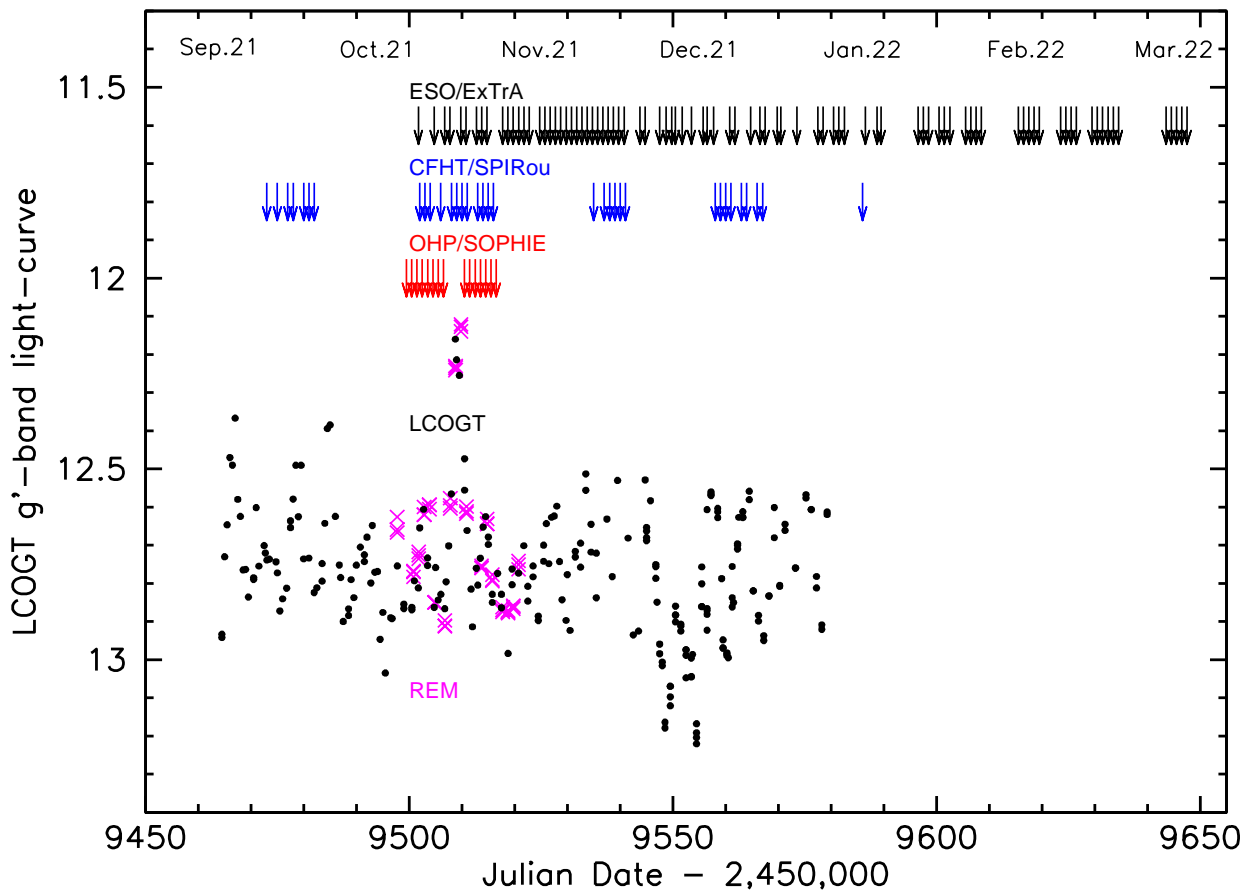}

\caption{Temporal sampling of the GM Aur campaign from September 6, 2021, to March 8, 2022. {\it Bottom:} g'-band light curve from LCOGT ({\it black dots}) and from REM ({\it magenta crosses}). The REM g'-band magnitudes are offset in this figure by +0.1 mag to match the LCOGT measurements. The mean photometric error on both the LCOGT and REM g'-band measurements is 0.025 mag. {\it Vertical arrows:} The vertical arrows show the dates of OHP/SOPHIE ({\it red}), CFHT/SPIRou ({\it blue}), and ESO/ExTrA ({\it black}) observations. The core of the campaign took place during October 2021 with contemporaneous measurements from the five instruments. }
              \label{obs_samp}%
    \end{figure*}

In this section, we describe the acquisition and data-reduction processes of photometric, spectroscopic, and spectropolarimetric datasets obtained during the large-scale campaign we performed on the cTTS GM Aur from September 6, 2021, to March 8, 2022, using CFHT/SPIRou, OHP/SOPHIE, ESO/ExTrA, LCOGT, and ESO/REM. A summary plot of the GM Aur observing campaign reported here is provided in Figure~\ref{obs_samp}. 

\subsection{LCOGT: Multiwavelength optical photometry}
\label{sec:lcogt}

GM Aur was observed at Las Cumbres Observatory Global Network \citep[LCOGT,][]{Brown13}  from September 6 to December 30, 2021. We acquired 850 images in the Sloan u'g'r'i'  filters over two runs with a sub-day cadence (LCO2021B-001, PI L. Rebull; CLN2021B-003, PI A. Bayo). The u' images were obtained with the Sinistro 1m telescopes of the network using an exposure time of 180 seconds and reading the 2Kx2K central window of the detector with a 2x2 binning, resulting in a 13x13 arcmin field of view on the sky. The g'r'i' images were obtained with the 0.4m SBIG telescopes, offering a field of view of 29.2x19.5 arcmin, with exposure times of 60, 20, and 20 seconds, respectively. We retrieved the BANZAI-reduced images from the LCOGT archive service and the noncalibrated photometric catalogs provided in the image headers for all detected stars in the field. 

In order to compute differential photometry, we considered two stars, HD 282625 and HD 282626, both located within 3 arcmin of GM Aur. The first star was used as a reference star to calibrate the differential light curve, and the second was used as a check star to assess that these are nonvariable sources. These field stars have spectral types F2 and F5, respectively, and are only slightly brighter than GM Aur. We confirmed that the two stars are nonvariable from their differential light curves, and we deduced a mean rms photometric error of 0.025 mag in the u'g'r' filters and 0.033 mag in the i' filter. We proceeded to compute the differential light curve between GM Aur and HD 282625 in all four filters. We adopted the mean magnitude of HD 282625 listed in the APASS and Pan-STARRS surveys, namely g'=11.331 mag, r'=10.916 mag, and i'=10.756 mag to calibrate the GM Aur light curve in the g'r'i' filters to within an accuracy of 0.02 mag. 

We were not able to find an estimate of the u'-band magnitude for the comparison stars in the literature. Instead, we assumed the intrinsic (u'-g') colors of an F2 star \citep{Covey07, Kraus07} for HD 282625, to which we applied interstellar reddening. From the observed versus intrinsic color indices of the HD 282625 (g'-r') and (r'-i') bands, we derived A$_V$ = 0.8$\pm$0.1 mag, using the R = 3.1 reddening law from \cite{Fiorucci03}. This procedure yielded an estimate of the reddened (u'-g') color of the comparison star from which we derived its u'-band magnitude. The photometric calibration of the GM Aur u'-band light curve is thus relatively indirect and probably not accurate to better than 0.1 mag\footnote{The table of photometric measurements is available electronically at CDS, Strasbourg.}.

\subsection{REM: Optical and near infrared photometry}
\label{sec:rem}

Observations were performed with the 60 cm robotic REM telescope located at the ESO La Silla Observatory (Chile), on 15 nights from JD 2,459,497 to JD 2,459,520 (October 9 to November 1, 2021). By means of a dichroic, REM simultaneously feeds two cameras at the two Nasmyth focal stations, one camera for the near-infrared (REMIR), and the other for the optical (ROSS2). The cameras have nearly the same field of view of about $10\arcmin\times 10\arcmin$ and use wide-band filters ($J$, $H$, and $K'$ for REMIR and Sloan/SDSS $g'$, $r'$, $i'$, and $z'$ for ROSS2). 

Exposure times were 60 s for ROSS2, which simultaneously acquires images in the four Sloan bands, and five ditherings of 3 s each were adopted for each filter of REMIR. For the ROSS2 camera, we generated master flats using the twilight flat-fields taken during the observing run, which are available in the REM archive. The latter were used to correct for pixel-to-pixel sensitivity variations, as well as for the vignetting and illumination of the field of view. After subtracting the dark-frame, each scientific image was divided by the proper master-flat, depending on the filter. The prereduction of the REMIR images is automatically done by the AQuA pipeline \citep{Testa04}, and the coadded and sky-subtracted frames, resulting from five individual ditherings, are made available to the observer.  

The adopted comparison stars are reported in Table~\ref{tab:sources} along with their $griz$ \citep{Tonry18} and $JHK'$ \citep{Cutri03} magnitudes. Aperture photometry for all the stars listed in Table~\ref{tab:sources} was performed with DAOPHOT by using the IDL\footnote{Interactive Data Language (IDL) is a registered trademark of  Harris Corporation.} routine \textsc{Aper}. For each frame and filter, we used the instrumental magnitudes of the stars listed in Table~\ref{tab:sources} to generate an artificial comparison, weighting them with the flux corresponding to their standard magnitude in a way similar to the ensemble photometry. This procedure also allowed us to evaluate a standard error based on the differences between the magnitudes calculated 
with different comparison stars.

The optical photometry gathered at REM is listed in Table~\ref{tab:remoptphot} of Appendix~\ref{app:rem}. In the common g'r'i' bands, we found it to agree well with that obtained at LCOGT with a tighter sampling rate,  and therefore, we did not use it further in the analysis below. The individual JHK' measurements are listed in Table~\ref{tab:remphot}. The median errors on JHK' measurements are 0.05, 0.05, and 0.06~mag, respectively. However, some measurements were affected by the nearby bright moon around JD 2,459,512, and had an error larger than 0.1 mag. We chose to discard these measurements. We eventually derived the following values for the median near-infrared magnitudes of the GM Aur system and their rms variations at the time of the observations: J=9.41 $\pm$ 0.10 mag, H=  8.71 $\pm$ 0.03 mag, and K'=8.40 $\pm$ 0.07 mag.  

\begin{table}
\tiny
\caption{REM JHK photometry. }             
\label{tab:remphot}      
\centering                          
\begin{tabular}{l l | l l | l l}        
\hline\hline                        
\noalign{\smallskip}
Julian date&J&Julian date&H&Julian date&K\\
(2,450,000+) & (mag) &  (2,450,000+)& (mag) &(2,450,000+)  & (mag)\\
\hline
\noalign{\smallskip}
9497.73017 & 9.4 & 9497.73222 & 8.73 & 9497.73431 & 8.43\\
9497.73084 & 9.41 & 9497.73289 & 8.69 & 9497.73497 & 8.39\\
9500.80130 & 9.39 & 9500.80335 & 8.62 & 9500.80543 & 8.37\\
9500.80197 & 9.36 & 9500.80402 & 8.67 & 9500.80609 & 8.36\\
9501.80968 & 9.41 & 9501.81171 & 8.67 & 9501.81379 & 8.39\\
9501.81034 & 9.38 & 9501.81238 & 8.69 & 9501.81445 & 8.4\\
9502.83416 & 9.57 & 9502.83554 & 8.69 & 9502.83764 & 8.56\\
9503.83843 & 9.66 & 9502.83621 & 8.71 & 9502.83830 & 8.49\\
9504.84267 & 9.72 & 9503.83980 & 8.72 & 9507.79811 & 8.34\\
9507.79393 & 9.43 & 9503.84047 & 8.72 & 9507.79877 & 8.35\\
9507.79459 & 9.4 & 9504.84406 & 8.75 & 9509.81133 & 8.23\\
9509.80714 & 9.28 & 9504.84473 & 8.74 & 9509.81198 & 8.27\\
9513.71118 & 9.36 & 9506.79183 & 8.75 & 9513.71545 & 8.26\\
9515.79210 & 9.35 & 9506.79250 & 8.73 & 9513.71611 & 8.3\\
9515.79276 & 9.41 & 9517.72696 & 8.72 & 9515.79695 & 8.33\\
9517.72494 & 9.42 & 9517.72762 & 8.7 & 9517.72906 & 8.42\\
9517.72559 & 9.42 & 9518.73394 & 8.7 & 9517.72972 & 8.4\\
9518.73191 & 9.43 & 9518.73461 & 8.69 & 9518.73604 & 8.4\\
9518.73256 & 9.44 & 9519.74207 & 8.71 & 9518.73670 & 8.43\\
9519.74005 & 9.41 & 9519.74274 & 8.71 & 9519.74416 & 8.43\\
9519.74071 & 9.41 & 9520.74628 & 8.68 & 9519.74482 & 8.41\\
9520.74424 & 9.38 & 9520.74695 & 8.68 & 9520.74837 & 8.42\\
9520.74490 & 9.37 &  --- & --- & 9520.74903 & 8.39\\
\hline                                   
\end{tabular}
\end{table}

\subsection{OHP SOPHIE: High-resolution optical spectroscopy}
\label{sec:ohp}


Observations were carried out from October 12 to 29, 2021, at Observatoire de Haute-Provence using the fiber-fed SOPHIE spectrograph \citep{Perruchot08} in high-efficiency mode, which delivers a spectral resolution of R $\sim$ 40,000 over the wavelength range 387-694 nm. We obtained 15 spectra over 18 nights, with an exposure time of 3600~s, yielding a signal-to-noise ratio ranging from 42 to 67 at 600 nm.  
The raw spectra were fully reduced at the telescope by the SOPHIE real-time pipeline \citep{Bouchy09}. The data products include a resampled 1D spectrum with a constant wavelength step of 0.01~\AA, corrected for barycentric radial velocity, an order-by-order estimate of the signal-to-noise ratio, and a measurement of the source radial velocity, \vrad, and projected rotational velocity, \vsini. The latter two quantities are derived from a cross-correlation analysis of nearly 7,000 spectral lines between the observed spectrum and a K5 spectral mask template \citep[e.g.,][]{Boisse10}. We list the values of these parameters in Table~\ref{ohp_obs}. The mean formal error provided by the SOPHIE pipe-line on the \vrad\ measurement is 0.013~\kms. This accuracy is well suited to investigating the significantly larger amplitude of photospheric line profile variability induced by surface spots and/or accretion flows in young stars \citep[e.g.,][]{Petrov01}. No error is provided by the pipeline for the \vsini\ measurements, and we assumed that an upper limit is given by the rms deviation of the  individual measurements, excluding JD 2,459,512 (see below), namely 0.38~\kms.

The SOPHIE spectrograph includes a second fiber that simultaneously records the spectrum of the nearby sky. Inspection of the cross-correlation function (CCF) of the sky fiber with a synthetic mask of spectral type G2 revealed that the signature of the moon becomes apparent at the expected barycentric Earth radial velocity from JD 2,459,505 onward because the growing moon approaches the target. The lunar contamination culminates on JD 2,459,512, as the bright moon is located about 10 degrees away from GM Aur, which explains the discrepant values measured for \vrad\ and \vsini\ on this date. Table~\ref{ohp_obs} suggests that except for JD 2,459,512, the contamination of the CCF by the moon only marginally impacts the \vrad\ and \vsini\ measurements. However, to be conservative, we only considered the \vrad\ and \vsini\ measurements obtained from the first six spectra of the observing run for the subsequent analysis, from  JD 2,459,499 to JD 2,459,504, where no lunar contamination is present. 

The journal of observations is given in Table~\ref{ohp_obs}. It lists the Julian date, the signal-to-noise ratio of individual spectra at 600 nm, the radial and rotational velocities derived from each spectrum, and the bisector span computed from the cross-correlation function \citep{Queloz01}.

\begin{table}
\caption{Journal of OHP/SOPHIE observations. }             
\label{ohp_obs}      
\centering                          
\begin{tabular}{l l l l l  }        
\hline\hline                        
\noalign{\smallskip}
Julian date  &  S/N  &  \vrad  &  \vsini  &  CCF span  \\
(2,450,000+) &  & \kms & \kms &  \kms \\
\hline                                   
\noalign{\smallskip}
9499.5168  &  52  &  14.03  &  12.12  &  0.39  \\
9500.5044  &  44  &  15.17  &  12.04  &  -0.25  \\
9501.6560  &  44  &  14.94  &  11.99  &  0.36  \\
9502.5564  &  65  &  15.14  &  12.3  &  -0.77   \\
9503.6296  &  59  &  14.91  &  12.24  &  -0.44   \\
9504.6045  &  67  &  14.5  &  12.78  &  0.21   \\
9505.5450  &  63  &  14.89  &  12.49  &  -0.22   \\
9506.4838  &  59  &  15.59  &  12.46  &  -0.67  \\
9510.5691  &  61  &  15.07  &  12.96  &  -0.33  \\
9511.5284  &  67  &  15.23  &  12.08  &  -0.75  \\
9512.5162  &  50  &  16.85$^\dagger$  &  9.74$^\dagger$  &  -2.19$^\dagger$  \\
9513.6282  &  52  &  15.13  &  11.76  &  0.09  \\
9514.5918  &  59  &  14.98  &  12.06  &  -0.85  \\
9515.6107  &  42  &  15.19  &  11.53  &  -1.02  \\
9516.5321  &  49  &  14.46  &  12.7  &  0.24  \\
\hline                                   
\end{tabular}

\noindent $^\dagger$ Uncertain value due to lunar contamination (see text).
\end{table}

\subsection{CFHT SPIRou: Near-infrared spectropolarimetry}
\label{sec:spirouobs}



Near-infrared spectropolarimetric observations of GM Aur were performed at CFHT using the SPIRou near-infrared spectropolarimeter. It has a spectral range covering from 0.95 to 2.50 $\mu$m in a single exposure at a spectral resolution of~70,000 \citep{Donati20a}. The observations were completed 
in the framework of the CFHT SPIRou Legacy Survey over four observing runs extending from September 15 to December 18, 2021. Each monthly run was scheduled around the full moon, and we aimed at obtaining one spectrum per night during the run. An additional single spectrum was obtained on January 6, 2022. We thus gathered 34 spectra, whose temporal sampling is shown in Fig.~\ref{obs_samp}. 
 
Each spectrum consists of four polarimetric sub-exposures\footnote{The polarimetric analysis of the dataset will be published in a companion paper (Zaire et al., in prep.).} for a total integration time of 2,200~s. Individual exposures were combined to yield a single spectrum with a signal-to-noise ratio ranging from $\sim$60 to 100. 
In one instance, on JD 2,459,503.08, the polarimetric sequence was aborted, and the spectrum consists of a single sub-exposure. The raw data were reduced within the SPIRou consortium, using version V6.132 of the APERO pipeline \citep{Cook22}. Spectra were cross-correlated with a K2 spectral mask template over about 6,700 spectral lines,  and the radial velocity of the object was derived with sub-\kms\ accuracy (0.08~\kms\  mean rms uncertainty) by fitting a Gaussian to the resulting CCF. The median \vrad\ amounts to 14.65 $\pm$ 0.27~\kms. Using TWA 9A, a WTTS of spectral type K5, as a template, the veiling was derived in the JHK bands following the procedure described in \cite{Sousa23} The median rms errors on veiling measurements in the JHK bands are 0.011, 0.025, and 0.027, respectively.  

The journal of observations is presented in Table~\ref{spirou_obs}. It lists the Julian date,
the signal-to-noise ratio at 2.16 $\mu$m, the photospheric radial velocity and its uncertainty, and the veiling in the JHK bands.

 
 \begin{table}
\caption{Journal of CFHT/SPIRou observations. }             
\label{spirou_obs}      
\centering                          
\begin{tabular}{l l l l l l l}        
\hline\hline                        
\noalign{\smallskip}
Julian date  &  S/N  &  \vrad  & $\sigma$\vrad & r$_J$ & r$_H$ & r$_K$ \\
(2,450,000+) &  & \multicolumn{2}{c}{\kms}   \\
\hline                                   
\noalign{\smallskip}
\multicolumn{6}{c}{September 2021}\\
\hline
\noalign{\smallskip}
9473.066   &  101  & 14.77  &  0.06  &  0.05  &  0.05  &  0.27  \\
9475.043   &  109  & 14.09  &  0.17  &  0.05  &  0.06  &  0.22  \\
9476.969   &  122  & 14.88  &  0.16  &  0.12  &  0.09  &  0.27  \\
9478.027   &  119  & 14.85  &  0.04  &  0.17  &  0.14  &  0.38  \\
9480.082   &  121  & 14.62  &  0.09  &  0.05  &  0.09  &  0.29  \\
9481.086   &  113  & 14.42  &  0.08  &  0.06  &  0.10  &  0.23  \\
9482.086   &  120  & 14.70  &  0.06  &  0.05  &  0.07  &  0.23  \\
\hline                                   
\noalign{\smallskip}
\multicolumn{6}{c}{October 2021}\\
\hline
\noalign{\smallskip}
9502.074   &  112  & 14.97  &  0.06  &  0.15  &  0.14  &  0.32  \\
9503.074$^\dagger$     &  64  & 14.48  &  ---  &  0.26  &  0.15  &  0.07  \\
9504.098   &  94  & 14.61  &  0.05  &  0.10  &  0.11  &  0.30  \\
9506.082   &  112  & 14.65  &  0.01  &  0.04  &  0.12  &  0.29  \\
9508.082   &  113  & 14.68  &  0.08  &  0.18  &  0.24  &  0.43  \\
9509.086   &  110  & 14.85  &  0.06  &  0.32  &  0.26  &  0.51  \\
9510.086   &  92  & 14.40  &  0.12  &  0.41  &  0.29  &  0.58  \\
9511.074   &  116  & 14.33  &  0.05  &  0.02  &  -0.04  &  0.28  \\
9513.090   &  98  & 14.70  &  0.25  &  0.09  &  0.13  &  0.29  \\
9514.051   &  105  & 14.66  &  0.16  &  0.18  &  0.12  &  0.36  \\
9515.078   &  108  & 14.66  &  0.03  &  0.10  &  0.11  &  0.28  \\
9516.102   &  104  & 14.73  &  0.16  &  0.10  &  0.09  &  0.26  \\
\hline                                   
\noalign{\smallskip}
\multicolumn{6}{c}{November 2021}\\
\hline
\noalign{\smallskip}
9535.102   &  102  & 14.27  &  0.09  &  0.13  &  0.07  &  0.25  \\
9537.098   &  101  & 14.95  &  0.15  &  0.11  &  0.08  &  0.26  \\
9538.039   &  104  & 14.22  &  0.10  &  0.05  &  0.11  &  0.22  \\
9538.984   &  95  & 14.40  &  0.10  &  0.12  &  0.08  &  0.28  \\
9539.988   &  107  & 14.51  &  0.03  &  0.14  &  0.13  &  0.27  \\
9541.012   &  110  & 13.96  &  0.10  &  0.14  &  0.14  &  0.38  \\
\hline                                   
\noalign{\smallskip}
\multicolumn{6}{c}{December 2021}\\
\hline
\noalign{\smallskip}
9557.980   &  110  & 14.42  &  0.11  &  0.22  &  0.19  &  0.50  \\
9558.949   &  110  & 14.12  &  0.03  &  0.21  &  0.19  &  0.45  \\
9559.961   &  95  & 14.71  &  0.11  &  0.09  &  0.11  &  0.35  \\
9560.965   &  94  & 15.04  &  0.06  &  0.07  &  0.12  &  0.35  \\
9563.078   &  106  & 14.73  &  0.06  &  0.21  &  0.06  &  0.43  \\
9564.059   &  102  & 14.34  &  0.10  &  0.32  &  0.11  &  0.43  \\
9566.051   &  96  & 14.37  &  0.13  &  0.07  &  0.03  &  0.34  \\
9566.953   &  95  & 14.93  &  0.03  &  0.06  &  0.02  &  0.25  \\
\hline                                   
\noalign{\smallskip}
\multicolumn{6}{c}{January 2022}\\
\hline
\noalign{\smallskip}
9585.941   &  112  & 14.30  &  0.06  &  0.11  &  0.12  &  0.25  \\
\hline                                   
\end{tabular}

\noindent $^\dagger$ Single sub-exposure.
\end{table}

\subsection{ESO ExTrA: Low-resolution near-infrared spectroscopy}


The ExTrA facility \citep{bonfils15}, located at La Silla Observatory in Chile, consists of three 60 cm telescopes and a single near-infrared (0.88 to 1.55~$\mu$m) fibre-fed spectrograph. We observed GM Aur on 89 nights between October 13, 2021, and March 8, 2022, using either one telescope (21 nights) or two telescopes simultaneously (68 nights). Five fiber units are located at the focal plane of each telescope, each consisting of two 8\arcsec\ aperture fibers. One fiber is used to observe a star and the other is used to observe the nearby sky background. We observed GM Aur with one fiber unit and used another fiber unit to simultaneously observe 2MASS J04535474+3021441 ($J=8.450 \pm 0.027$ mag) as a comparison star to compute differential photometry. We used the higher-resolution mode of the spectrograph (R$\sim$200) and 300-second exposures. We obtained between 1 and 30 exposures per night for a total of 1898 spectra with a median signal-to-noise ratio of 105 at 1.05~$\mu$m for GM Aur. The ExTrA data were corrected for dark current, extracted using the flat-field, corrected for sky background emission, and were wavelength calibrated using custom data reduction software. Median spectra of GM Aur were computed for each night and telescope, yielding a total of 157 spectra with a median signal-to-noise ratio of 179 and a standard deviation of 62. 

We computed differential photometry of GM Aur relative to the comparison star by integrating the individual ExTrA spectra over the J-filter passband. The UKIRT-WFCAM filter transmission curves were retrieved from the SVO Filter Profile Service\footnote{\url{http://svo2.cab.inta-csic.es/theory/fps/}} \citep{rodrigo12,rodrigo20}. 
We multiplied each corrected individual spectrum of GM Aur and the comparison star by the filter transmission curve, integrated the flux, and obtained a magnitude difference from the flux ratio. We computed a differential magnitude measurement for each night as the mean and standard deviation of the individual measurements taken on that night. We then derived the J-band magnitude of GM Aur from the 2MASS magnitude of the comparison star. We obtained a median value and 68.3\% confidence interval of $J = 9.417\pm0.061$ mag for the ExTrA observations. The results are listed in Table~\ref{tab:extraewJ} of Appendix~\ref{app:extraewJ}.

\section{Results}
\label{sec:results}

\subsection{Multicolor photometry}
\label{sec:resphot}


  \begin{figure*}
   \centering
   \includegraphics[width=0.95\hsize]{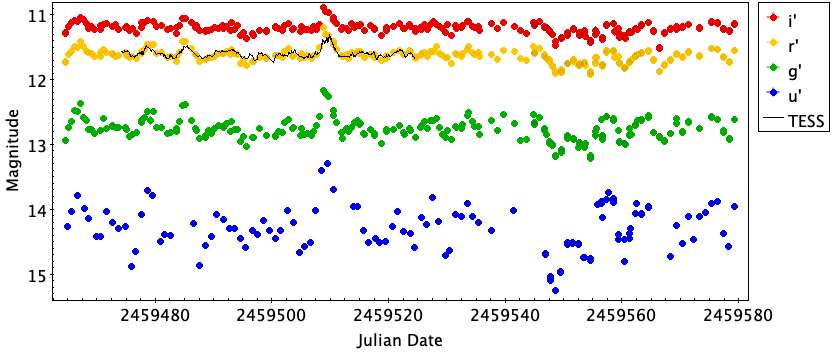}
    \includegraphics[width=0.95\hsize]{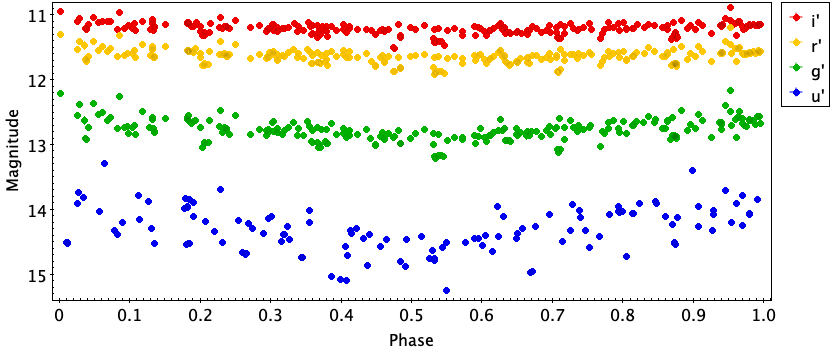}
     \includegraphics[width=0.95\hsize]{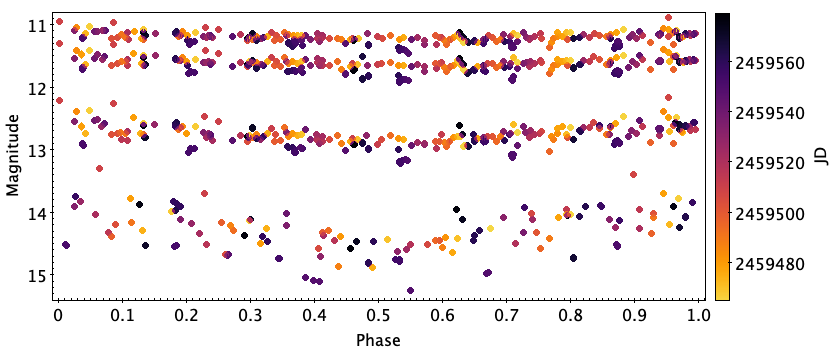}
\caption{GM Aur light curves. {\it Top:} GM AUr's u'g'r'i' light curves from LCOGT observations that extended over nearly four months.  The mean photometric error is 0.025 mag in the g'r'i' bands and 0.033 mag in the u' band. A scaled TESS light curve ({\it black}) obtained contemporaneously is overplotted onto the LCOGT r'-band light curve. {\it Middle:} LCOGT light curves folded in phase with a period of 6.04 days with the ephemeris of Eq.\ref{eq:ephem}. {\it Bottom:} Same as above, with a color scale for data points that reflects the Julian date. The amplitude of the light variations appears to increase slightly toward the end of the observing run. In all panels, the error bars on the measurements are smaller than the symbol size.}
              \label{fig:lcogt_lc}%
    \end{figure*}

  \begin{figure}
   \centering
   \includegraphics[width=\hsize]{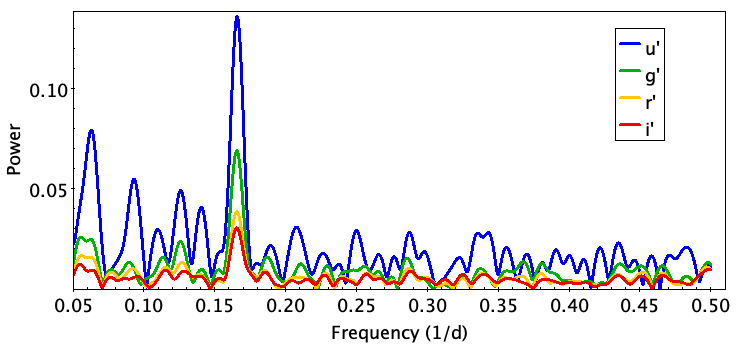}
    \includegraphics[width=\hsize]{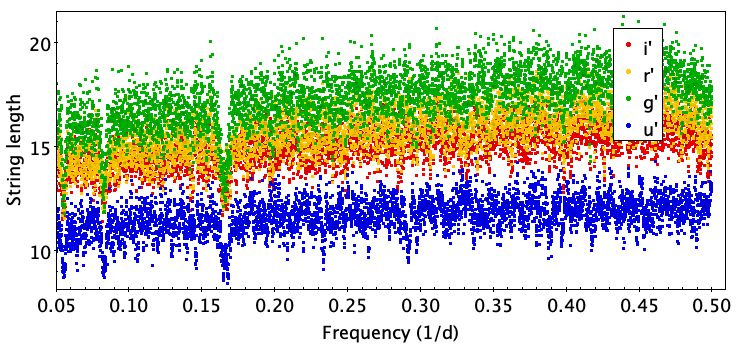}
\caption{Period search results. {\it Top:} CLEAN periodogram analysis of the u'g'r'i' light curves. A peak occurs at the frequency of 0.166 day$^{-1}$, which corresponds to a period of 6.04 $\pm$ 0.15 days.  {\it Bottom:} String-length analysis of the u'g'r'i' light curves. A clear minimum appears for a period of 6.03 $\pm$ 0.15 days.}
              \label{lcogt_per}%
    \end{figure}

    \begin{figure}
   \centering
   \includegraphics[width=0.9\hsize]{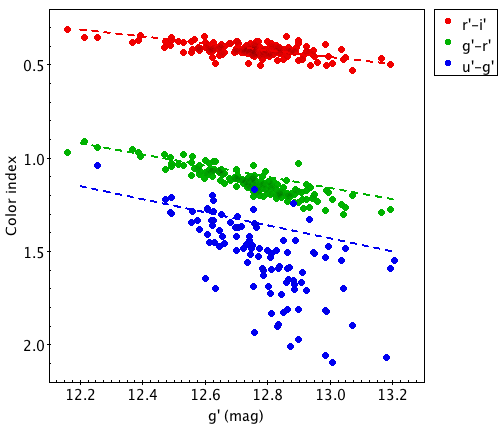}
\caption{Color-magnitude relation. The (u'-g'), (g'-r'), and (r'-i') colors are plotted as a function of the system brightness in the g'-band. The dashed lines indicate the expected ISM reddening slope computed from \cite{Fiorucci03} with R=3.1.}
              \label{lcogt_colors}%
    \end{figure}
   
The LCOGT u'g'r'i' light curves of GM Aur span nearly four months and are shown in Fig.~\ref{fig:lcogt_lc}. A TESS light curve extending over 50 days, from JD 2,459,474 to JD 2,459,524, is also shown. We derive a mean magnitude and rms variability of u'=14.32 $\pm$ 0.35 mag, g'=12.77 $\pm$ 0.17 mag, r'=11.64 $\pm$ 0.11 mag, and i'=11.21 $\pm$ 0.09 mag, respectively. The variability amplitude is much larger than the photometric errors in all bands; it amounts to 0.025 mag in the g'r'i' bands and 0.033 mag in the u' band, and it decreases toward longer wavelengths, from u' to i'. We ran two period-search algorithms on the light curves: a CLEAN periodogram analysis \citep{Roberts87}, which is conceptually similar to a Fourier transform, and the String-Length method \citep{Dworetsky83}, which finds the period that minimizes the average distance between consecutive points in the phased light curve. Both yielded the same period at all wavelengths, namely P=6.04 $\pm$ 0.15 days, with the uncertainty being estimated from the standard deviation of a Gaussian fitted to the periodogram peak.  The results of the period search are shown in Fig.~\ref{lcogt_per}. The light curves folded in phase at the 6.04~d period are shown in Fig.~\ref{fig:lcogt_lc}. They clearly show the modulation of the brightness level at this period, particularly in the u' band, where the amplitude is largest. We ascribe this low-level modulation to surface spots and the P=6.04 d period to stellar rotation. This estimate agrees within the error bars with the previously reported periods for GM Aur, namely P=6.1 d by \cite{Percy10} and P=6.02 d by \cite{Artemenko12}. We therefore adopt the following ephemeris: \begin{equation} HJD (d) = 2,459,460.80 + 6.04\times E,\label{eq:ephem}\end{equation} which defines the rotational phase \phirot=(HJD-2,459,460.80)/P$_{rot}$ (modulo P$_{rot}$), where phase zero (\phirot = 0) is chosen as the epoch of maximum optical brightness in the rotational cycle. 

Superimposed onto spot modulation, additional signs of intrinsic variability are visible. This is the case, for instance, of an apparent brightness event that is centered on JD 2,459,509 and lasted for a few days, as well as more pronounced wide dips toward the end of the observing period, from JD 2,459,544 on. Interestingly, the first half of the light curve exhibits several day-long brightening events, while the last third is dominated by wide dimming events. We note that most of the brightening events, with g'$\leq$12.5 mag, tend to occur at rotational phases shorter than 0.15 or longer than 0.85, that is, close to the maximum brightness of the spot modulation. If the photometric variations of the system are modulated by the visibility of a bright accretion spot at the stellar surface, these brightening events seen around the time of maximum accretion shock visibility most likely reflect varying accretion on the stellar surface. The wide dips in the last part of the light curve exhibit the same periodicity and phase as the spot modulation. They reach a minimum brightness close to phase 0.5, with an amplitude that steadily decreases from 0.6 mag to 0.3 mag in the g' band over a timescale of a few weeks, from JD 2,459,550 to JD 2,459,570. 

Fig.~\ref{lcogt_colors} shows the color behavior of the system. As the system fades, it becomes redder, with a color slope larger than expected from ISM-like extinction at least in the (u'-g') and (g'-r') color indices. According to \cite{Venuti15}, a large color slope at short wavelengths is characteristics of accretion-driven photometric variability. However, both the large scatter seen in the (u'-g') color index at low brightness levels and the changing shape of the light curve during the semester suggest that several sources of variability might be present, such as a combination of accretion and obscuration events. 

\subsection{High-resolution optical spectroscopy}


We took advantage of the wide wavelength range covered by the SOPHIE spectrograph at high spectral resolution to derive the stellar parameters of GM Aur, to measure optical veiling, and to investigate the emission line profiles and their variability across the optical range. These results are described in the next subsections.  
    
\subsubsection{Stellar parameters} 
\label{sec:starprop}


\begin{table}
\caption{Properties of the GM Aur system.  }   
\label{tab:starprop}      
\centering                          
\begin{tabular}{ll}        
\hline                     
\hline
Parameter & Value \\
\hline
SpT & K4-K5\\
A$_V$ & 0.3 $\pm$ 0.3 mag\\
\teff & 4287 $\pm$ 35~K\\
\lstar/\lsun & 0.9 $\pm$ 0.2\\
\rstar/\rsun & 1.7 $\pm$ 0.2 \\
\mstar/\msun & 0.95 $\pm$ 0.13 \\
\macc &  0.7 $\pm$ 0.3 10$^{-8}$ \msunyr\\
EW(LiI) &  420 $\pm$ 23 m\AA \\
\vsini & 14.9 $\pm$ 0.3~\kms\\
\prot & 6.04 $\pm$ 0.15~d\\
\rcor &  8.3 $\pm$  0.5~\rstar (0.064~au)\\
\hline                                   
\end{tabular}
\end{table}

To derive the stellar parameters (\teff, \vsini, \vrad, and \vmic), we fit synthetic ZEEMAN spectra \citep{Landstreet88, Wade01, Folsom12} to the average of the first six SOPHIE spectra, which are not contaminated by the moon. Synthetic spectra were computed from MARCS atmosphere models \citep{Gustafsson08} and the VALD line list database \citep{Ryabchikova15}. We applied a $\chi^2$ minimization procedure based on a Levenberg-Marquart algorithm over seven wavelength windows ranging from 455 to 649~nm, excluding the regions affected by tellurics, emission, or molecular lines. 
We set the macroturbulent velocity to 2.0 \kms \ and  the surface gravity \logg\ to 4.0, and we assumed  solar metallicity. These are typical parameters for low-mass TTSs. During the fitting procedure, we applied the mean veiling value derived below for the GM Aur mean optical spectrum ($r_{0.55}$ = 0.3, see Sect.~\ref{sophieveiling}) to the synthetic spectra. Finally, we averaged the results obtained from the various wavelength windows, except for one window that yielded values higher than 2$\sigma$ from the mean. We thus derived \teff\ = 4287 $\pm$ 35 K, \vrad\  = 14.94 $\pm$ 0.14 \kms, \vmic = 3.3 $\pm$ 0.4 \kms, and \vsini\ = 14.9 $\pm$ 0.3 \kms. 
We also used the ROTFIT package \citep{Frasca03, Frasca06} applied to the mean GM Aur's SOPHIE spectrum, from which we derive \teff\ = 4505 $\pm$ 53 K, \vrad\  = 15.37 $\pm$ 0.26 \kms, and \vsini\ = 13.0 $\pm$ 0.7 \kms. 

While all these values are consistent within 3$\sigma$, the large \teff\ difference derived from ZEEMAN and ROTFIT illustrates model-dependent uncertainties that are likely related to the use of different model templates (MARCs for ZEEMAN vs. BTSetll for ROTFIT) and possibly to wavelength-dependent systematics induced by starspots \citep{Gangi22, Flores22}. For consistency with similar observing campaigns that we previously performed on young stars \citep[e.g.][]{Pouilly20, Pouilly21, Bouvier20b, Alencar18}, we adopted the results of the ZEEMAN fitting. This estimate also agrees better with the K6 spectral type derived by \cite{Herczeg14}, from which they deduced \teff = 4115~K. When the \teff-SpT conversion tables from \cite{Herczeg14} and \cite{Pecaut13} are used, the \teff\ value derived above corresponds to a K4-K5 spectral type, which agrees fairly well with previous estimates obtained from optical and near-infrared spectroscopy \citep[K5-K6; e.g.,][]{Espaillat10, Herczeg14}.

The \vrad\ and \vsini\ values can be compared to those derived from the uncontaminated CCF of the first six SOPHIE spectra, namely $<$\vrad$>$ = 14.79 $\pm$ 0.40 \kms\ and $<$\vsini$>$ = 12.25 $\pm$ 0.26 \kms. The quoted uncertainties are the rms of the six measurements and therefore include intrinsic variability of the CCF profiles due to spot modulation, for example. The two estimates of \vrad\ agree within the errors as well as with the median \vrad\ value derived from the SPIRou spectra (see Section~\ref{sec:spirouobs}), while the \vsini\ value derived from the CCF is significantly lower than the value deduced from spectral fitting. We suspect that the discrepancy may arise from the color-dependent FWHM-\vsini\ relation used for SOPHIE CCFs, which is calibrated on main-sequence stars \citep{Boisse10} and may not be fully adequate for pre-main-sequence objects.

From the average g'r'i' colors reported for GM Aur in Section~\ref{sec:lcogt}, the intrinsic colors of a K4-K5 dwarf listed by \cite{Kraus07} and \cite{Covey07}, and using ISM extinction coefficients from \cite{Fiorucci03}, we derive a visual extinction on the line of sight A$_V$ = 0.3 $\pm$ 0.3 mag. This agrees with previous determinations \citep[e.g.,][]{Herczeg14}. The REM photometry reported above yields a median J-band magnitude of 9.42 $\pm$ 0.11 mag, which is close to that of 2MASS (J=9.34 mag). We dereddened the median J-band magnitude with A$_j$ = 0.28 $\times$ A$_{V}$ = 0.084 $\pm$ 0.084 mag and used the J-band bolometric correction listed in \cite{Pecaut13} for a K4-K5 dwarf, BC$_J$=1.55 $\pm$ 0.03 mag, to derive the stellar luminosity, \lstar = 0.9 $\pm$ 0.2 \lsun, assuming the Gaia distance of 157.9 $\pm$ 1.2 pc \citep{Gaia21}. We thus derive a stellar radius \rstar = 1.7 $\pm$ 0.2 \rsun\ from Stefan's law, and a stellar mass \mstar = 1.05 $\pm$ 0.05 \msun\ from the \cite{Siess00} pre-main-sequence evolution models, while CESAM models \citep{Marques13} yield  \mstar = 0.88 $\pm$ 0.12 \msun\ (E. Al\'ecian, priv. comm.). We therefore adopt \mstar = 0.95 $\pm$ 0.13 \msun, in agreement with \cite{Baraffe15} models. This estimate is also consistent within the errors with the dynamical mass estimate, M$_{dyn}$ = 1.00 $\pm$ 0.02 \msun, reported by \cite{Guilloteau14}, which was later revised to M$_{dyn}$ = 1.14 $\pm$ 0.02 \msun\ by \cite{Simon19}.  Table~\ref{tab:starprop} summarizes the derived stellar parameters.


Finally, we combined the \vsini\ and rotational period measurements with the stellar radius estimate to derive the stellar inclination $\sin i$ = \prot\ $\times$ \vsini\ / (2 $\pi$ \rstar) = 1.05 with the values listed in Table~\ref{tab:starprop}.  Accounting for 1$\sigma$ uncertainties on the stellar parameters, we derive a lower limit of $i_\star \geq$ 63\degr\ for the stellar rotational axis onto the line of sight.  This value is significantly higher than the inclination inferred from high-resolution ALMA images of the outer disk of GM Aur observed at millimeter wavelength, which yield $i_{disk}$ = 53.2 deg \citep{Huang20}.  It agrees better, however, with the inclination value derived for the disk seen in scattered light with adaptive optics on a scale of a few dozen au, for which \cite{oh16} obtained  $i_{disk}$ = 64 $\pm$ 2 deg. On the much smaller scale of 0.013 au, \cite{Bohn22} measured an inner-disk inclination $i_{disk}$ = 68$^{+16}_{-28}$ deg from long-baseline K-band interferometry using VLTI/GRAVITY.  They did not find evidence for an inner and outer disk misalignment in this system. As outlined by \cite{Appenzeller13}, the uncertainty on the determination of stellar inclinations from rotation measurements rapidly increases at large angles and is prone to systematic errors. For GM Aur, inferring the stellar inclination from the disk inclination might therefore be more reliable than estimating it from rotation measurements, owing in particular to the significant uncertainty on the stellar radius. Nevertheless, all independent measurements indicate a moderate to high inclination for the system. 


\subsubsection{Veiling} 
\label{sophieveiling}

  \begin{figure}
   \centering
   \includegraphics[width=\hsize]{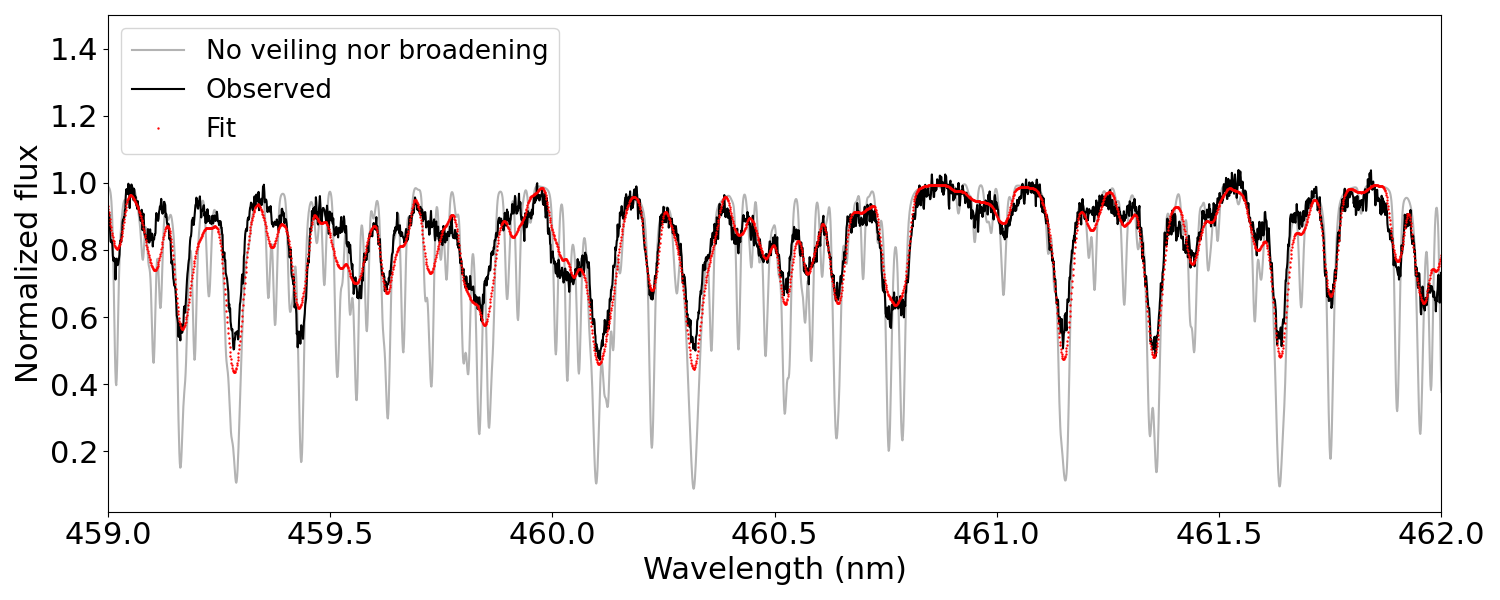}
      \includegraphics[width=\hsize]{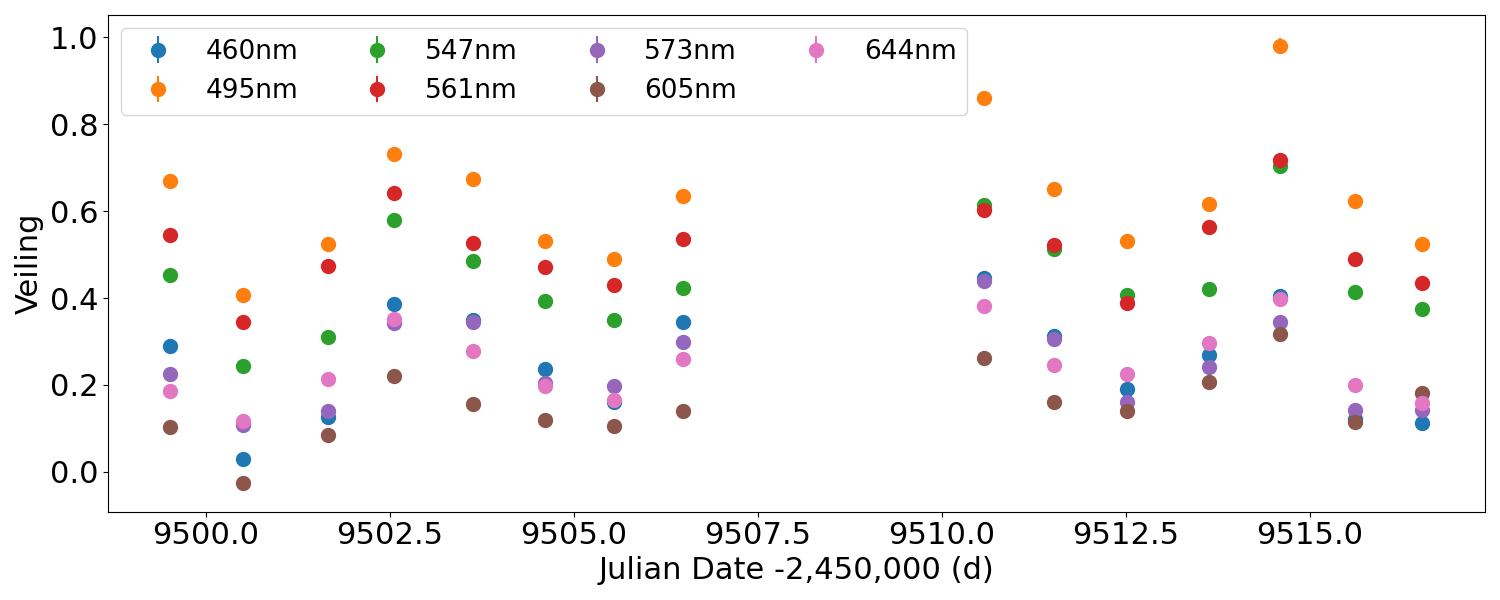}
\caption{Optical veiling measurements from the OHP/SOPHIE spectra. {\it Top:} Part of a spectral window showing the template spectrum ({\it gray}), the observed spectrum ({\it black}), and the velocity broadened and veiled template spectrum ({\it red}) that fits the observed spectrum. {\it Bottom:}  Veiling measured in several spectral windows (see text) as a function of Julian date. The central wavelength of the spectral windows is indicated in the top left corner of the panel.}
              \label{fig:ohpveiling}%
    \end{figure}

  \begin{figure}
   \centering
   \includegraphics[width=\hsize]{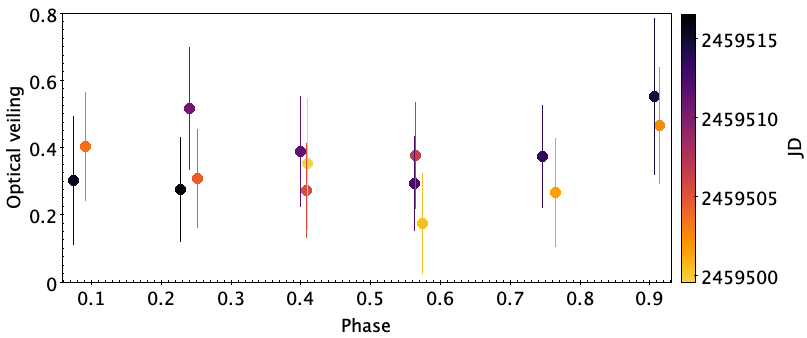}
\caption{Mean optical veiling  plotted as a function of rotational phase. The color code indicates the Julian date.}
              \label{fig:ohpveiling_phase}%
    \end{figure}

   \begin{figure}
   \centering
      \includegraphics[width=0.495\hsize]{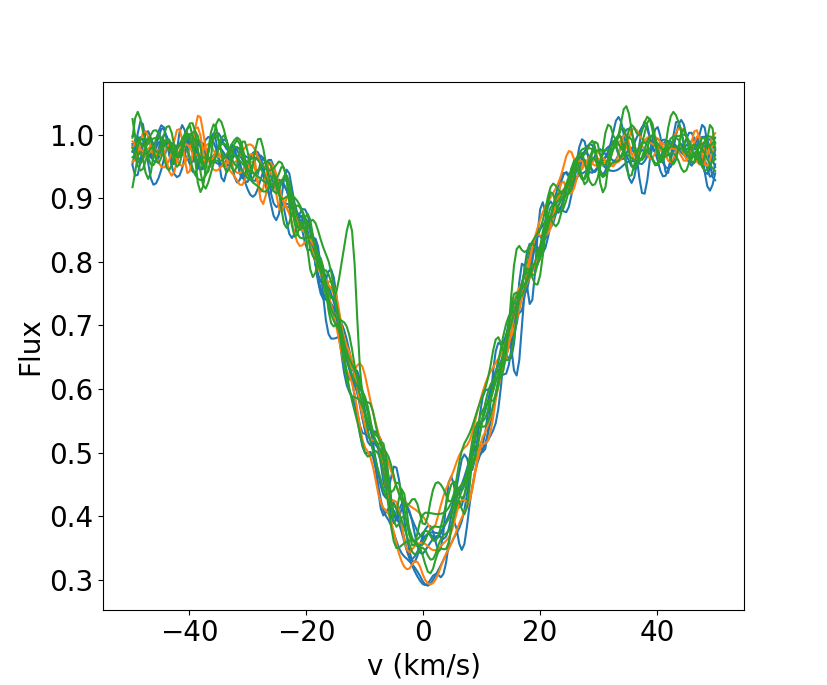}
      \includegraphics[width=0.495\hsize]{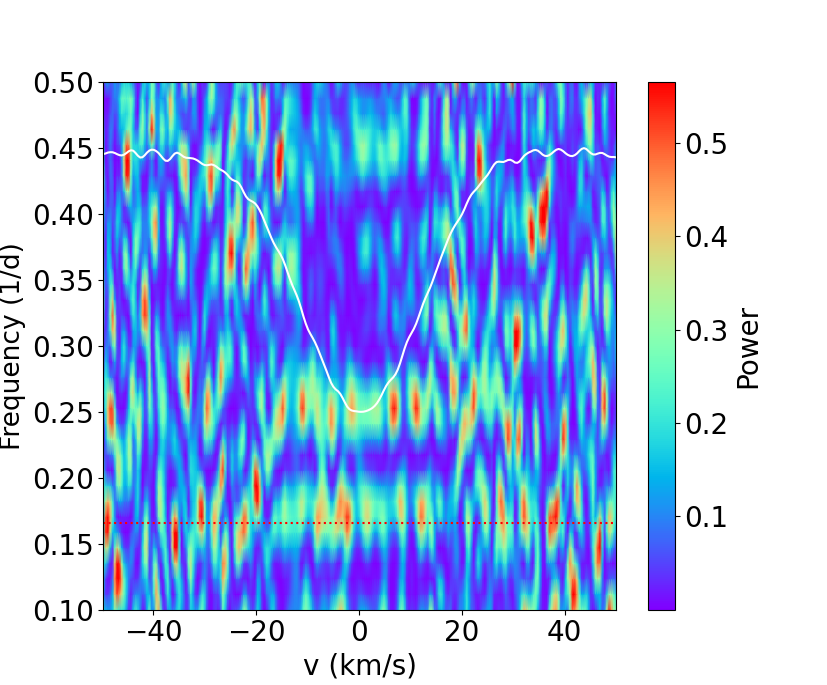}
   \caption{LiI 6707 \AA\ line profile.  {\it Left:} The 34 line profile measurements from SOPHIE spectra are shown superimposed. The color code corresponds to successive rotational cycles. {\it Right:}  2D periodogram across the line profile. The dotted horizontal line drawn at a frequency of 0.166 day$^{-1}$ indicates the stellar rotational period. The white curve displays the mean line profile. The color code reflects the periodogram power. }
              \label{fig:li}%
    \end{figure}

At optical wavelengths, accreting T Tauri stars exhibit an additional source of continuum flux, which  presumably arises from the accretion shock at the stellar surface. This optical excess continuum  partly veils the stellar photospheric spectrum and is therefore referred to as "veiling" \citep{Hartigan95}.
We measured the optical veiling from the high-resolution OHP/SOPHIE spectra by comparing the photospheric spectrum of GM Aur to that of the unveiled nonaccreting template V819 Tau,  a WTTS with \teff = 4250 $\pm$ 50 K, \vrad\ = 16.6 \kms, and \vsini\ = 9.5 \kms\ \citep{Donati15}. We retrieved an archival CFHT/ESPaDOnS spectrum of V819 Tau, which we resampled at the spectral resolution of OHP/SOPHIE spectra, translated into the radial velocity of GM Aur, and rotationally broadened using the rotational function from \cite{Gray73} to match the rotational velocity of GM Aur. The template spectrum was then fit to the GM Aur spectra over the same wavelength windows as discussed in Section~\ref{sec:starprop} by adjusting the veiling using the following formula:

\begin{equation}
    I = \frac{I_0 + r}{1 + r},
\end{equation}

\noindent where $I$ is the veiled spectrum, $I_0$ is the spectrum without veiling, and $r$ is the continuum veiling. \cite{Rei18} showed that an additional line-dependent veiling component may be present in the strongest photospheric lines. We therefore retained only weak to moderate lines with an EW between 0.01 and 0.1~\AA\  to perform the fit. The veiling measured for each spectrum in each spectral window is shown in Figure~\ref{fig:ohpveiling}. Systematic offsets are clearly seen between the veiling values measured in the different spectral windows. These offsets may partly reflect a wavelength-dependent veiling, as veiling seems to decrease toward longer wavelengths for all but the bluest spectral window, but it may also result from systematic errors depending on the specific sample of lines included in each window. We therefore computed an average veiling value, $r_{0.55}$, over all spectral windows for individual GM Aur spectra. This value is listed in Table~\ref{tab:ewohp} with its associated uncertainty. The optical veiling is moderate, ranging from 0.17 to 0.55 at 5500 \AA. Similar mean values were derived from ROTFIT, namely $r$ = 0.5, 0.4, and 0.3 at 4500, 6000, and 6500 \AA, respectively, using a library of 400 spectral templates with spectral types FGKM from the OHP/ELODIE database, which yielded a best $\chi^2$ fit for spectral types ranging from K3.5 to K5.  Figure~\ref{fig:ohpveiling_phase} shows the mean optical veiling plotted along the rotational phase. A hint of higher veiling values towards phases 0 and 1, and minimum values around phase 0.5-0.6 appears. A periodogram analysis of the mean veiling variation did not yield significant results, however. 

The lack of significant veiling variability is confirmed by the examination of the LiI~6707~\AA\  photospheric line profile shown in Fig.~\ref{fig:li}. The shape and depth of the line appear to remain quite stable over the duration of the observing run, and a periodogram analysis across the line profile \citep{Giampapa93} revealed no signs of periodic modulation. This suggests that the source of optical veiling remains at least partly in view throughout the rotational cycle. This might arise from a high-latitude accretion shock.


\subsubsection{Emission lines} 
\label{ressophielines}

\begin{table}
\caption{Optical line EW and veiling measurements from the OHP/SOPHIE spectra.}             
\label{tab:ewohp}      
\centering                          
\begin{tabular}{l l l l l l l}        
\hline\hline                        
\noalign{\smallskip}
Julian date & \multicolumn{3}{c}{EW} & \multicolumn{2}{c}{Veiling} \\
&  \ha  &  \hb & \hei &r$_{0.55}$ & \it{rms} \\ 
(2,450,000+) & \AA & \AA & \AA  \\
\hline                                   
\noalign{\smallskip}
9499.51683266 & 92.3 & 9.2 & 0.31  &  0.35  &  0.19 \\
9500.50445956 & 81.1 & 6.8 & 0.37  &  0.17  &  0.15 \\
9501.65604543 & 86 & 10.3 & 0.5  &  0.26  &  0.16 \\
9502.55640645 & 73.3 & 7.4 & 0.45  &  0.46  &  0.17 \\
9503.62969684 & 79.6 & 8.5 & 0.4  &  0.4  &  0.15 \\
9504.60450624 & 88.1 & 10.5 & 0.44  &  0.3  &  0.14 \\
9505.54502964 & 85.8 & 10.2 & 0.23  &  0.27  &  0.13 \\
9506.48382743 & 86.4 & 9 & 0.33  &  0.37  &  0.15 \\
9510.56919423 & 102 & 15.1 & 0.74  &  0.51  &  0.18 \\
9511.52849738 & 91.1 & 12.6 & 0.34  &  0.38  &  0.16 \\
9512.51625045 & 68.3$^\dagger$  & 3.7$^\dagger$  & 0.18$^\dagger$   &  0.29$^\dagger$   &  0.13$^\dagger$  \\
9513.62825832 & 82.6 & 10.1 & 0.36  &  0.37  &  0.15 \\
9514.59185217 & 81.5 & 9 & 0.5  &  0.55  &  0.23 \\
9515.61072618 & 76.8 & 7.7 & 0.42  &  0.3  &  0.19 \\
9516.53216217 & 74.7 & 9.6 & 0.36  &  0.27  &  0.15 \\
\hline                                   
\end{tabular}

\noindent $^\dagger$ Uncertain value due to lunar contamination (see text).
\end{table}
   \begin{figure*}
   \centering
   \includegraphics[width=0.33\hsize]{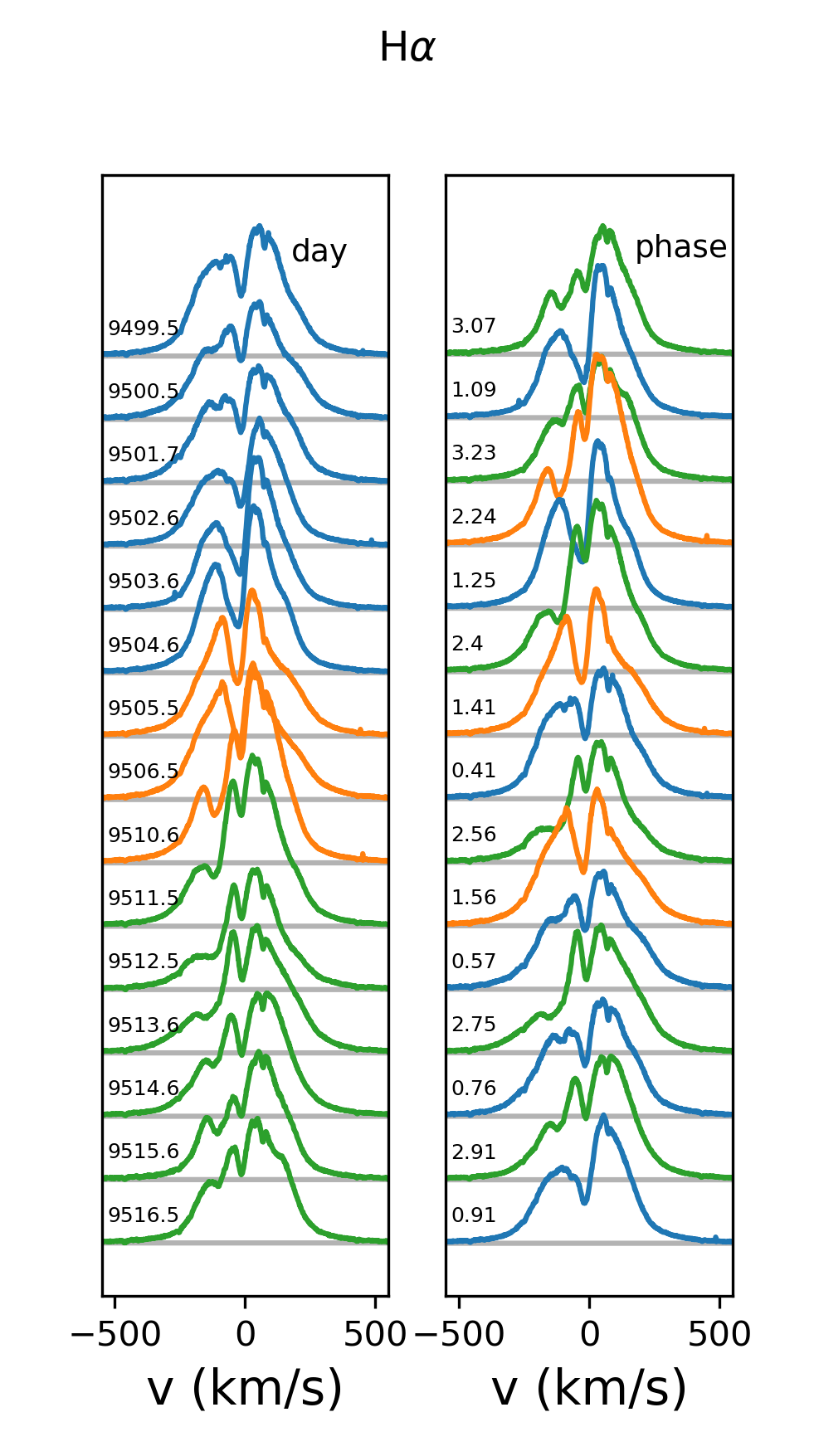}
      \includegraphics[width=0.33\hsize]{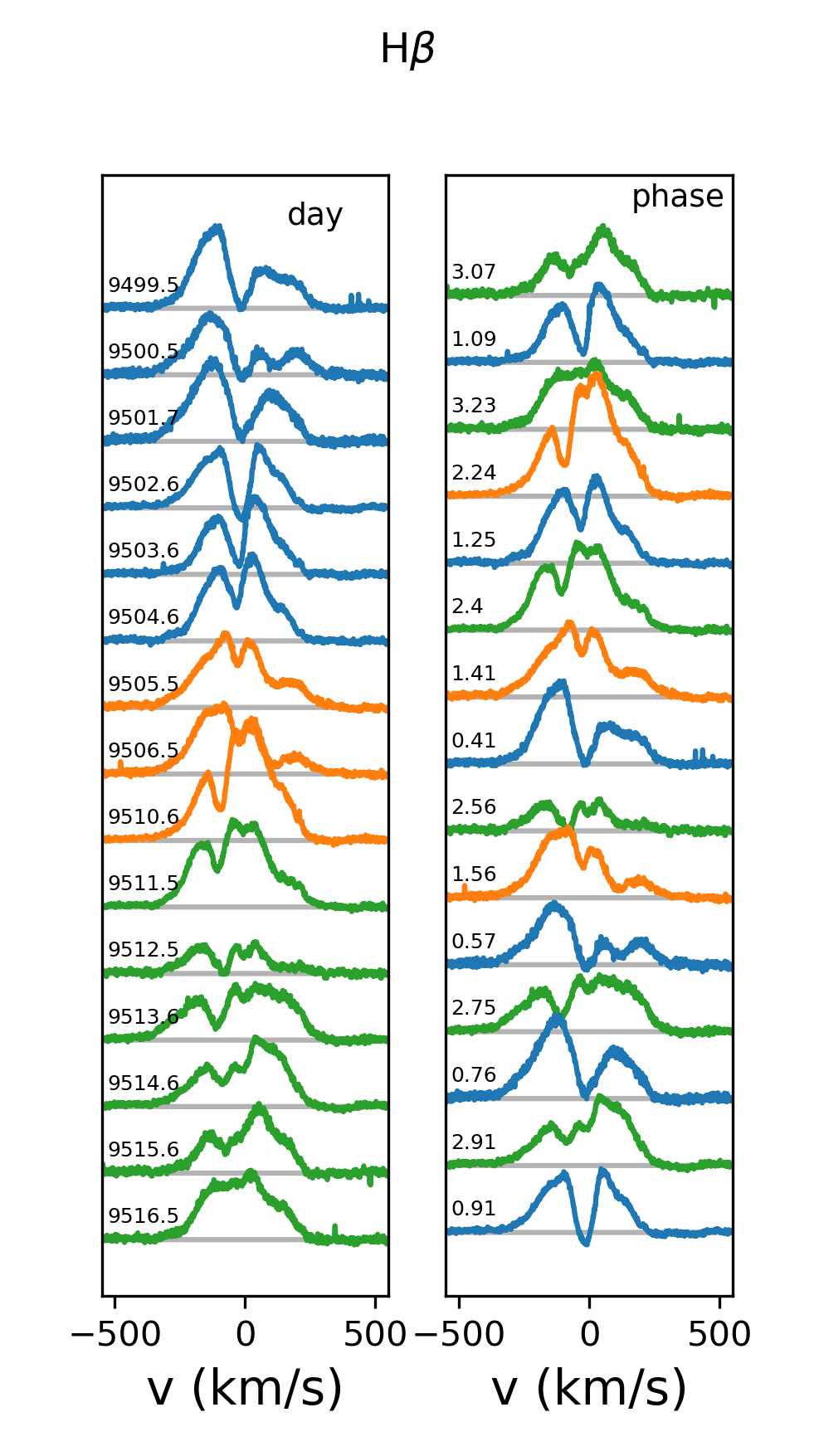}
      \includegraphics[width=0.33\hsize]{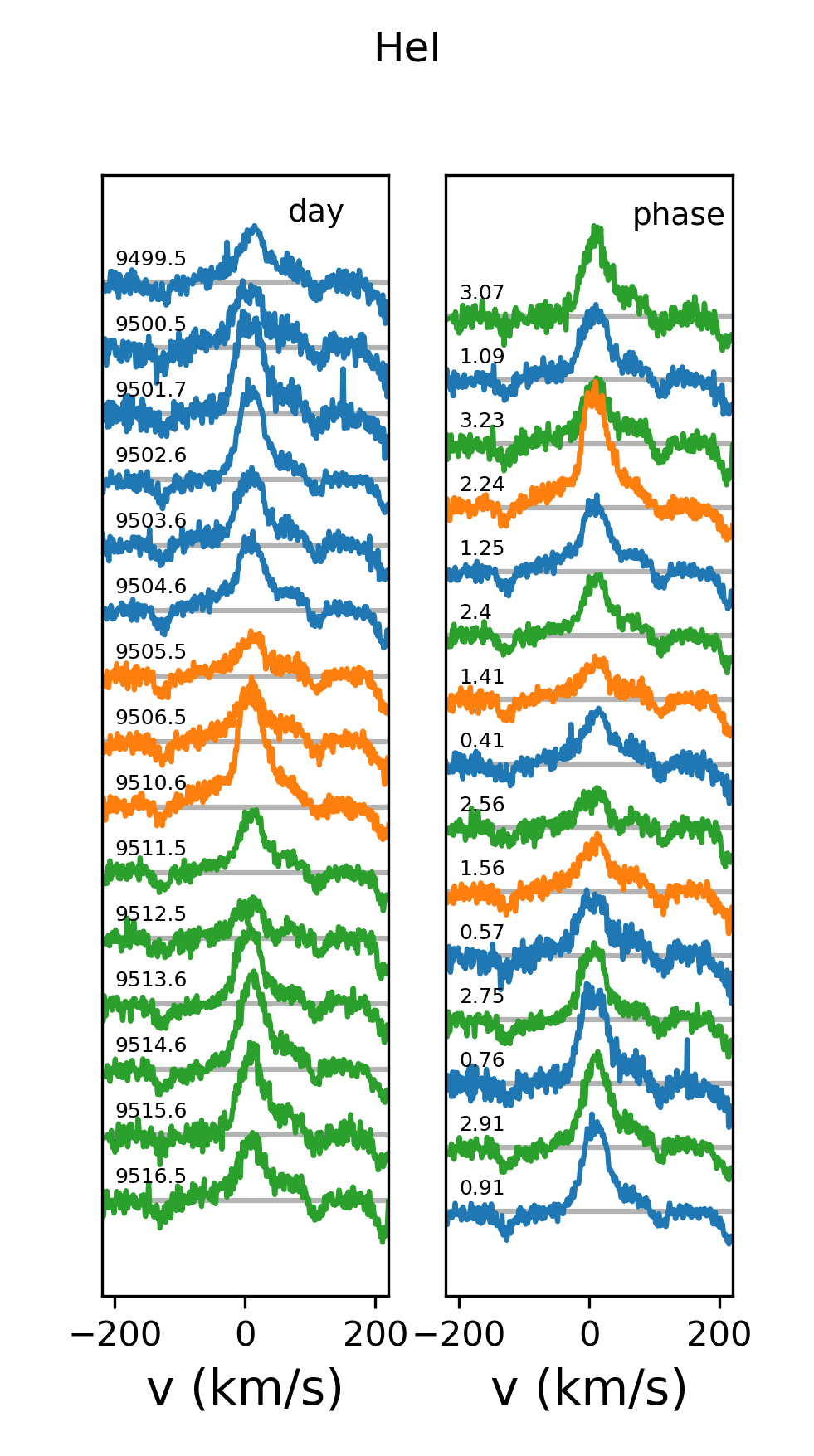}

   \includegraphics[width=0.33\hsize]{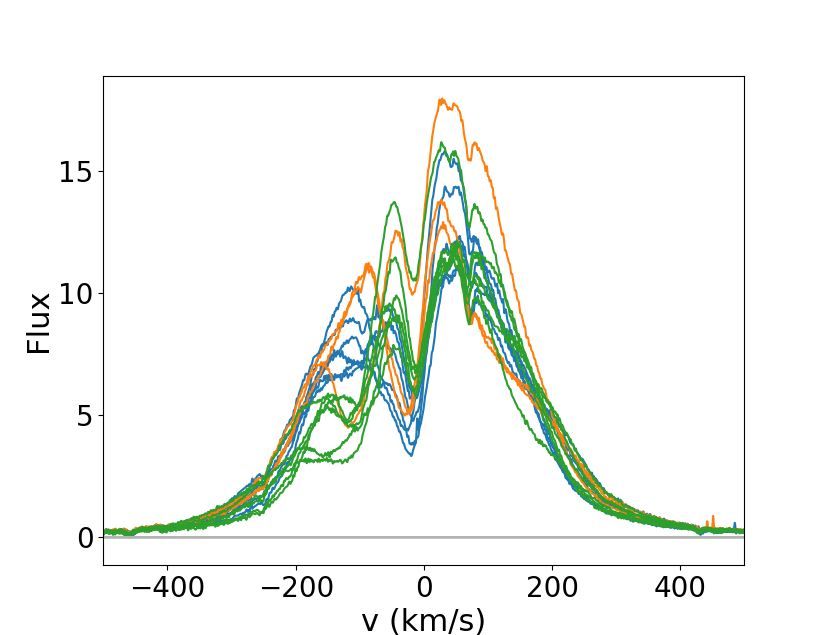}
      \includegraphics[width=0.33\hsize]{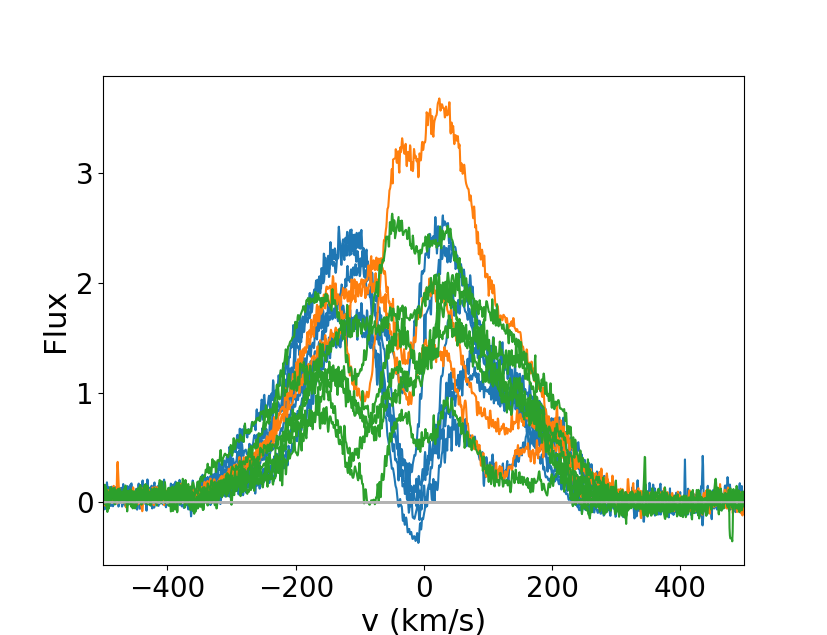}
      \includegraphics[width=0.33\hsize]{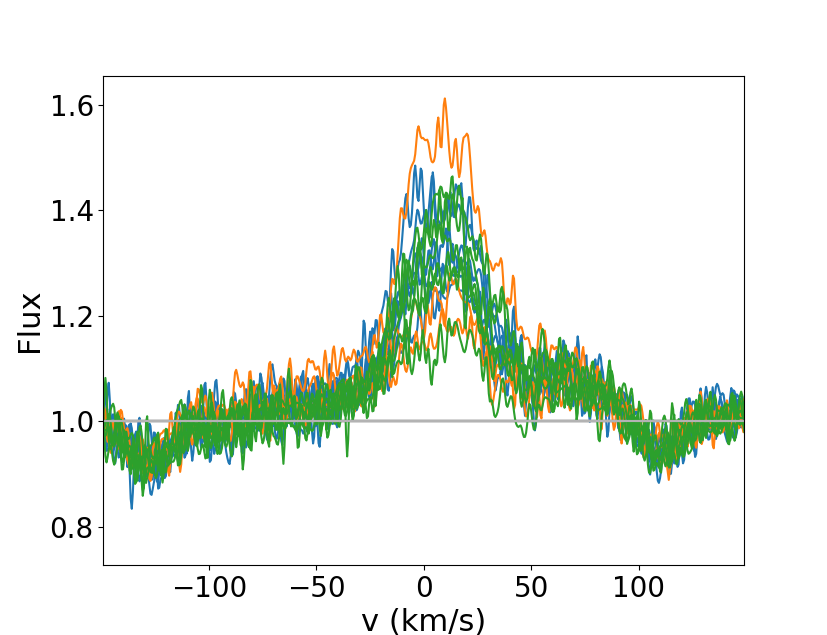}

   \includegraphics[width=0.33\hsize]{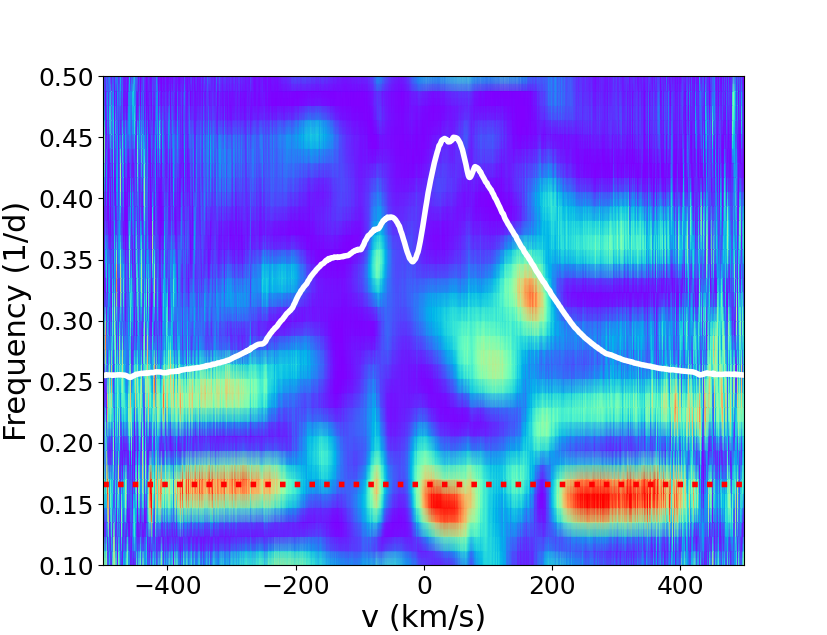}
      \includegraphics[width=0.33\hsize]{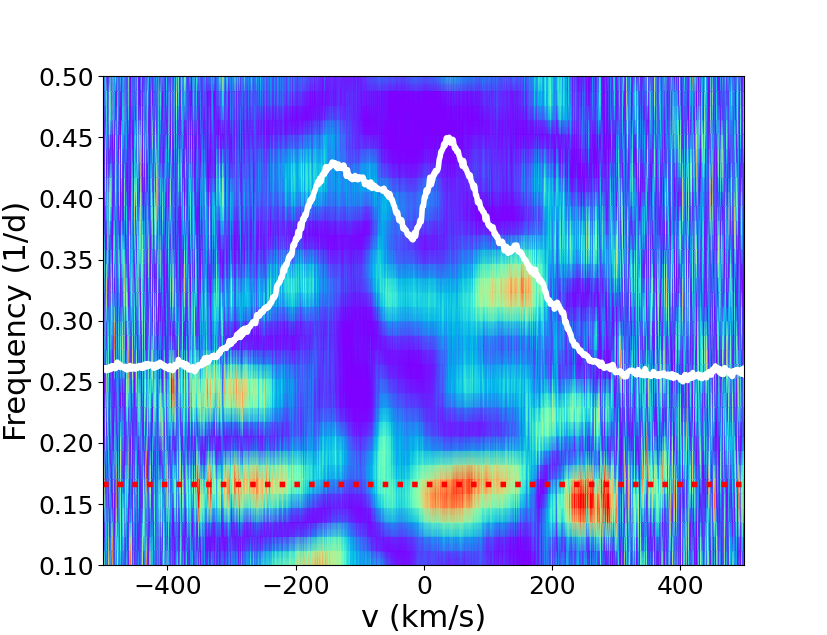}
      \includegraphics[width=0.33\hsize]{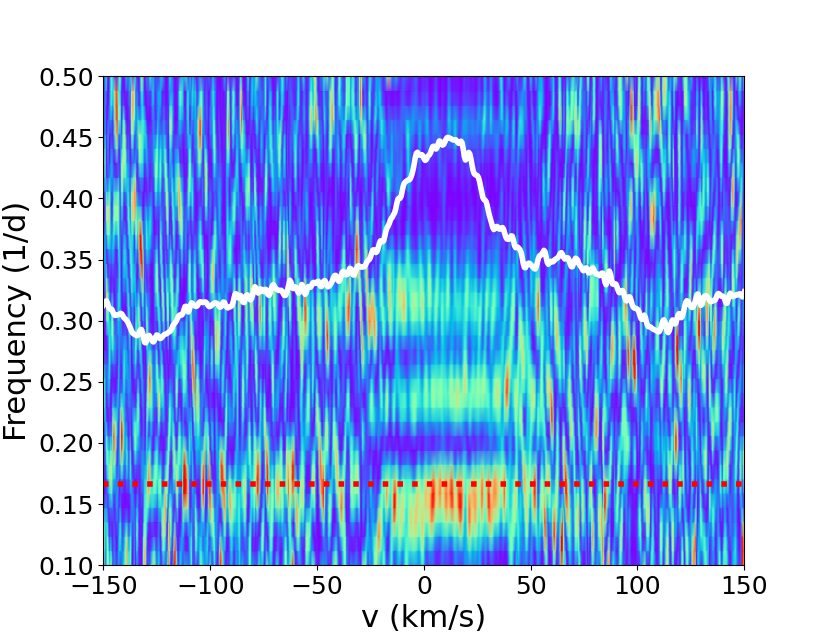}
      
   \caption{Optical line profile variability. {\it Top:} Series of optical line profiles \ha, \hb, and \hei\  plotted as a function of Julian date ({\it left subpanels}) and rotational phase ({\it right subpanels}). The color code corresponds to successive rotational cycles. {\it Middle:}  Same profiles, superimposed to illustrate their variability.  {\it Bottom:} 2D periodograms across the line profiles. The color code reflects the periodogram power, from zero ({\it blue}) to 0.6 ({\it red}). The dotted horizontal red line drawn at a frequency of 0.166 day$^{-1}$ indicates the stellar rotational period. The white curve is the mean line profile.}
              \label{fig:sophieprof}%
    \end{figure*}

The main emission lines seen in the optical spectrum of GM Aur, namely \ha, \hb, and \hei\ 5876~\AA, are displayed in Figure~\ref{fig:sophieprof}\footnote{The [OI] 6300 \AA\ line is also seen in emission in the spectrum. However, it is significantly contaminated by sky, which cannot be easily corrected for due to the different response of the object and sky fibers of the SOPHIE spectrograph. The other emission lines seen in the high signal-to-noise ratio mean SOPHIE spectrum are Ca II H\&K, \hc, \hd, \he, and He I 6678 \AA, the latter being weak and affected by a deep photospheric FeI line.}. The Balmer lines show a broad emission peak and a slightly blueshifted absorption component, whose depth varies from being hardly discernable to reaching below the continuum in the \hb\ profile. The wide, slightly blueshifted absorption component peaks at -30 to -20 \kms and covers a velocity range from about -90 to +40 \kms\ in the \hb\ line profile. Additional absorption components appear at higher blueshifted velocities, peaking from -110 to -90 \kms, and extending down to about -160~\kms. These blueshifted absorption components cause most of the variability in the  Balmer line profile. The \hb\ profile also exhibits significant variability over the red wing, up to velocities of +200~\kms. However, none of the profiles exhibits redshifted absorption components reaching below the continuum level. 

Owing to the complex line shapes, equivalent widths (EW) of the \ha, \hb, and \hei\ 5876~\AA\ lines were computed by directly integrating below the line profile. The results are listed in Table~\ref{tab:ewohp}. The measurement accuracy is 10\% or better for EW(\ha) and EW(\hb), and about 20\% for EW(\hei) due to the more uncertain continuum location. We note that on JD 2,459,512, the EWs measurements are systematically lower than during the rest of the observations, which might be due to lunar contamination, as discussed in Section~\ref{sec:ohp}.  The average and rms values we obtain are EW(\ha) = 83 $\pm$ 8~\AA, EW(\hb) = 9.3 $\pm$ 2.5~\AA, and EW(\hei) = 0.40 $\pm$ 0.13~\AA. 


The \hei\ 5876~\AA\ line profile is roughly symmetric and consists of a narrow component (FWHM $\sim$ 30-40~\kms) superimposed on a broad pedestal, as previously reported by \cite{McGinnis20}. The peak intensity of the narrow component varies significantly, while the broad component appears relatively stable (see Fig.~\ref{fig:sophieprof}). We fit the HeI line profile with a two-component Gaussian model to extract the properties of the narrow (NC) and broad (BC) components. The EWs were derived from the Gaussian fit of the components. The radial velocity, FWHM, and EW of the NC and BC, as well as their uncertainty derived from the covariance matrix of the Levenberg-Marquart fitting  algorithm, are listed in Table~\ref{tab:heincbc}.

   \begin{figure}[t]
   \centering
      \includegraphics[width=\hsize]{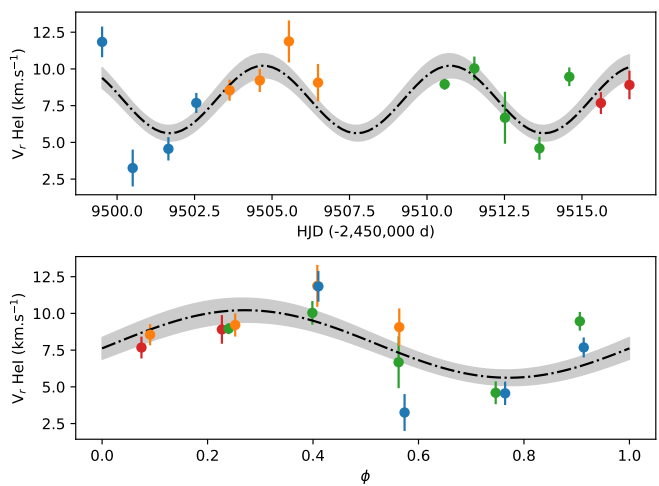}
   \caption{Radial velocity of the narrow component of the \hei\ 5876~\AA\ line profile in the stellar rest frame plotted as a function of Julian date ({\it top}) and rotational phase computed from the ephemeris of Eq.\ref{eq:ephem} ({\it bottom}). The color code corresponds to successive rotational cycles. The fit by a geometrical accretion shock  model (see text) is shown ({\it dash-dotted curve}) together with its 1$\sigma$ uncertainty ({\it gray area}).}
              \label{fig:ohpheivrad}%
    \end{figure}
    
   \begin{table*}
\caption{Properties of the narrow (NC) and broad (BC) components of the HeI 5876~\AA\ line profile and their 1$\sigma$ error.   }             
\label{tab:heincbc}      
\centering                          
  \scriptsize
\begin{tabular}{l l l l lllllll l l}        
\hline\hline                        
\noalign{\smallskip}
Julian date & \multicolumn{4}{c}{\vrad} & \multicolumn{4}{c}{FWHM}  & \multicolumn{4}{c}{EW} \\
 (2,450,000+) & \multicolumn{4}{c}{\kms} & \multicolumn{4}{c}{\kms}  & \multicolumn{4}{c}{\AA} \\
 & NC & err & BC & err &  NC & err & BC & err & NC &err & BC & err\\
\hline                                   
\noalign{\smallskip}
9499.51683  &  11.8  &  1.0  &  14.8  &  1.8  &  27.7  &  3.6  &  100.7  &  6.7  &  0.07  &  0.02  &  0.30  &  0.05 \\
9500.50445  &  3.3  &  1.3  &  14.5  &  3.1  &  38.7  &  4.5  &  106.6  &  9.6  &  0.13  &  0.03  &  0.30  &  0.07 \\
9501.65604  &  4.6  &  0.8  &  20.0  &  3.7  &  40.7  &  2.7  &  91.0  &  7.4  &  0.28  &  0.04  &  0.29  &  0.08 \\
9502.55640  &  7.7  &  0.7  &  20.0  &  2.5  &  36.2  &  2.4  &  81.2  &  5.5  &  0.22  &  0.03  &  0.27  &  0.06 \\
9503.62969  &  8.6  &  0.7  &  12.9  &  2.3  &  37.6  &  2.8  &  113.2  &  9.4  &  0.18  &  0.03  &  0.30  &  0.06 \\
9504.60450  &  9.2  &  0.8  &  10.1  &  1.8  &  34.0  &  2.9  &  111.4  &  7.3  &  0.13  &  0.02  &  0.36  &  0.06 \\
9505.54503  &  11.9  &  1.4  &  8.7  &  2.5  &  28.2  &  4.9  &  101.5  &  9.3  &  0.06  &  0.02  &  0.22  &  0.05 \\
9506.48382  &  9.1  &  1.3  &  11.9  &  1.9  &  31.3  &  4.6  &  104.3  &  7.7  &  0.08  &  0.02  &  0.31  &  0.06 \\
9510.56919  &  9.0  &  0.4  &  11.0  &  1.3  &  36.3  &  1.4  &  120.7  &  5.4  &  0.29  &  0.02  &  0.51  &  0.06 \\
9511.52849  &  10.0  &  0.8  &  14.5  &  1.8  &  30.9  &  3.0  &  97.3  &  7.0  &  0.11  &  0.02  &  0.27  &  0.05 \\
9512.51624  &  6.7  &  1.8  &  7.6  &  4.5  &  38.6  &  7.7  &  99.3  &  21.6  &  0.08  &  0.04  &  0.14  &  0.09 \\
9513.62825  &  4.6  &  0.8  &  20.0  &  4.6  &  38.3  &  2.6  &  89.3  &  9.2  &  0.21  &  0.03  &  0.18  &  0.06 \\
9514.59185  &  9.5  &  0.6  &  20.0  &  1.8  &  33.6  &  2.3  &  90.1  &  5.0  &  0.19  &  0.03  &  0.36  &  0.06 \\
9515.61072  &  7.7  &  0.7  &  20.0  &  2.0  &  31.4  &  2.5  &  84.4  &  4.9  &  0.15  &  0.03  &  0.32  &  0.05 \\
9516.53216  &  8.9  &  1.0  &  15.5  &  2.0  &  29.4  &  3.3  &  102.5  &  7.1  &  0.10  &  0.02  &  0.32  &  0.05 \\
\hline                     
\hline                                   
\end{tabular}
\end{table*}

We derived the radial velocity of the narrow component and found it to be  variable and redshifted by $\sim$5-10 \kms\ relative to the stellar velocity. Figure~\ref{fig:ohpheivrad} shows the HeI NC radial velocity curve plotted as a function of Julian date and rotational phase. As previously reported by \cite{McGinnis20}, HeI NC \vrad\ appears to be rotationally modulated with a full amplitude of $\sim$6~\kms, as expected if the NC component of the HeI 5876~\AA\ line profile were produced in a high-latitude accretion shock at the stellar surface. We fit the observed NC \vrad\ curve with the geometrical accretion shock model described in \cite{Pouilly21}. The model computes the variation of the HeI NC radial velocity as the combination of the rotational modulation of the accretion shock and the intrinsic inflow velocity. The free parameters of the model are the inflow velocity, the latitude of the accretion shock, the phase at which it faces the observer, and the stellar inclination. The HeI NC \vrad\ curve is best reproduced with an accretion shock located at a latitude of 83 $\pm$ 1.5$\degr$ that faces the observer at phase 0.2 $\pm$ 0.08 and has a radial post-shock velocity of 18.3 $\pm$ 1.0 \kms\ in the stellar rest frame. Because the model now includes the inflow velocity, the HeI NC radial velocity curve does not reach the mean velocity at the time when the spot faces the observer, as it would for the case of static stellar spots. The stellar inclination  we derive from the model is $i$ = 64 $\pm$ 2.2$\degr$, which is consistent with the inner disk inclination derived from K-band VLTI/GRAVITY data (see Section~\ref{sec:starprop}). 
According to the model, the accretion shock faces the observer close to the origin of the phase, which is consistent with the photometric behavior described above (see Sect.~\ref{sec:resphot}). The \hei\ line profile is also strongest over rotational phases ranging from 0.75 to 1.09 (excluding the probable accretion burst occuring at JD  2,459,510, see Sect.~\ref{sec:resphot}), and this is also when the highest veiling values are measured in the JHK bands (see Fig.~\ref{fig:spirouveilew} in Appendix~\ref{app:nirveil}). These two results further support a maximum visibility of the accretion shock close to \phirot = 0.

A bidimensional periodogram analysis \citep{Giampapa93} of the Balmer and \hei\ 5876~\AA\ line profiles reveals a periodic modulation of part of the profiles (see Fig.~\ref{fig:sophieprof}). The intensity of the narrow component of the \hei\ line profile is modulated at a frequency of $\sim$0.15 d$^{-1}$, corresponding to a period of 6.7 d, with a large uncertainty due to the limited temporal sampling of the spectral series, however, that translates into a poor frequency resolution. As  the \hei\ line profile NC component arises in the accretion shock \citep{Beristain01}, it is expected to be modulated at the stellar rotational period or close to it in case of latitudinal differential rotation at the stellar surface. In the Balmer line profiles, a modulation close to the stellar period also appears over three distinct locations: in the highly redshifted wing over the velocity range $\sim$200-400~\kms, at slightly redshifted velocities between 0 and $\sim$50~\kms, and at highly blueshifted velocities from -400 to -200~\kms. While the maximum power of the periodogram in the blue wing appears at the stellar rotation period, it seems to drift to longer periods, similar to that seen in the \hei\ NC component, toward the red wing. If the red wing modulation of Balmer line profiles is caused by the absorption of shock emission by a funnel flow crossing the line of sight, this might be an indication of differential rotation along the funnel flow. Noticeably, no sign of periodic variability is seen in the Balmer line profiles in the velocity channels in which the variable blueshifted absorption components arise.  This suggests that these components either result from sporadic ejection processes  or that they vary on periods longer than ten days. 

Correlation matrices \citep{Johns95} between line profiles were computed and are presented in Appendix~\ref{app:cm}. They display the degree to which temporal flux variations in a pair of spectral lines are correlated. Matrices can be computed for a single profile (autocorrelation), for instance,  \ha$\star$\ha, to investigate how the different parts of the profile vary with respect to each other, or between two profiles (cross-correlation), for example, \ha$\star$\hb, to compare the intensity variations of different lines. The \ha$\star$\ha\ and \hb$\star$\hb\ matrices shown in Fig.~\ref{fig:cmhahb} are quite similar. 
The blue and red wings of the profiles vary in a correlated way. 
The high-velocity red wings are anticorrelated with the emission peak region. This may be a sign that high-velocity redshifted absorption components appear when the peak intensity is higher, although the absorptions do not reach below the continuum. Many absorption features are superimposed on the emission line component. These features do not present a periodicity and are not correlated with each other, nor with the emission part of the profile. The \hb$\star$\ha\ matrix mimics the matrices of  \ha$\star$\ha\ and the \hb$\star$\hb, showing that both lines are formed in the same region, as expected from magnetospheric accretion models \citep[e.g.,][]{Muzerolle01}.


\subsection{High-resolution near-infrared spectroscopy}


  \begin{figure*}
   \centering
      \includegraphics[width=0.9\hsize]{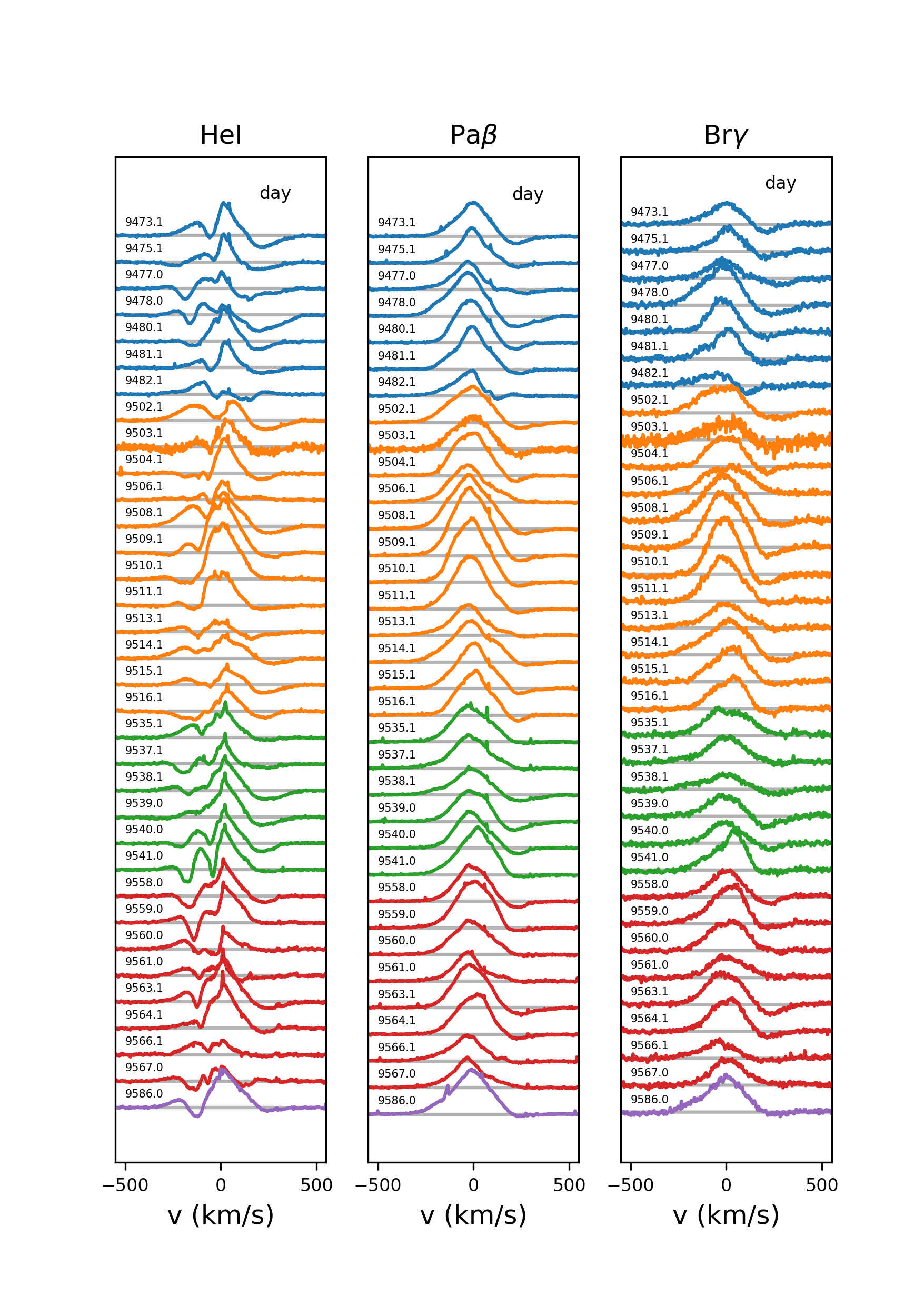}    
   \caption{Near-infrared line profiles: \hei\ ({\it left}), \pab\ ({\it center}), and \brg\ ({\it right}), plotted with arbitrary offsets as a function of Julian date. Each color represents a SPIRou run, namely September (blue), October (orange), November (green), December 2021 (red), and January 2022 (purple).  The October SPIRou run is contemporaneous to the OHP/SOPHIE observations.}
              \label{fig:spirou_prof}%
    \end{figure*}
    
    
   \begin{figure*}
   \centering
   \includegraphics[width=0.3\hsize]{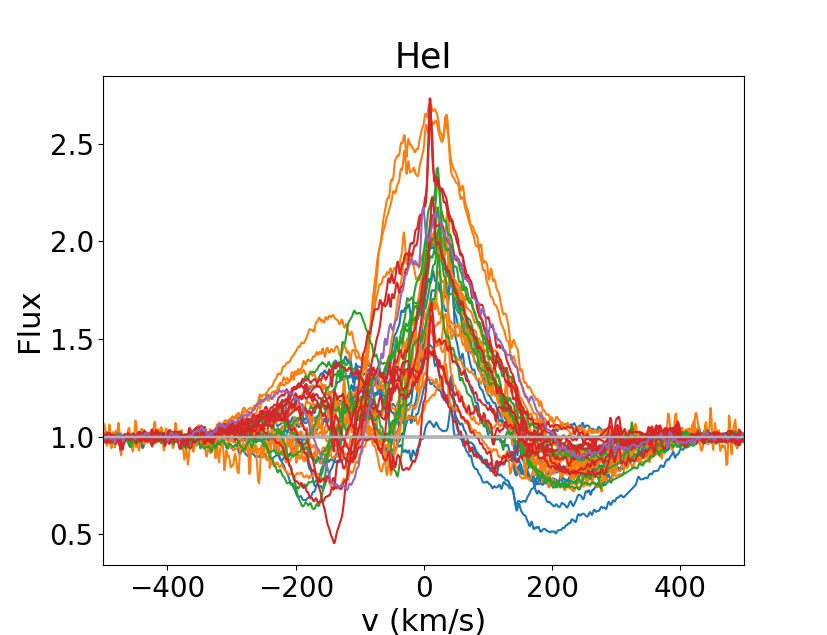}
    \includegraphics[width=0.3\hsize]{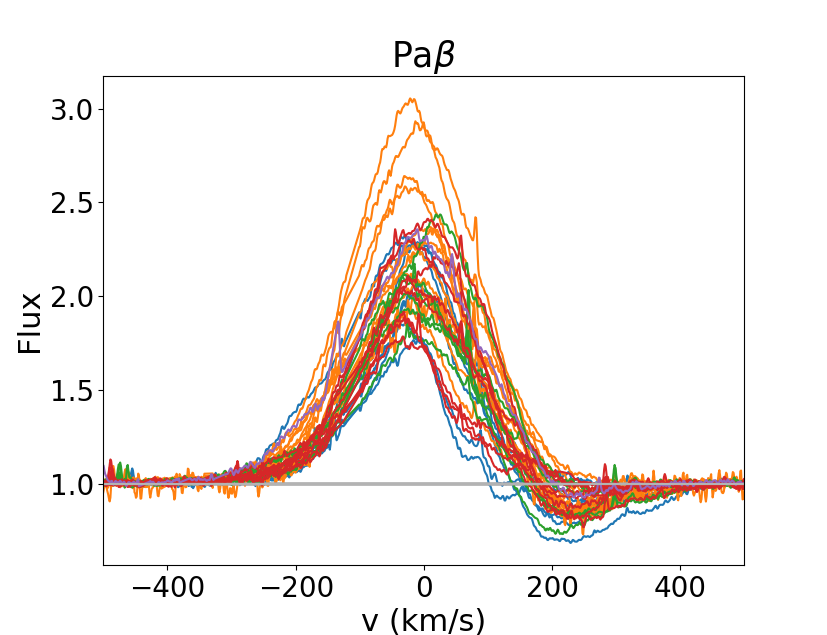}
    \includegraphics[width=0.3\hsize]{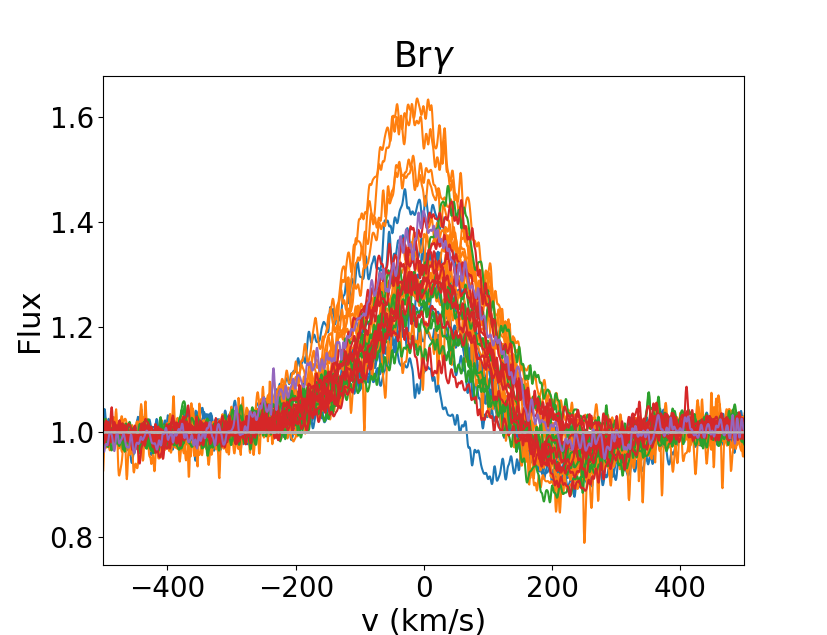}
   \includegraphics[width=0.3\hsize]{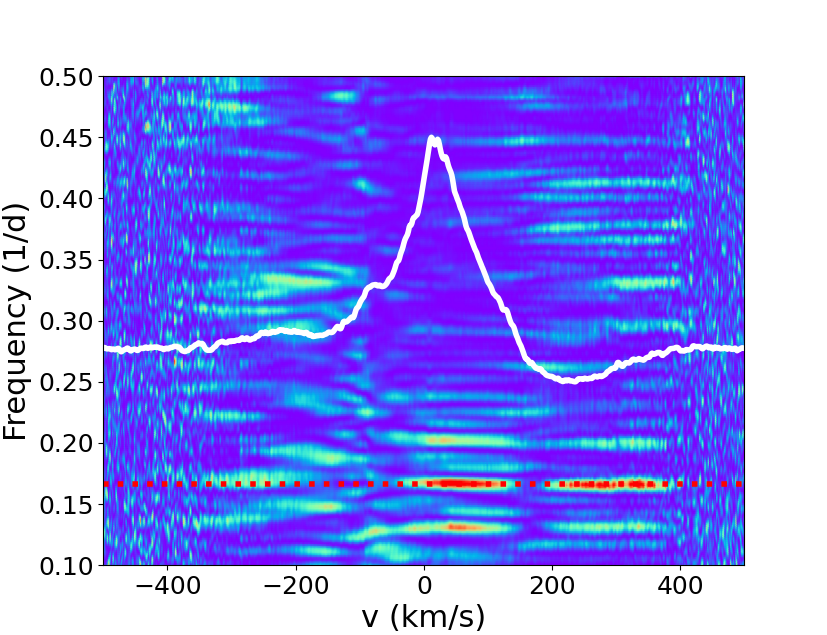}
    \includegraphics[width=0.3\hsize]{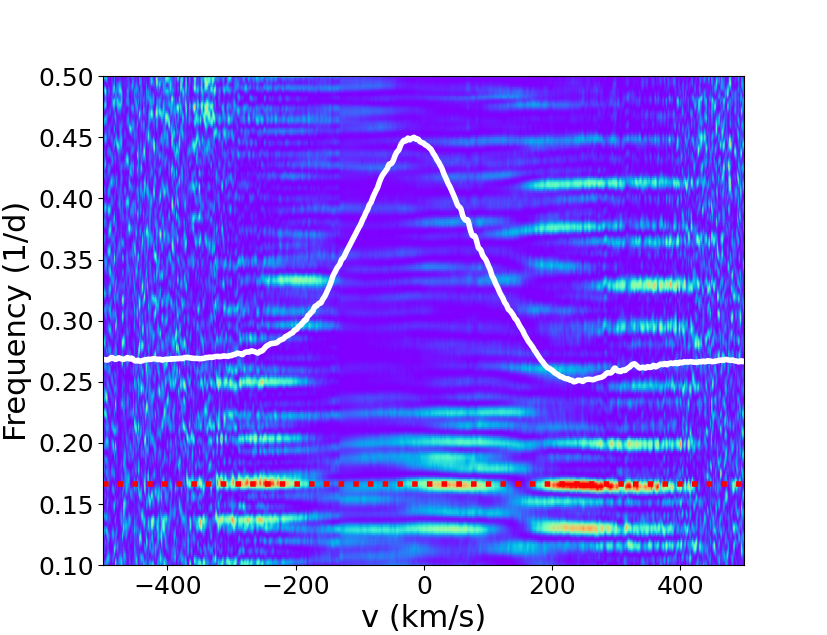}
    \includegraphics[width=0.3\hsize]{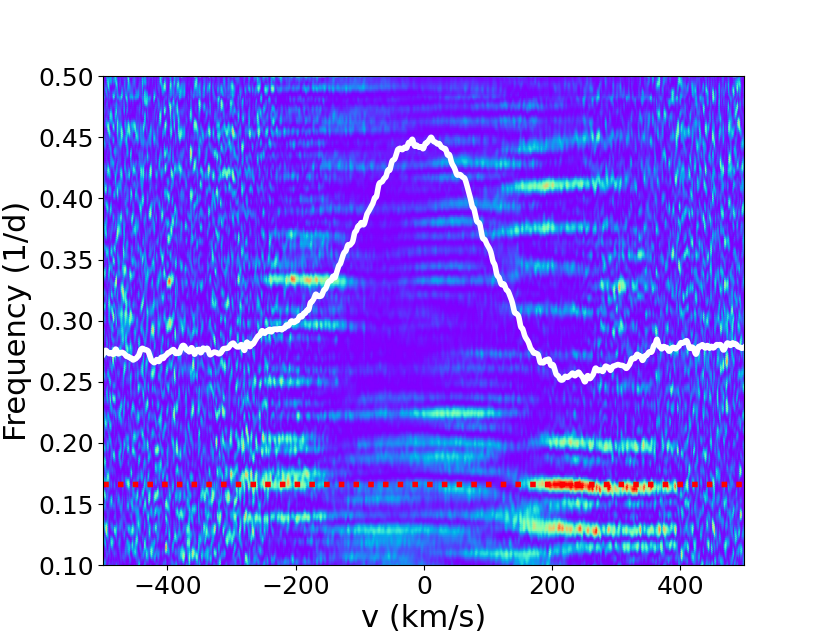}
   \caption{Near-infrared line profile variability. {\it Top:} Series of residual near-infrared line profiles \hei\ ({\it left}), \pab\ ({\it center}), and \brg\ ({\it right}), plotted superimposed to illustrate their variability. Each color represents a SPIRou run as in Fig.~\ref{fig:spirou_prof}.  {\it Bottom:} 2D periodograms across the line profiles. The color code reflects the periodogram power from zero ({\it blue}) to 0.5 ({\it red}). The dotted red horizontal line drawn at a frequency of 0.166 day$^{-1}$ indicates the stellar rotational period. The white curve displays the mean line profile.}
              \label{fig:spirou_p2d}%
    \end{figure*}
    
The main emission lines seen in the high-resolution near-infrared spectrum of GM Aur, namely \hei\ 10830~\AA, \pab, and \brg, are depicted in Figures~\ref{fig:spirou_prof} and \ref{fig:spirou_p2d}\footnote{\pac\ and \pad\ also appear in SPIRou spectra. Their shape and variability behavior is similar to that of \pab\ and \brg. \brd\ is also included in the SPIRou wavelength range, but lies at the intersection of spectral orders. In all SPIRou spectra, we also detect a weak H2 2.12 $\mu$m line, with EW = 0.22 $\pm$ 0.04 \AA, and FWHM = 1.40 $\pm$ 0.05 \AA\ ($\simeq$ 20 \kms). This line is located at the stellar rest velocity (\vrad = 0.88 $\pm$ 0.75 \kms).}. The near-infrared hydrogen lines exhibit broad emission profiles, with FWHM $\sim$200~\kms, whose peaks are slightly blueshifted compared to the stellar velocity, and are located at -30 and -10 \kms\ for \pab\ and \brg, respectively. The profiles are roughly symmetric, but have pronounced high-velocity redshifted absorptions that extend up to +400~\kms \ and reach below the continuum level.  \pab\ and \brg\  exhibit inverse P Cygni (IPC) profiles of varying depth in most SPIRou spectra, which is in stark contrast to the optical hydrogen line profiles, \ha\ and \hb, which do not exhibit such features (see Sect.~\ref{ressophielines}).  


Correlation matrices for the near-infrared hydrogen line profiles are shown in Fig.~\ref{fig:cmpabbrg}. The \pab$\star$\pab, \brg$\star$\brg, and \pab$\star$\brg\ matrices are quite similar. The emission part of the profiles is overall correlated, and anti-correlated with the high-velocity redshifted absorption component.  This indicates that the redshifted absorption is deeper when the emission line is more intense. Cross-correlation matrices between optical and near-infrared line profiles observed in the same night during the October runs are shown in Fig.~\ref{fig:cmhapab}. The \brg\ and the \pab\ emission components correlate well with the main \ha\ and \hb\ emission components (-100 \kms  < v <  200 \kms), which suggests that at least part of the line emission forms in the same extended region. The \pab\ redshifted absorption, and to a lesser extent, the \brg\ redshifted absorption, correlates well with the \ha\ and \hb\ blue and red high-velocity wings. This indicates that they all form in the same region close to the star.  The near-infrared profiles are strongly anticorrelated with the highly variable blueshifted absorption components that appear at velocities of about -100 \kms\ in the Balmer line profiles and presumably trace outflows. Except for this feature, the overall correlated variations of optical and near-infrared line profiles suggest that they are all good diagnostics of the accretion flow, as previously suggested by \cite{Alcala14}, for example. Detailed modeling of the line profile shapes and variability is needed, however, to ascertain the exact location from which they arise \citep{Tessore23}.

The \hei\ 10830~\AA\ line profile is more complex. It is usually dominated by an emission peak at low velocities that becomes quite weak at specific times, however. The profile also often displays highly redshifted absorption features, similar to the IPC components seen in the \pab\ and \brg\ lines. Unlike the near-infrared hydrogen profiles however, the \hei\ line additionally exhibits various absorption components in the blue wing of the profile. At least two systems of blueshifted absorptions can be identified: a low-velocity system extending between -20 to -80~\kms, and a high-velocity system ranging from -100 to -250~\kms. Figures~\ref{fig:spirou_prof} and \ref{fig:spirou_p2d} show that the two systems are quite variable. These two absorption systems are reminiscent of those observed in the \ha\ and \hb\ lines, although they occur at slightly bluer velocities in the \hei\ profile, with an offset of about -40 \kms\ compared to the optical profiles. 

Correlation matrices involving the \hei\ 10830 \AA\ line are displayed in Fig.~\ref{fig:cmheipab}. The \hei\ autocorrelation matrix shows several correlated components. Over the velocity channels extending from -100 \kms\ up to 200 \kms, the line flux varies in a correlated fashion, but this region of the line does not correlate with the rest of the profile. Similarly, the redshifted absorption region, which extends from 200 \kms\ to 400 \kms, varies as a whole and does not correlate with the rest of the profile. The blue wing of the profile presents many short-duration absorption components superimposed on the emission component, and, probably due to this, each velocity bin varies independently of the other. Comparison of the \hei\ 10830 \AA\ line profile to optical and near-infrared hydrogen profiles indicates that the emission core and peak intensity are correlated, as are the high-velocity redshifted absorptions seen in the \hei, \pab, and \brg\ line profiles. However, the \hei\ blue wing shows a strong anticorrelation with the core of the \hb\ profile.  

The periodogram analysis of the line profiles is shown in Figure~\ref{fig:spirou_p2d}. In the three lines, the IPC components appear to be periodically modulated at the stellar rotation period. As the SPIRou dataset extends over four months, this suggests that a quite stable structure, presumably the accretion funnel flow, gives rise to this spectral feature\footnote{The long time coverage of the SPIRou observations enabled us to explore periods up to 100 days. However, we did not find significant periods longer than those reported here.}. The redshifted absorption component is deeper when the emission line is more
intense, which agrees with the assumption that the two components trace the densest part of the accretion
column close to the star. The low-velocity red wing of the \hei\ line profile also shows a periodic modulation at the same period, which might trace the visibility of the accretion shock, as seen in the optical \hei\ line profile. The periodogram power in the blue wing of the lines is weaker, except perhaps over the high-velocity channels of the \pab\ line. In particular, the variable blueshifted absorption systems seen in the \hei\ line profile are not periodically modulated on this long timescale. We performed a similar analysis of each SPIRou run individually, and the results are shown in Appendix~\ref{app:spirouruns}. Fig.~\ref{fig:spirou_sep} to \ref{fig:spirou_dec} reveal that the most conspicuous high-velocity redshifted absorptions in the \hei, \pab, and \brg\ line profiles occur preferentially around \phirot=0.  During the SPIRou October run, which was simultaneous with the OHP/SOPHIE observations, we realized that the highly blueshifted velocity channels of the near-infrared lines appear to be modulated at the stellar rotation period (see Fig.~\ref{fig:spirou_oct}). Hence, while the modulation of high-velocity blueshifted absorptions seen in the \hei\ profile disappear on a timescale of several months, the modulation may survive over a few rotational periods. Finally, similar to what is seen in the Balmer line profiles over a much shorter time span (see Section~\ref{ressophielines}), there is a marginal indication from the periodograms of the near-infrared lines that the modulation period drifts from the blue to the red wing of the profiles. This might be a sign of differential rotation in the source of the variability. 
     
The EW of the \heiir\ 10830~\AA, \pab, and \brg\ lines was computed on the residual profiles using two methods. The first method consists of adjusting a Gaussian fit to the line profile, and the second method is integrating below the line profile. The two methods were applied to the \pab\ and \brg\ lines that most often exhibit a strong Gaussian-like emission and, at times, a pronounced redshifted absorption. The difference between the EW derived from the Gaussian fit and from the profile integration then provides an estimate of the strength of the redshifted absorption component. The \heiir\ line exhibits a complex  profile, with pronounced absorptions appearing in both the blue and red wings, and it cannot be fit by a Gaussian. We report here \heiir\ EWs measured through profile integration only. EW measurements are usually accurate to within 5\% because of the well-defined adjacent stellar continuum level. 

\begin{figure}
  \centering
  \includegraphics[width=0.49\textwidth]{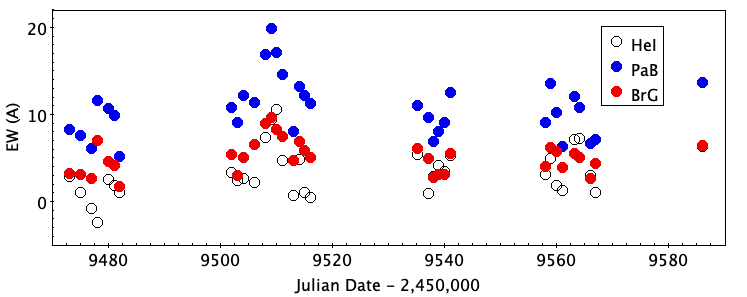}
    \includegraphics[width=0.49\textwidth]{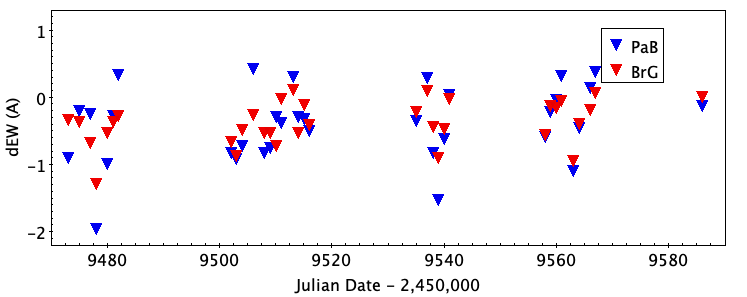}
  \caption{EW of near-infrared lines. {\it Top:} EW of the \hei, \pab,\, and \brg\ lines measured on SPIRou spectra plotted as a function of Julian date. The measurement uncertainties are about the symbol size. {\it Bottom:} Difference between the EWs measured by direct line profile integration (i.e., including the redshifted absorption component) and the EW derived from the Gaussian fitting of the emission component only plotted as a function of Julian date. This differential quantity measures the strength of the redshifted absorption component in the \pab\ and \brg\ line profiles. The more negative the differential quantity, the deeper the redshifted absorption.} \label{fig:spirou_EW}
\end{figure}

The results are listed in Table~\ref{tab:ewspirou}, and the evolution of EW during the observing campaign is illustrated on Fig.~\ref{fig:spirou_EW}. While the EW variations of the three lines are usually correlated, there are notable exceptions. For instance, during the first SPIRou run around JD 2,459,478, the \hei\ line goes into absorption, driven by a broad redshifted absorption component that reaches half the continuum value, while the \pab\ and \brg\ lines reach a local intensity maximum, even though they also exhibit a pronounced redshifted absorption (see Fig.~\ref{fig:spirou_sep}). In contrast, during the second SPIRou run, the variations of the three lines are well correlated. Line EWs and near-infrared veiling measurements (listed in Table~\ref{spirou_obs}) are both shown as a function of Julian date and rotational phase in Fig.~\ref{fig:spirouveilew} (Appendix~\ref{app:nirveil}). Both quantities appear to be rotationally modulated, with maximum values occurring close to \phirot=0. 
    
\begin{figure}
  \centering
  \includegraphics[width=0.49\textwidth]{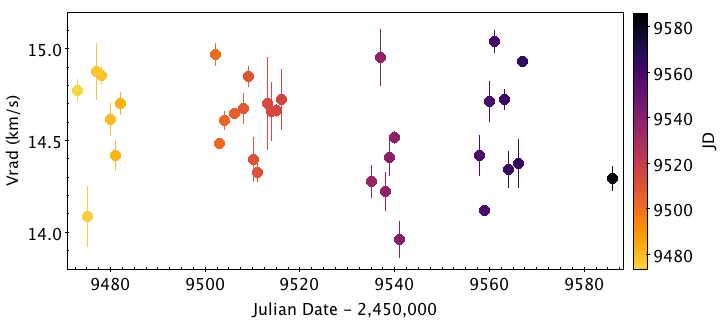}
  \includegraphics[width=0.49\textwidth]{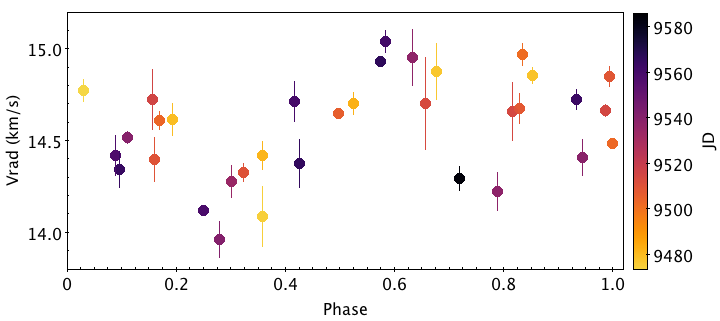}
  \caption{Radial velocity variations measured in the near-infrared photospheric lines of the SPIRou spectra. {\it Top:} Radial velocity as a function of Julian date. {\it Bottom:} Radial velocity curve folded in phase at the stellar rotational period P=6.04~days. The color code reflects the Julian date.} \label{fig:spirouvrad}
\end{figure}

Finally, the photospheric radial velocity curve derived from SPIRou spectra is shown in Figure~\ref{fig:spirouvrad}. \vrad\ is found to vary between 13.96 and 15.04~\kms, with a median value of 14.65~\kms. A CLEAN periodogram analysis reveals a period P=5.94 $\pm$ 0.11~d, and the String-Length method yields P = 5.98 $\pm$ 0.12~d, both consistent with the 6.04 $\pm$ 0.15~d photometric period derived in Section~\ref{sec:resphot}. The photospheric radial velocity curve folded in phase at the stellar rotation period is shown in Fig.~\ref{fig:spirouvrad}. \vrad\ exhibits a roughly sinusoidal variations in rotational phase, which suggests that it is modulated by a surface spot. It reaches the median velocity going blueward around \phirot = 0, as expected from a stellar spot facing the observer at this phase. Because this is also the phase of maximum brightness of the system (see Section~\ref{sec:resphot}),  this suggests that a hot spot modulates the photospheric \vrad\ curve. 
The \vrad\ curve derived from the near-infrared photospheric lines (see Fig.~\ref{fig:spirouvrad}) is inverted compared to the \vrad\ curve derived for the \hei\ 5876 \AA\ NC line profile (see Fig.~\ref{fig:ohpheivrad}). Similar antiphase radial velocity variations between absorption and emission lines have been reported for other accreting T Tauri stars and were interpreted as caused by the modulation of the line profiles by a hot spot at the stellar surface \citep{Petrov01, Petrov11, Gahm13}. Alternatively, the photospheric radial velocity variations may also be produced by a cool spot that coexists with the accretion shock at the same location on the stellar surface, as Doppler images of accreting T Tauri stars suggest \citep[e.g.,][]{Donati10, Donati11, Donati19,Donati20b}. 


\begin{table}
  \centering
  \setlength{\tabcolsep}{4pt}
  \scriptsize
  \renewcommand{\arraystretch}{0.9}
\caption{ Near-infrared line EW measurements from the CFHT/SPIRou spectra. }             
\label{tab:ewspirou}      
\centering                          
\begin{tabular}{l l l l l l l l}        
\hline\hline                        
\noalign{\smallskip}
Julian date & \multicolumn{7}{c}{EW (\AA)}  \\
 (2,450,000+) &  \heiir$_{int}$  &  \pab$_g$ & \pab$_{int}$  & \pab$_{diff}$ &  \brg$_g$ & \brg$_{int}$ & \brg$_{diff}$ \\ 
\hline                                   
\noalign{\smallskip}
9473.066 & 2.88 & 8.29 & 7.41 & -0.87 & 3.26 & 2.94 & -0.32 \\
9475.043 & 0.98 & 7.58 & 7.40 & -0.18 & 3.07 & 2.73 & -0.34 \\
9476.969 & -0.76 & 6.05 & 5.82 & -0.23 & 2.60 & 1.95 & -0.65 \\
9478.027 & -2.45 & 11.58 & 9.64 & -1.93 & 6.97 & 5.70 & -1.26 \\
9480.082 & 2.57 & 10.74 & 9.76 & -0.97 & 4.56 & 4.05 & -0.51 \\
9481.086 & 1.86 & 9.90 & 9.65 & -0.25 & 4.15 & 3.80 & -0.35 \\
9482.086 & 0.98 & 5.18 & 5.53 & 0.35 & 1.72 & 1.47 & -0.25 \\
9502.074 & 3.33 & 10.82 & 10.02 & -0.80 & 5.37 & 4.73 & -0.63 \\
9503.074 & 2.43 & 9.02 & 8.12 & -0.89 & 3.02 & 2.17 & -0.85 \\
9504.098 & 2.58 & 12.18 & 11.48 & -0.70 & 5.09 & 4.63 & -0.45 \\
9506.082 & 2.19 & 11.36 & 11.80 & 0.44 & 6.54 & 6.29 & -0.24 \\
9508.082 & 7.33 & 16.93 & 16.12 & -0.81 & 9.00 & 8.48 & -0.51 \\
9509.086 & 9.45 & 19.84 & 19.11 & -0.73 & 9.68 & 9.16 & -0.51 \\
9510.086 & 10.53 & 17.07 & 16.80 & -0.27 & 8.27 & 7.57 & -0.70 \\
9511.074 & 4.73 & 14.56 & 14.20 & -0.36 & 7.48 & 7.48 & 0.00 \\
9513.090 & 0.73 & 8.00 & 8.32 & 0.31 & 4.68 & 4.82 & 0.13 \\
9514.051 & 4.82 & 13.19 & 12.92 & -0.27 & 6.88 & 6.37 & -0.51 \\
9515.078 & 1.03 & 12.18 & 11.88 & -0.30 & 5.81 & 5.71 & -0.09 \\
9516.102 & 0.51 & 11.26 & 10.78 & -0.48 & 5.02 & 4.62 & -0.39 \\
9535.102 & 5.37 & 11.00 & 10.67 & -0.33 & 6.03 & 5.83 & -0.20 \\
9537.098 & 0.96 & 9.65 & 9.96 & 0.30 & 4.93 & 5.04 & 0.11 \\
9538.039 & 2.84 & 6.93 & 6.13 & -0.80 & 2.74 & 2.31 & -0.42 \\
9538.984 & 4.10 & 7.99 & 6.49 & -1.50 & 3.12 & 2.23 & -0.88 \\
9539.988 & 3.45 & 9.12 & 8.52 & -0.59 & 3.08 & 2.63 & -0.45 \\
9541.012 & 5.32 & 12.48 & 12.53 & 0.05 & 5.46 & 5.46 & 0.00 \\
9557.980 & 3.13 & 9.12 & 8.55 & -0.57 & 3.96 & 3.42 & -0.53 \\
9558.949 & 4.88 & 13.61 & 13.41 & -0.20 & 6.25 & 6.14 & -0.10 \\
9559.961 & 1.81 & 10.18 & 10.16 & -0.02 & 5.77 & 5.63 & -0.14 \\
9560.965 & 1.29 & 6.34 & 6.67 & 0.33 & 3.85 & 3.82 & -0.02 \\
9563.078 & 7.11 & 12.10 & 11.03 & -1.07 & 5.48 & 4.56 & -0.92 \\
9564.059 & 7.27 & 10.85 & 10.41 & -0.44 & 5.09 & 4.72 & -0.37 \\
9566.051 & 2.98 & 6.62 & 6.78 & 0.15 & 2.62 & 2.44 & -0.17 \\
9566.953 & 1.05 & 7.17 & 7.57 & 0.39 & 4.39 & 4.47 & 0.08 \\
9585.941 & 6.31 & 13.67 & 13.56 & -0.11 & 6.47 & 6.50 & 0.02 \\
\hline                                   
\end{tabular}

Note: EW$_g$ is obtained from a Gaussian fit to the line profile, while EW$_{int}$ is measured from line profile integration. EW$_{diff}$ is the difference between EW$_{int}$  and EW$_g$. 
\end{table}


\subsection{Low-resolution near-infrared spectroscopy}
\label{sec:extra_res}


\begin{figure}
  \centering
  \includegraphics[width=0.49\textwidth]{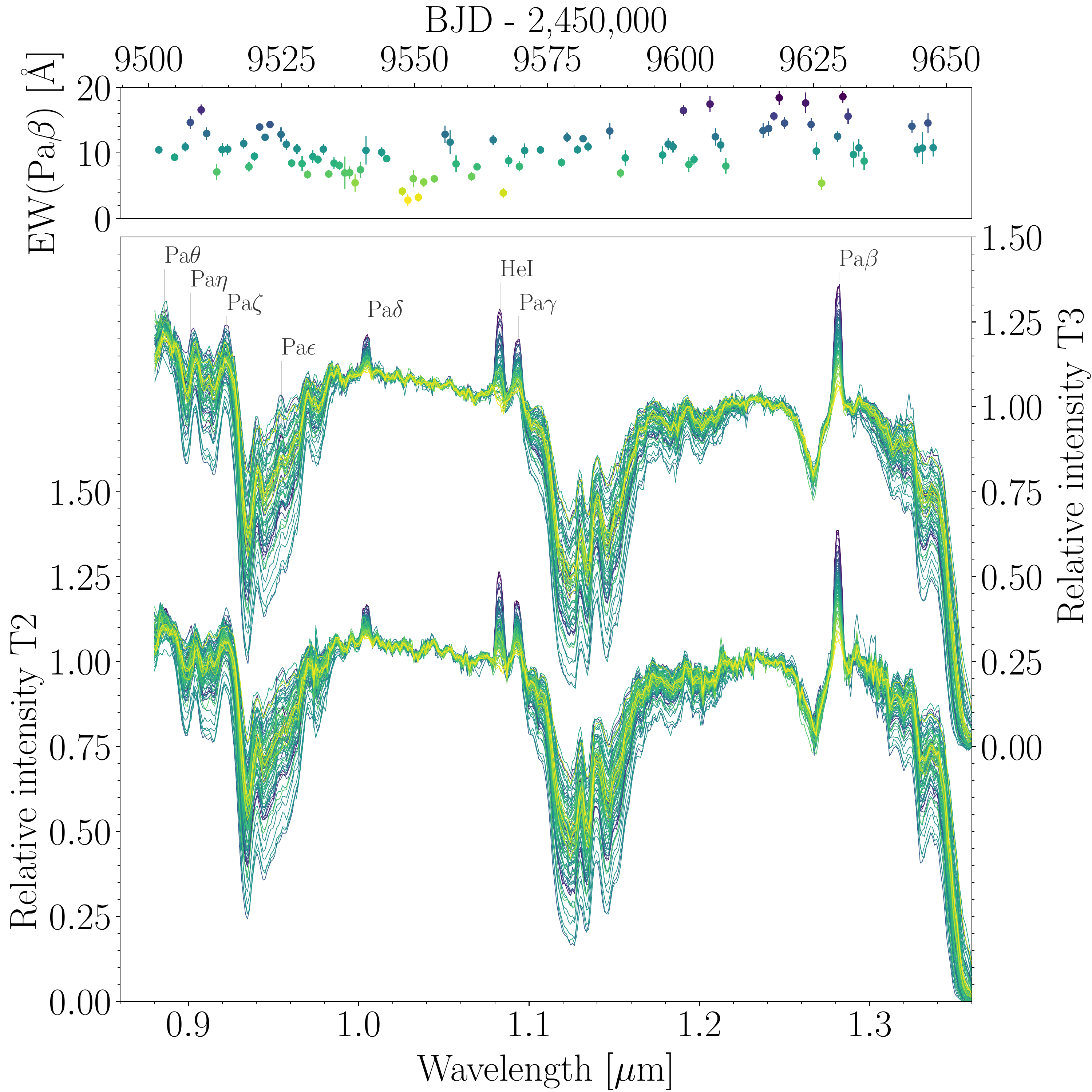}
  \caption{Median spectrum for each night from the ExTrA-T2 telescope (lower part, 88 spectra) and ExTrA-T3 telescope (upper part, 69 spectra). The main emission lines are indicated. The color is proportional to the EW of the Pa$\beta$ line shown in the top panel as a function of Julian date.} \label{fig:ExTrA_night}
\end{figure}

The median nightly spectra obtained from the ExTrA-T2 and ExTrA-T3 telescopes are shown in Figure~\ref{fig:ExTrA_night}. We computed the EW of the \hei\ 10830~\AA, \pab, \pac, and \pad\  lines from the ExTrA spectra using {\sc \tt specutils}\footnote{\url{https://github.com/astropy/specutils}}. We analyzed each telescope independently because the point spread function (and therefore the resolution) of the spectrograph depends on the position on the detector. First, we fit a Gaussian to each line on the median spectra to measure the line center and full width, the latter we defined as amounting to six times the standard deviation of the Gaussian fit in order to isolate the line profile from the nearby continuum. The local continuum around each line was modeled with a third-degree polynomial, adjusted on three line widths centered on the line, but excluding the line. Then, for each line in each of the 1898 individual spectra, we computed the EW by integrating the flux over the full line width using the parameters and the normalization region defined from the median spectra. Finally, we computed the median of the individual measurements for each night, regardless of the telescope. The median is less affected by outliers than the mean, and the differences between the mean and the median values are within the error bars. Table~\ref{tab:extraewJ} in Appendix~\ref{app:extraewJ} lists the results. 

In order to estimate the reliability of the procedure, we compared EWs derived from ExTrA and from SPIRou spectra for the \pab\ line using 20 measurements obtained on each instrument less than one day apart. The results show an excellent correlation, with a slight tendency for the EW measured from ExTrA to exceed those measured from SPIRou, with a mean difference of 0.7$\pm$1.1 ~\AA. A similar result is obtained for the \heiir\ line, with a mean difference of 0.31~\AA\ and an rms of 1.1~\AA\ between ExTrA and SPIRou estimates. The comparison with high-resolution measurements thus validates the EWs obtained from low-resolution spectra.    

\begin{table}
  \centering
  \setlength{\tabcolsep}{4pt}
  \renewcommand{\arraystretch}{0.9}
  \caption{Minimum, median, and maximum near-infrared line EW measurements from the ExTrA spectra.}
\begin{tabular}{ccccc}
\hline
\hline
\noalign{\smallskip}
&\multicolumn{4}{c}{EW (\AA)}\\
  & \hei & \pab & \pac & \pad \\
\hline
\noalign{\smallskip}
Min. & -1.9 & 2.8 & 3.4 &  0.2 \\
Med. & 2.8 & 10.5 & 7.0 & 2.1  \\
Max. & 11.8 & 18.6 & 11.4 &  4.8 \\
\hline
\end{tabular}
\label{table:EW_ExTrA_mean}
\end{table}

\begin{figure*}
  \centering
    \includegraphics[width=0.6\textwidth]{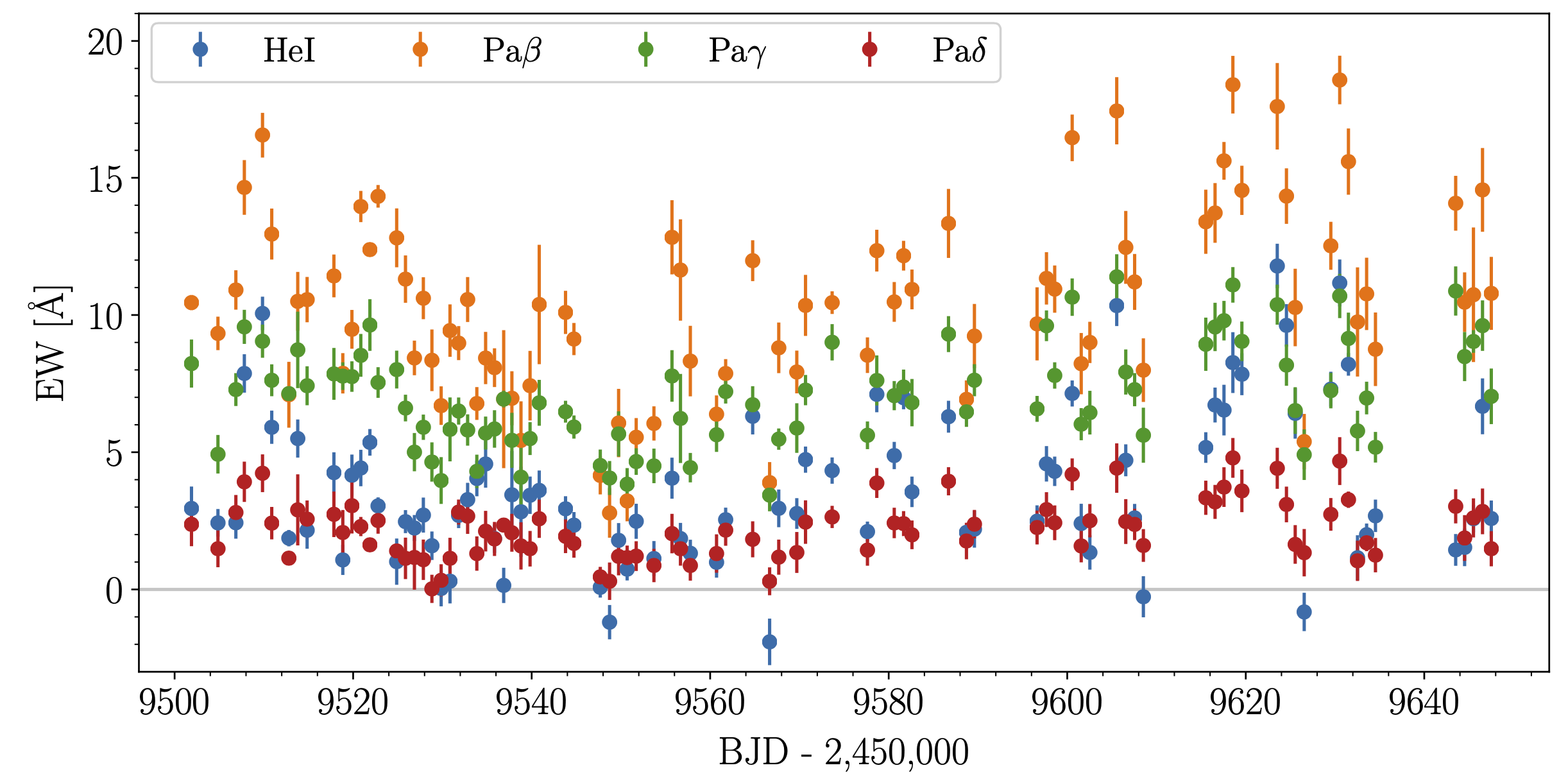}
    \includegraphics[width=0.3\textwidth]{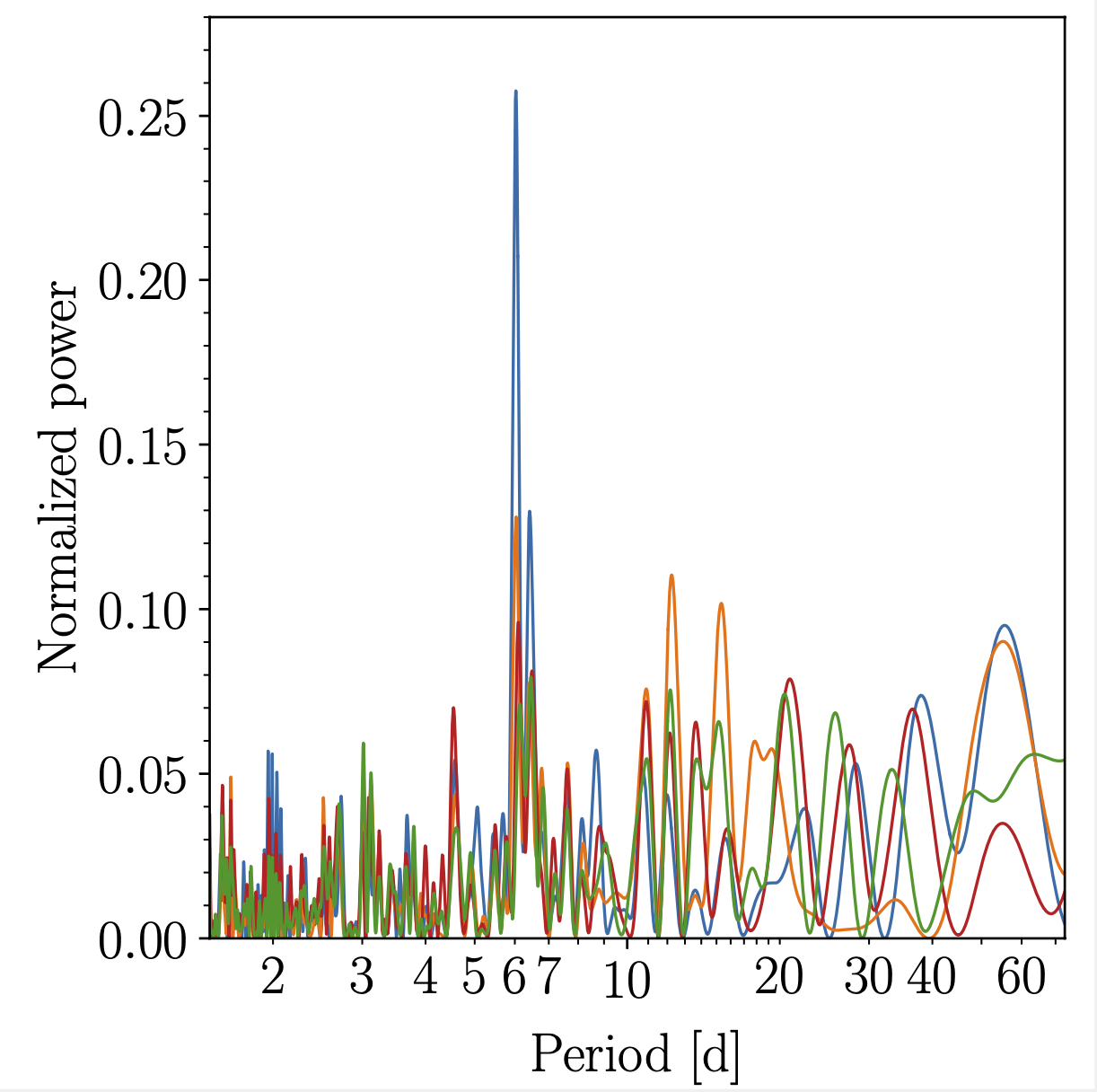}
  \caption{EW variability of the near-infrared line profiles. {\it Left:} EW measurements plotted as a function of Julian date for the \hei, \pab, \pac, and \pad\ lines from the ExTrA spectra. {\it Right:} GLS periodogram of the EW measurements. The period on the x-axis is displayed on a log scale.} \label{fig:EW_ExTrA}
\end{figure*}


\begin{figure}
  \centering
  \includegraphics[width=0.49\textwidth]{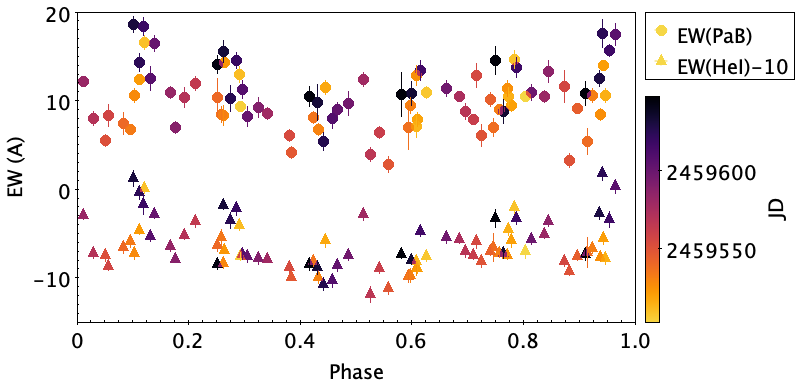}
   \includegraphics[width=0.49\textwidth]{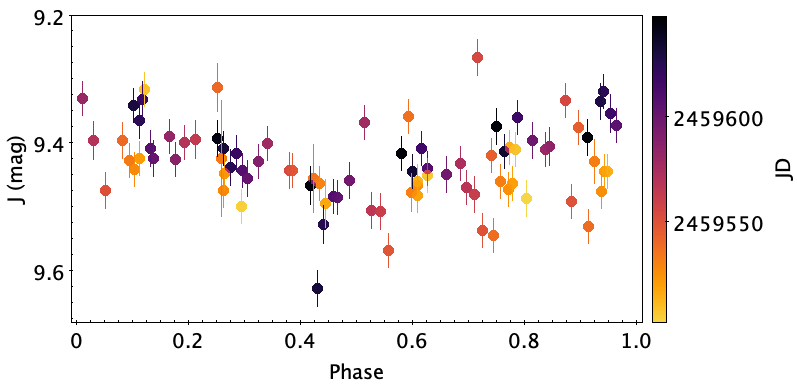}

  \caption{EWs of near-infrared line profiles and J-band magnitude variability. {\it Top:} EW of the near-infrared \hei\ and \pab\ lines folded in phase at the stellar rotational period. EW(\hei) is offset by -10~\AA\ for clarity. {\it Bottom:} J-band light curve, deduced from ExTrA spectra, folded in phase at the stellar period. The brightness modulation is similar to that observed at optical wavelengths (see Fig.~\ref{fig:lcogt_lc}).} \label{fig:EW_ExTrA_phased}
\end{figure}

\begin{figure}
  \centering
  \includegraphics[width=0.45\textwidth]{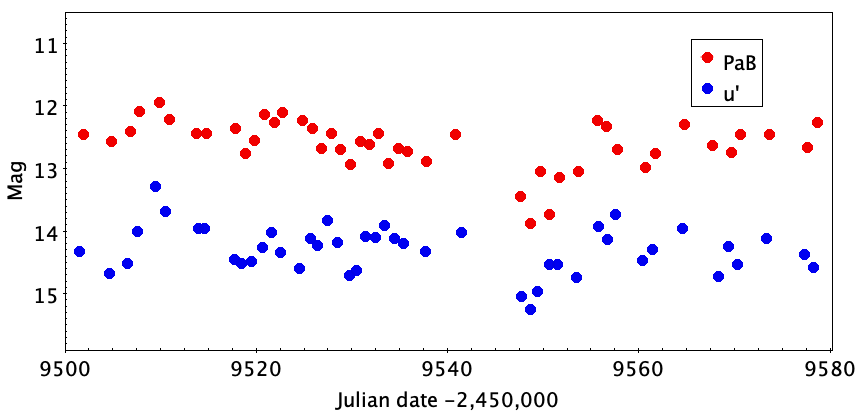}
  \caption{Mid-term variation of EW(\pab) ({\it red}) compared to the system u'-band light curve ({\it blue}) for measurements taken less than one day apart. To facilitate the comparison, the EW measurements are plotted on a magnitude scale and are offset, namely -2.5 $\log$ EW(\pab) + 5.  } \label{fig:ewpab_up}
\end{figure}

The line variability is illustrated on Figure~\ref{fig:ExTrA_night}, where the median nightly spectra are superimposed. 
The median and extreme values of EWs measured for the \hei, \pab, \pac, and \pad\ lines are listed in Table~\ref{table:EW_ExTrA_mean}, 
and the night-by-night measurements are listed in Table~\ref{tab:extraewJ}. We note that the \hei\ line at times appears to be in absorption at this low spectral resolution. 

Figure~\ref{fig:EW_ExTrA} shows the EW measurements plotted as a function of time and its generalized Lomb-Scargle periodogram \citep[GLS;][]{zechmeister09}. The EW of the four lines is found to be modulated with a period of 6.028 $\pm$ 0.087 days, consistent with the stellar rotation period, where the error estimate is the standard deviation of a Gaussian fit to the periodogram peak. Fig.~\ref{fig:EW_ExTrA_phased} shows the \hei\ and \pab\ line EWs folded in phase at the stellar rotation period. The modulation of the line strength clearly appears, with maximum flux around phase zero. As shown in the same figure, these variations follow the modulated brightness level of the system in the J band. 

Longer-term EW variations of higher amplitudes are also clearly seen in Fig.~\ref{fig:EW_ExTrA}. These variations seem to be correlated with the multicolor photometry presented in Section~\ref{sec:resphot},  
in particular, during the brightness event centered on JD 2,459,509 and the wide dip around JD 2,459,549. In Figure~\ref{fig:ewpab_up}, a clear correlation appears between the photometric variations in the u' band and the \pab\ line variations. If the brightening of the system in the u' band is linked to accretion, this correlation suggests that most of the emission line flux is connected to the same process.

\subsection{Mass-accretion rate}


The combination of optical and near-infrared spectroscopy offers a number of emission lines from which we can estimate the mass accretion rate onto the star, using the empirical relations between line luminosity and accretion luminosity proposed by \cite{Alcala17}. Combining the range of \ha\ and \hb\ EWs reported above with the nearby continuum fluxes computed from the r' and g' magnitudes corrected for extinction, we obtain the line fluxes and luminosities as follows:
$F_{line}= F^o_\lambda \times EW(line) \times 10^{-0.4 (m_\lambda-A_\lambda)}$ and $L_{line}=4 \pi d^2 F_{line}$, where $F^o_\lambda$ is the flux of a zero-magnitude star\footnote{$F^o_\lambda$ = 2.43 10$^{-9}$  and 5.27 10$^{-9}$ erg s$^{-1}$ cm$^{-2}$ \AA$^{-1}$ in the r' and g' bands, respectively.},  $m_\lambda$ and $A_\lambda$ are the magnitude and extinction in the photometric band of interest,  and $d$ is the distance to the star. The accretion luminosity was derived from the line luminosity using the relation reported by \cite{Alcala17}, and \macc\ was deduced from \lacc\ assuming  a magnetospheric radius of 5~\rstar\ \citep[see][Eq.(1)]{Alcala17}. Taking the uncertainties on all involved quantities into account, we thus derive \macc\ = 0.7 $\pm$ 0.3 and 0.5 $\pm $0.4 $ \times$ 10$^{-8}$~\msunyr\ from the \ha\ and \hb\ line fluxes, respectively. 

We performed a similar analysis using the extensive measurements of \pab\ EWs obtained from the ExTrA spectra. From the extreme values, namely EW(\pab) = 2.8 --18.6~\AA, and the mean REM J-band magnitude of the system during the observing period, J = 9.41 $\pm$ 0.10, we derive \macc\ = 0.3 -- 2.0 $\times$ 10$^{-8}$~\msunyr, with a median value of \macc\ = 1.0 $ \times$ 10$^{-8}$~\msunyr. Finally, we derived additional estimates of \macc\ from the \brg\ line flux, using the median and extreme values of EW(\brg) measured on SPIRou spectra, namely 5.1, 1.7, and 9.7~\AA, which we calibrated with the mean REM K'-band magnitude of the system, K'= 8.40. Using the relations reported by \cite{Alcala17} between line and accretion luminosity, we thus derived \macc\ = 0.2-1.8 $\times$ 10$^{-8}$~\msunyr, with a median value of \macc\ = 0.8 $\times$10$^{-8}$~\msunyr. 

The dispersion in the \macc\ estimates partly results from intrinsic \macc\ variability. For instance, the SOPHIE spectra were obtained during a brightening of the system that occurred around JD 2,459, 512, at a time at which all optical and infrared lines were relatively strong. We thus derive a relatively high value of \macc\  from these spectra compared, for example, to the smaller \macc\ estimate derived from the \brg\  line in the SPIRou spectra that were obtained during more quiescent phases of the system. We cannot exclude, however, that some of the dispersion may also arise from systematic uncertainties in the empirical line-to-accretion luminosity relations over the optical and near-infrared wavelength ranges. In any case, the various estimates agree globally, with \macc\ typically varying between 0.2 and 2.0 $\times$10$^{-8}$~\msunyr. 






\section{Discussion}
\label{sec:discussion}



\begin{figure*}
  \centering
  \includegraphics[width=\textwidth]{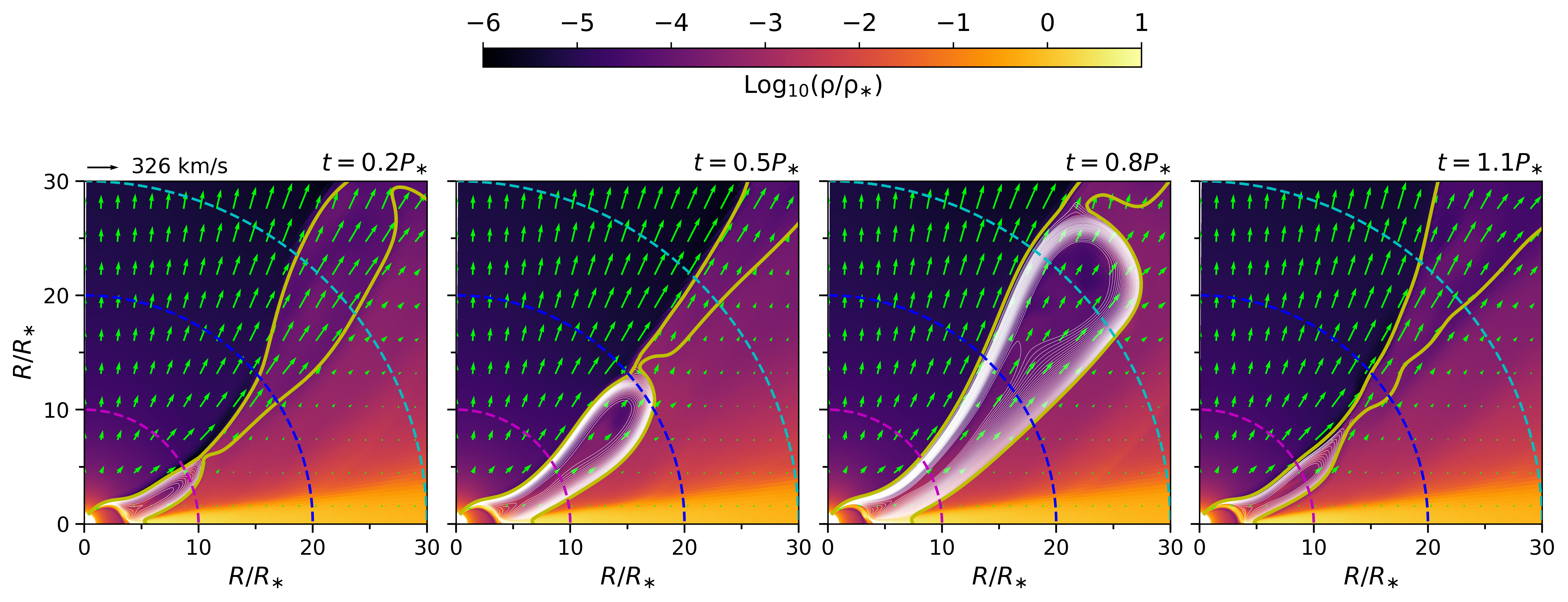}
  \caption{Magnetospheric ejection model. Development of a magnetospheric ejection computed from star-disk interaction MHD simulations. The snapshots shown here are extracted from model 3 of \cite{Pantolmos20}, where the magnetospheric truncation radius amounts to 54\% of the corotation radius. The three snapshots are shown at 0.2, 0.5, 0.8, and 1.1 \prot. The white curves indicate expanding magnetic field lines that give rise to the ejection of plasmoids. The color scale indicates density. The green arrows show the velocity field of the stellar and disk winds and of MEs at their interface.} 
  \label{fig:me}
\end{figure*}

\begin{figure}
  \centering
  \includegraphics[width=0.4\textwidth]{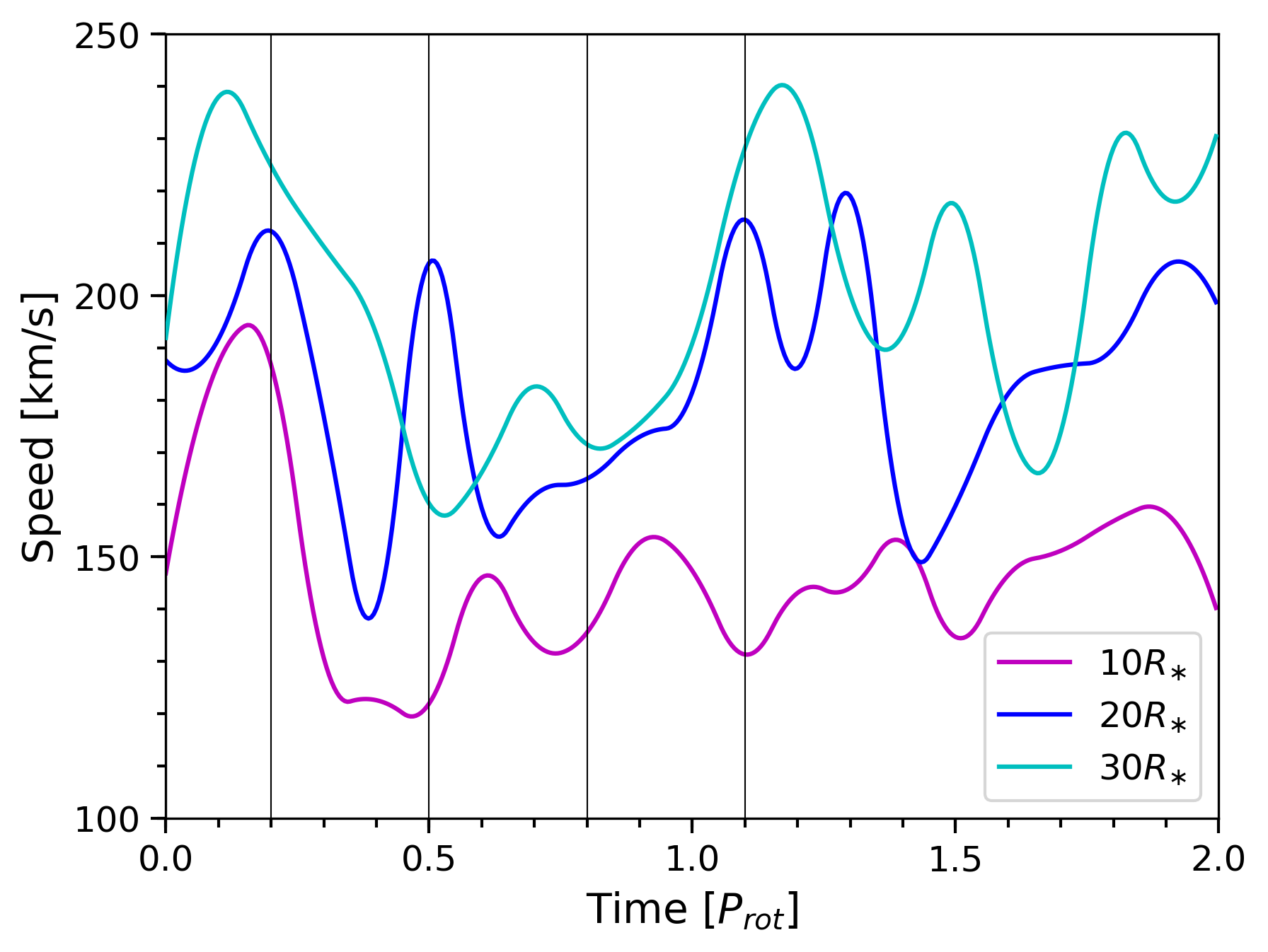}
  \caption{Magnetospheric ejection model. Velocity of the gas in the ejected plasmoid at a distance of 10 {\it (magenta}), 20 ({\it blue}), and 30 ({\it cyan}) stellar radii as a function of time in units of the stellar rotational period. Successive ejections of plasmoids are featured. The vertical lines correspond to the snapshots shown in Fig.\ref{fig:me}.}
  \label{fig:mespeed}
\end{figure}

The monitoring campaign we performed on GM Aur reveals significant but relatively low-level temporal variability over a timescale of six months. The photometric variations are mild, as might be expected for this moderately accreting young system (\macc\ $\sim$ 0.8 $\times$ 10$^{-8}$ \msunyr), with amplitudes ranging from 1.5 mag in the u' band to 0.3 mag in the i' band, and 0.1 mag in the J band. The brightness of the system varies smoothly because it is modulated by the visibility of surface spots at the stellar rotational period of 6.04~d. Except for the \hei\ 10830 \AA\ line profile, the spectral appearance of the system does not change drastically on a timescale of months, and the veiling is low and stable, amounting to about 0.3 in the optical range. The strength of the main emission lines (\ha, \hb, \hei\ 5876\AA, \pab, and \brg) varies by a factor of 2 to 3 over the course of the semester. In contrast, the \hei\ 10830 \AA\ line profile exhibits extreme variability and is sometimes barely noticeable in emission. The line profile shapes are strongly variable on a timescale of days, and the development of both blueshifted and redshifted absorption components is superimposed onto a broad emission component. Blueshifted absorption components  indicate outflows \citep[e.g.,][]{Edwards03, Kwan07}, and redshifted components probe funnel flows \citep[e.g.,][]{Edwards06, Fischer08}. Remarkably, the redshifted absorption features that reach below the continuum level, that is, inverse P Cygni profiles (IPC) \cite[see][]{Calvet92}, which are seen in the high-resolution near infrared line profiles (\hei\ 10830 \AA, \pab, \brg), are steadily modulated by stellar rotation over an extended observing time span of 3 months. Rotational modulation is also clearly detected in the strength of the near-infrared lines and is continuously observed at low resolution for more than five months. Assuming that most of the line flux arises from the magnetospheric accretion region, as suggested by the periodic appearance of the IPCs and the correlation between line flux and u'-band excess, this indicates a stable large-scale accretion structure on this timescale. 

In contrast, high-velocity blueshifted absorption components are neither periodic nor stable on this timescale. While they are ubiquitous in the optical line profiles, most notably \ha\ and \hb, and are also quite conspicuous in the \hei\ 10830 \AA\ line profile, their signature evolves on a timescale of a few days, sometimes drifting in velocity before disappearing altogether. As an example, Fig.~\ref{fig:spirou_nov} shows the evolution of the blueshifted absorption components in the \hei\ line profile over the course of the SPIRou November run. On JD 9537, a high-velocity blueshifted feature appears in the profile and drifts toward lower velocities over the next several days, from -200 \kms\ to -120 \kms. Soon after, on JD 9540, a new high-velocity component appears and follows the same trend. A similar behavior is seen in the high-velocity blueshifted component appearing in the \hei\ 10830 \AA\ line profile during the SPIRou September run, which drifted from -240 \kms\ to -150 \kms\ on a timescale of five days, from JD 9475 to JD 9480 (see Fig.~\ref{fig:spirou_sep}). This suggests episodic outflows lasting for a few days only. We see no sign of a steady, constant velocity wind in the \hei\ line profiles over the observing period, nor do we find evidence for a rotational modulation of the outflow signatures. The only stable outflow signature seen in the optical and near-infrared line profiles consists of a narrow, low-velocity  blueshifted absorption in the \ha\ profile that peaks at -20 \kms\ and remains visible over more than two weeks.

Observations thus suggest that we witness a globally stable accretion structure and a succession of short-lived episodic outflows. The contrasting behavior of accretion and outflow diagnostics observed on a timescale of months thus raises the question whether the two processes are physically connected on the scale of a few stellar radii that we probe here. To examine this issue, we investigated the aftermath of the brightening event GM Aur underwent around JD 2,459,509. On this date, the system exhibited a significant brightening at optical and near-infrared wavelengths, most notably in the u' band ($\sim$1 mag), as well as some of the strongest line fluxes and highest optical and near-infrared veiling values measured during the campaign. A simultaneous TESS light curve of the system recorded the brightening event (see Fig.~\ref{fig:lcogt_lc}), which started on JD 2,459,508 and ended on JD 2,459,511, and exhibited a flat peak lasting for two days with a 20\% flux increase. We therefore interpret this episode as an accretion burst that occurred around the rotational phase \phirot=0, corresponding to the maximum visibility of the accretion shock, and lasted for several days.  From the measured continuum level and Balmer line EW during the burst, we derive an increase of a factor of 2 in the accretion rate, reaching $\sim$2$\times$10$^{-8}$ \msunyr. Inspection of the optical and near-infrared line profiles during and after this event reveals high-velocity blueshifted absorption components that appear in the Balmer and \hei\ 10830 \AA\ line profiles. On JD 2,459,509 a new blueshifted absorption component appears in the \hei\ infrared line at a velocity of  -280 \kms and extends down to -350 \kms\ before it reaches the continuum level again, and it drifts to lower velocities ($\sim$ -150 \kms) over the next few days. The closest optical spectrum was recorded only toward the end of the accretion burst, on JD 2,459,510.6. In this spectrum, the \ha\ and \hb\ profiles feature a new blueshifted absorption component at a velocity of -110 \kms \ that lasts for a few days. The contemporaneous occurrence of an accretion burst rapidly followed by outflow signatures in the line profiles therefore suggests that the accretion and ejection processes are physically connected on small scales.

We propose that the most likely scenario that accounts for these episodic events is magnetospheric ejections, possibly triggered by magnetic reconnections in the accreting magnetosphere. Magnetospheric ejections \citep{Zanni13, Sauty22}, or nonstationary conical winds \citep{Romanova09}, are caused by the expansion and reconnection of the field lines that connect the star with the disk. The inflation process is the result of the star-disk differential rotation and the consequent build-up of toroidal magnetic field pressure \citep[e.g.,][]{Goodson97}. Quasi-periodic ejections of plasmoids are predicted to occur throughout the magnetospheric inflation cycle on a timescale of about the rotational period \citep{Hayashi96}. The speed and variability of the outflows likely depend on various parameters, such as the magnetic field strength and topology, the thermal disk pressure, nonideal MHD effects, and the interaction of the magnetospheric-ejection region with the surrounding outflows, that is, stellar and disk winds  \citep{Miller97, Romanova09, Zanni13}. Previous monitoring campaigns on young stellar objects have reported evidence for magnetospheric inflation cycles \citep{Bouvier03, Alencar18}. We show a 2.5D MHD simulation of the interaction between an inner accretion disk and a dipolar magnetosphere from \cite{Pantolmos20} in Fig. 18. The figure illustrates the ejection of plasmoids along expanding field lines at the distance of a few stellar radii from the stellar surface. The timescale for successive ejections is about that of the stellar rotation period. The outflow speed, shown in Fig~\ref{fig:mespeed}, reaches more than 200 \kms\ in the early phases of the ejection, then decelerates to about 150~\kms\ on a timescale of days (0.3$\times$\prot), and finally vanishes altogether. A direct comparison of the model to observations is not straightforward. The terminal speed we derive from observations is higher than predicted by the simulation, and its temporal evolution on a timescale of a few days might be dominated by projection effects as the system rotates and does not trace the evolution of the plasmoid velocity field. Moreover, the sporadic ejections we observe do not appear to have the quasi-periodic character of the model. Both the wind speed and ejection timescale may, however, depend on numerical effects. In any case, the behavior of the magnetospheric ejection model qualitatively matches the dynamics of the high-velocity blueshifted absorptions seen in the line profiles of GM Aur on a timescale of days to weeks. Moreover, in the scenario of a magnetospheric inflation cycle, magnetic reconnection leads to an accretion burst and simultaneously triggers an ejection episode \citep[e.g.,][]{Goodson99}. This is quite reminiscent indeed of what is suggested by the variability of GM Aur.  
  
We also noted a change in the GM Aur  light curve whose first part is dominated by successive low-level brightening events, up to the major accretion burst described above, while it exhibits luminosity dips toward the end of the observations. It is unclear whether the contrasting behavior of the system luminosity observed before and after the main accretion burst is a consequence of the burst itself, perhaps inducing a structural change in the star-disk interaction process. It is conceivable that the rearrangement of the magnetic topology after the inflation or reconnection event that led to the burst has impacted the large-scale geometry of the star-disk interaction region. Either a modest increase in the inner disk scale-height \citep{Nagel17} or a reduction of the extent of the truncation radius is prone to trigger a periodic obscuration of the central star by circumstellar dust, that is, a dipper phenomenon \citep{Cody14}, especially in young systems seen at high inclination \citep{Mcginnis15, Bodman17}. The magnetic topology and  the mass accretion rate may thus have slightly evolved over the six-month time span of the campaign, as suggested by the long-term variations of the emission line EWs shown in Fig.~\ref{fig:EW_ExTrA}. However, the results we report here clearly indicate that the large-scale geometry of the star-disk interaction was not drastically modified over this timescale. This is evidenced by the phase stability of the modulated light curve, the smooth variability of the emission line profiles around their mean shape, and the strictly periodic appearance of IPC profiles over at least three months. All these accretion diagnostics support a globally stable magnetospheric accretion structure during the campaign.    

Finally, it is interesting to compare the line profile shape and variability reported here to those reported by \cite{McGinnis20}, which were obtained in 2011, that is, ten years prior to our observing campaign. The shapes of the \ha\ and \hb\  profiles are quite different in the two studies. In \cite{McGinnis20}, these are pure emission profiles with a triangular shape, without any significant absorption components, and they exhibit little variability over the timescale of a week (see their Figures S5, S6). Here, the same profiles appear to be much more structured, with highly variable absorption components on the same timescale. The \hei\ 5876\AA\ line profile variability also differs between the two studies, but in the opposite direction. In both studies,  the line profile consists of a broad and a narrow component. However, in \cite{McGinnis20}, the intensity of the broad component clearly varies, especially on the blue wing, while we found it to be quite stable here. It seems that the behavior of the system was different between the two epochs. The \ha\ and \hb\ line EWs are indeed systematically higher in \cite{McGinnis20} and were comparable to the highest values we measured here during the JD 2,459,509 accretion burst. It is therefore likely that GM Aur was in a state of more active accretion during the 2011 observations. This is consistent with the more triangular shape and lack of structure of the Balmer emission line profiles, which are predicted to become more optical thick as the funnel flow density increases \citep{Muzerolle01}. It is also consistent with the higher level of variability seen in  the blue wing of the broad component of the \hei\ 5876 \AA\ line profile that may betray the existence of a hot accretion-driven wind at times of enhanced accretion \citep{Beristain01}. Long-term changes in the system behavior driven by a varying mass accretion rate and/or a change in the magnetic topology are therefore likely to occur. Over the six-month span of our observing campaign, the significant variation observed in the EW of the \pab\ line profile, which by the end of the campaign reaches similar levels to those measured during the JD 2,459,509 accretion burst (see Fig.~\ref{fig:EW_ExTrA}), suggests that such changes may occur on a timescale of a few weeks to months.

\section{Conclusion}
\label{sec:conclusion}

By combining optical and near-infrared high-resolution spectroscopic time series, seconded by a long-term monitoring of the photometric variability of the system and low-resolution near-infrared spectrophotometry, we were able to characterize the accretion and ejection process occurring in the young system GM Aur on a timescale ranging from days to months. We report a stable accretion pattern according to which the large-scale magnetic field of the star controls the accretion of gas from the inner disk onto the central star along funnel flows. The appearance of inverse P Cygni profiles that signal the crossing of funnel flows on the line of sight is remarkably periodic at the stellar rotation period of 6.04 days. Similarly, the photometric and line flux variations, both driven by the visibility of the accretion shock located at the foot of the main accretion column, are modulated at the same period. While the amplitude varies, the phase of variability of all these accretion diagnostics remains stable over the 30 rotational periods covered by the campaign. This suggests that the underlying magnetic topology that controls the non-axisymmetric accretion flow, presumably an inclined dipole on the large scale, did not undergo major changes over a timescale of six months. In stark contrast, high-velocity blueshifted absorption components that indicate outflows appear at random times in the emission line profiles. They are not rotationally modulated, and their signatures last for a few days only. We argue that these transient outflows associated with a stable accretion pattern are best accounted for by magnetospheric ejection models, as predicted by MHD simulations. 
Thus, by probing the dynamics of the star-disk interaction region, these results show that the physical connection between accretion and ejection processes that has long been established on large scales also appears to be valid on the much smaller sub-au scales.

\begin{acknowledgements}

We thank the referee for a prompt and detailed report. This study is based on observations obtained at the Canada-France-Hawaii Telescope (CFHT) which is operated by the National Research Council (NRC) of Canada, the Institut National des Sciences de l'Univers of the Centre National de la Recherche Scientifique (CNRS) of France, and the University of Hawaii. The observations at the CFHT were performed with care and respect from the summit of Maunakea which is a significant cultural and historic site; based on observations made at Observatoire de Haute Provence (CNRS), France; based on data collected under the ExTrA project at the ESO La Silla Paranal Observatory. ExTrA is a project of Institut de Plan\'etologie et d'Astrophysique de Grenoble (IPAG/CNRS/UGA), funded by the European Research Council under the ERC Grant Agreement n. 337591-ExTrA. We thank \'Agnes K\'osp\'al for providing a reduced TESS light curve of GM Aur. Funding for the TESS mission is provided by NASA’s Science Mission directorate. This project has received funding from the European Research Council (ERC) under the European Union’s Horizon 2020 research and innovation programme (grant agreement no. 742095; SPIDI: Star-Planets-Inner Disk-Interactions, http://www.spidi-eu.org). We acknowledge funding from the French National Research Agency (ANR) under contract number ANR-18-CE31-0019 (SPlaSH).
SHPA acknowledges financial support from CNPq, CAPES and Fapemig. JFD acknowledges funding from the European Research Council (ERC) under the H2020 research \& innovation programme (grant agreement no. 740651 NewWorlds). AF acknowledges support by the PRIN-INAF 2019 STRADE (Spectroscopically TRAcing the Disk dispersal Evolution) and by the Large Grant INAF YODA (YSOs Outflow, Disks and Accretion). JFG was supported by funda\c c\~ao para a Ci\^encia e Tecnologia (FCT) through the research grants UIDB/04434/2020 and UIDP/04434/2020. This work benefited from discussions with the ODYSSEUS (HST AR-16129) and PENELLOPE teams. Some of the plots presented in this paper were built using TOPCAT \citep{Taylor05}.  \end{acknowledgements}

\bibliographystyle{aa} 
\bibliography{gmaur_rev0} 

\begin{appendix} 

\section {REM comparison stars and optical photometry}
\label{app:rem}

Table~\ref{tab:sources} lists the comparison stars we used to calibrate the REM optical photometry (see Section~\ref{sec:rem}). Table~\ref{tab:remoptphot} lists the derived g'r'i'z' magnitudes for GM Aur. 

\begin{table*}
\caption{Literature data for comparison stars in the field of GM Aur on the REM cameras.}
\begin{center}
\begin{tabular}{lccccccccc}   
\hline\hline
\noalign{\smallskip}
Id & Name  & 2MASS &  $g$  &  $r$   &  $i$   &  $z$   &  $J$   &  $H$   &  $K'$  \\             
   &       &       & (mag) &  (mag) &  (mag) &  (mag) &  (mag) &  (mag) &  (mag) \\
\hline
\noalign{\smallskip}

*2   &  HD 282625     & J04551650+3022369 & 11.310  & 10.914  & 10.772  & 10.707 &  9.781  &  9.554 &  9.487 \\   
*3   &  TIC 96533048  & J04551015+3021333 & 12.361  & 11.399  & 10.936  & 10.668 &  9.295  &  8.716 &  8.563 \\   
*4   &  HD 282626     & J04550536+3020382 & 11.747  & 11.375  & 11.238  & 11.174 & 10.239  & 10.052 &  9.987 \\   
*5   &  \dots         & J04551400+3020168 & 13.279  & 12.637  & 12.312  & 12.109 & 10.986  & 10.601 & 10.501 \\   
*6   &  \dots         & J04550078+3020090 & 13.351  & 12.288  & 11.604  & 11.135 &  9.564  &  8.828 &  8.608 \\   
*7   &  \dots         & J04550253+3019228 & 14.529  & 13.494  & 12.892  & 12.496 & 10.994  & 10.264 & 10.086 \\   
\noalign{\smallskip}
\hline  
\end{tabular}
\end{center}
Notes: $griz$ magnitudes from Pan-STARRS \citep{Tonry18}; 
$JHK'$ magnitudes from 2MASS \citep{Cutri03}.\\
\label{tab:sources}
\end{table*}

\begin{table}
\centering
  \setlength{\tabcolsep}{4pt}
  \scriptsize
    \renewcommand{\arraystretch}{0.9}
\caption{REM g'r'i'z' photometry} 
\begin{tabular}{lllllllll}   
\hline\hline
\noalign{\smallskip}
Julian date & g' & err & r' & err & i' & err & z' & err \\
(2,450,000+)   &  (mag) &  (mag) &  (mag) &  (mag) &  (mag) &  (mag) &  (mag) \\
\hline
\noalign{\smallskip}
9497.72986  &  12.525  &  0.045  &  11.566  &  0.008  &  11.161  &  0.006  &  10.861  &  0.012 \\
9497.73190  &  12.559  &  0.042  &  11.587  &  0.011  &  11.171  &  0.007  &  10.872  &  0.017 \\
9497.73396  &  12.566  &  0.051  &  11.601  &  0.019  &  11.193  &  0.004  &  10.892  &  0.007 \\
9500.80099  &  12.67  &  0.016  &  11.589  &  0.005  &  11.174  &  0.016  &  10.827  &  0.004 \\
9500.80302  &  12.667  &  0.02  &  11.592  &  0.001  &  11.175  &  0.013  &  10.838  &  0.006 \\
9500.80505  &  12.682  &  0.01  &  11.608  &  0.007  &  11.182  &  0.019  &  10.853  &  0.005 \\
9501.80935  &  12.617  &  0.02  &  11.58  &  0.01  &  11.168  &  0.01  &  10.871  &  0.007 \\
9501.81138  &  12.626  &  0.024  &  11.582  &  0.01  &  11.169  &  0.007  &  10.871  &  0.006 \\
9501.81344  &  12.635  &  0.019  &  11.585  &  0.013  &  11.173  &  0.008  &  10.868  &  0.004 \\
9502.83317  &  12.5  &  0.022  &  11.544  &  0.003  &  11.144  &  0.012  &  10.825  &  0.01 \\
9502.83523  &  12.52  &  0.024  &  11.548  &  0.003  &  11.149  &  0.01  &  10.834  &  0.011 \\
9502.83729  &  12.519  &  0.021  &  11.55  &  0.004  &  11.149  &  0.012  &  10.843  &  0.011 \\
9503.83744  &  12.494  &  0.035  &  11.522  &  0.007  &  11.129  &  0.014  &  10.812  &  0.009 \\
9503.83947  &  12.494  &  0.032  &  11.515  &  0.004  &  11.127  &  0.013  &  10.802  &  0.009 \\
9503.84153  &  12.505  &  0.033  &  11.521  &  0.004  &  11.121  &  0.013  &  10.816  &  0.01 \\
9504.84169  &  12.75  &  0.028  &  11.656  &  0.008  &  11.223  &  0.004  &  10.902  &  0.02 \\
9504.84375  &  12.749  &  0.029  &  11.653  &  0.008  &  11.222  &  0.007  &  10.897  &  0.019 \\
9504.84578  &  12.75  &  0.027  &  11.654  &  0.009  &  11.223  &  0.006  &  10.89  &  0.02 \\
9506.78944  &  12.81  &  0.041  &  11.67  &  0.01  &  11.239  &  0.009  &  10.907  &  0.006 \\
9506.79150  &  12.812  &  0.031  &  11.671  &  0.006  &  11.243  &  0.012  &  10.906  &  0.01 \\
9506.79353  &  12.797  &  0.034  &  11.679  &  0.008  &  11.233  &  0.011  &  10.886  &  0.009 \\
9507.79360  &  12.503  &  0.045  &  11.53  &  0.014  &  11.134  &  0.008  &  10.814  &  0.01 \\
9507.79566  &  12.495  &  0.043  &  11.534  &  0.011  &  11.131  &  0.009  &  10.822  &  0.015 \\
9507.79772  &  12.477  &  0.044  &  11.519  &  0.013  &  11.127  &  0.009  &  10.82  &  0.012 \\
9508.80257  &  12.135  &  0.048  &  11.284  &  0.006  &  10.957  &  0  &  10.661  &  0 \\
9508.80461  &  12.141  &  0.042  &  11.277  &  0.001  &  10.953  &  0  &  10.65  &  0.005 \\
9508.80664  &  12.13  &  0.037  &  11.278  &  0.004  &  10.952  &  0.002  &  10.653  &  0.01 \\
9509.80682  &  12.039  &  0.042  &  11.218  &  0  &  10.929  &  0.013  &  10.627  &  0.008 \\
9509.80888  &  12.021  &  0.035  &  11.209  &  0.001  &  10.924  &  0.017  &  10.63  &  0.013 \\
9509.81092  &  12.027  &  0.033  &  11.217  &  0.001  &  10.929  &  0.014  &  10.618  &  0.012 \\
9510.86553  &  12.498  &  0.025  &  11.493  &  0.001  &  11.105  &  0.016  &  10.788  &  0.008 \\
9510.86759  &  12.511  &  0.024  &  11.494  &  0.002  &  11.123  &  0.013  &  10.787  &  0.008 \\
9510.86963  &  12.517  &  0.022  &  11.495  &  0.001  &  11.109  &  0.008  &  10.801  &  0.019 \\
9513.71086  &  12.66  &  0.017  &  11.618  &  0.002  &  11.202  &  0.011  &  10.866  &  0.013 \\
9513.71300  &  12.659  &  0.017  &  11.615  &  0.001  &  11.209  &  0.013  &  10.867  &  0 \\
9513.71503  &  12.653  &  0.016  &  11.601  &  0.001  &  11.216  &  0.009  &  10.868  &  0.011 \\
9514.78750  &  12.531  &  0.034  &  11.536  &  0.001  &  11.169  &  0.013  &  10.841  &  0.004 \\
9514.78954  &  12.543  &  0.028  &  11.534  &  0.001  &  11.158  &  0.013  &  10.832  &  0.013 \\
9514.79157  &  12.544  &  0.026  &  11.535  &  0.003  &  11.17  &  0.01  &  10.819  &  0.008 \\
9515.79177  &  12.677  &  0.016  &  11.617  &  0.003  &  11.184  &  0.002  &  10.861  &  0.025 \\
9515.79383  &  12.693  &  0.017  &  11.622  &  0.006  &  11.191  &  0.007  &  10.868  &  0.017 \\
9515.79586  &  12.689  &  0.019  &  11.621  &  0.008  &  11.184  &  0  &  10.854  &  0.02 \\
9517.72459  &  12.762  &  0.019  &  11.641  &  0.007  &  11.252  &  0.008  &  10.898  &  0.009 \\
9517.72665  &  12.774  &  0.014  &  11.642  &  0.01  &  11.249  &  0.002  &  10.904  &  0.004 \\
9517.72869  &  12.769  &  0.016  &  11.638  &  0.01  &  11.247  &  0.006  &  10.9  &  0.009 \\
9518.73156  &  12.774  &  0.021  &  11.663  &  0.002  &  11.228  &  0.012  &  10.898  &  0.008 \\
9518.73362  &  12.778  &  0.019  &  11.659  &  0.002  &  11.222  &  0.012  &  10.887  &  0.008 \\
9518.73568  &  12.771  &  0.017  &  11.659  &  0.001  &  11.227  &  0.013  &  10.892  &  0.011 \\
9519.73968  &  12.76  &  0.051  &  11.693  &  0.012  &  11.223  &  0.01  &  10.879  &  0.006 \\
9519.74174  &  12.758  &  0.051  &  11.683  &  0.011  &  11.218  &  0.008  &  10.877  &  0.006 \\
9519.74378  &  12.766  &  0.051  &  11.682  &  0.011  &  11.214  &  0.01  &  10.882  &  0.011 \\
9520.74391  &  12.641  &  0.016  &  11.597  &  0.001  &  11.176  &  0.012  &  10.846  &  0.008 \\
9520.74595  &  12.651  &  0.016  &  11.596  &  0.001  &  11.177  &  0.015  &  10.866  &  0.007 \\
9520.74801  &  12.662  &  0.014  &  11.603  &  0.001  &  11.177  &  0.011  &  10.867  &  0.008 \\
\noalign{\smallskip}
\hline  
\end{tabular}
\label{tab:remoptphot}
\end{table}

\section{Line profile correlation matrices}
\label{app:cm}

\begin{figure*}
  \centering
  \includegraphics[width=0.3\textwidth]{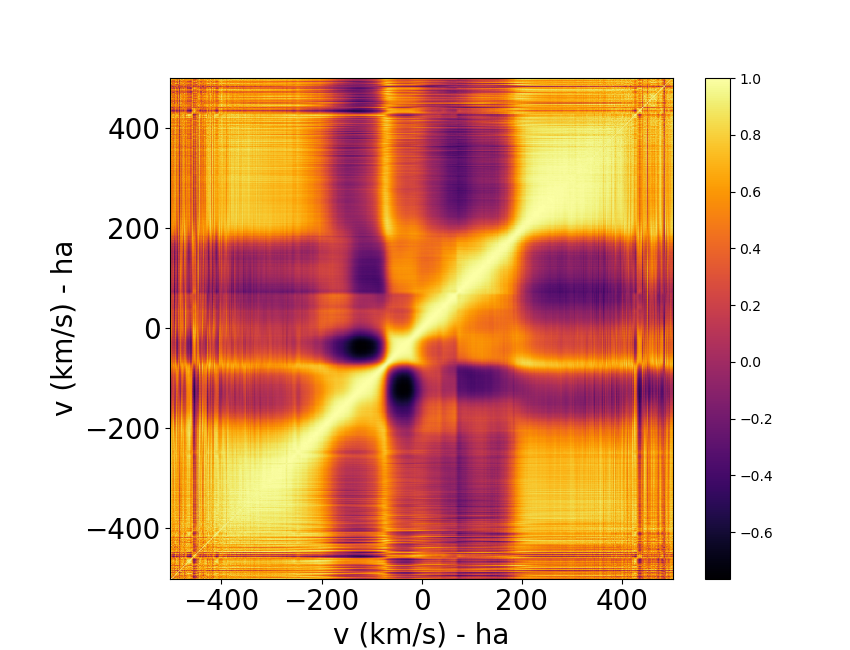}
  \includegraphics[width=0.3\textwidth]{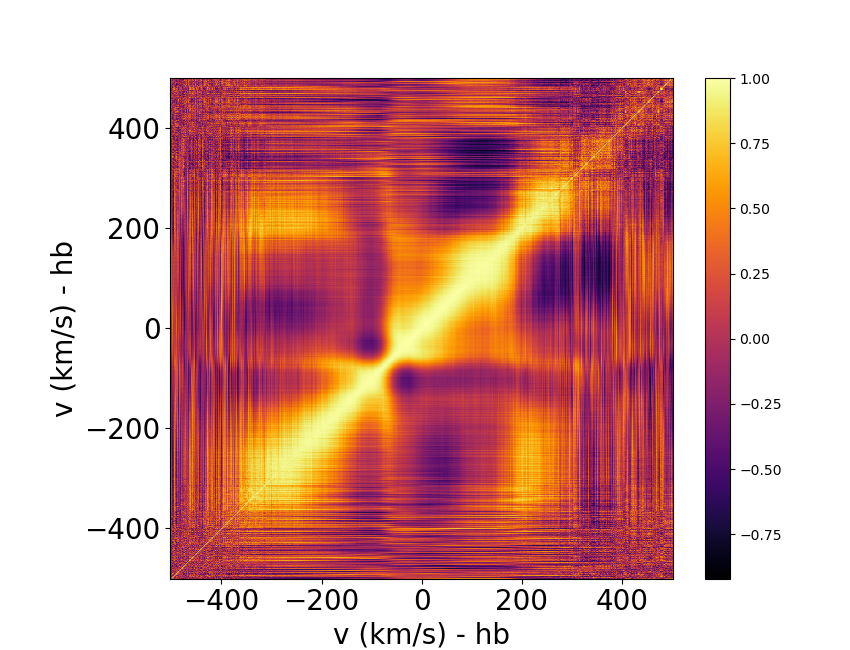}
  \includegraphics[width=0.3\textwidth]{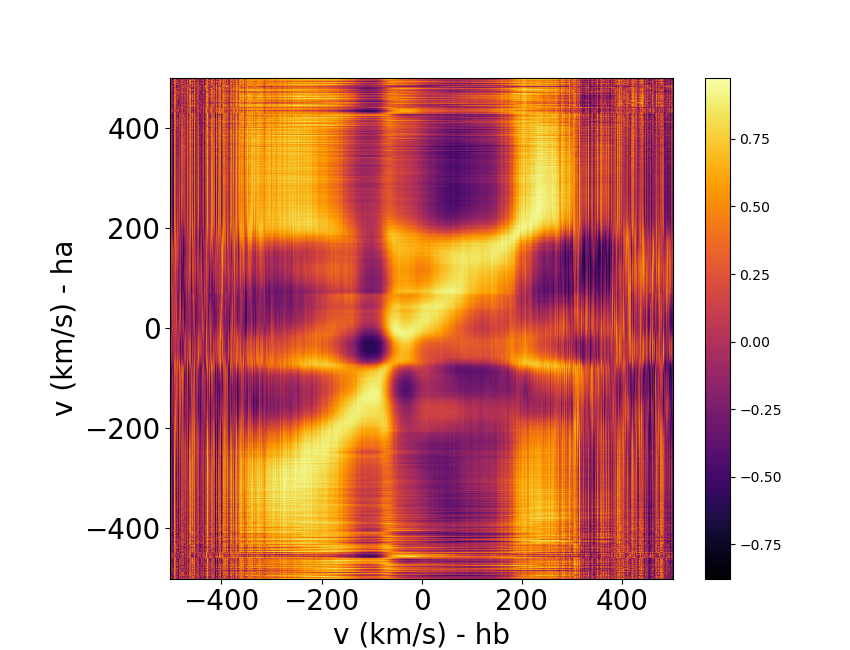}
  \caption{ Correlation matrices for hydrogen optical line profiles computed from the 15 OHP/SOPHIE spectra obtained during the campaign: \ha$\star$\ha\ ({\it left}), \hb$\star$\hb\ (\it{center}), \textup{and} \ha$\star$\hb\ (\it{right}).} 
  \label{fig:cmhahb}
\end{figure*}

\begin{figure*}
  \centering
  \includegraphics[width=0.3\textwidth]{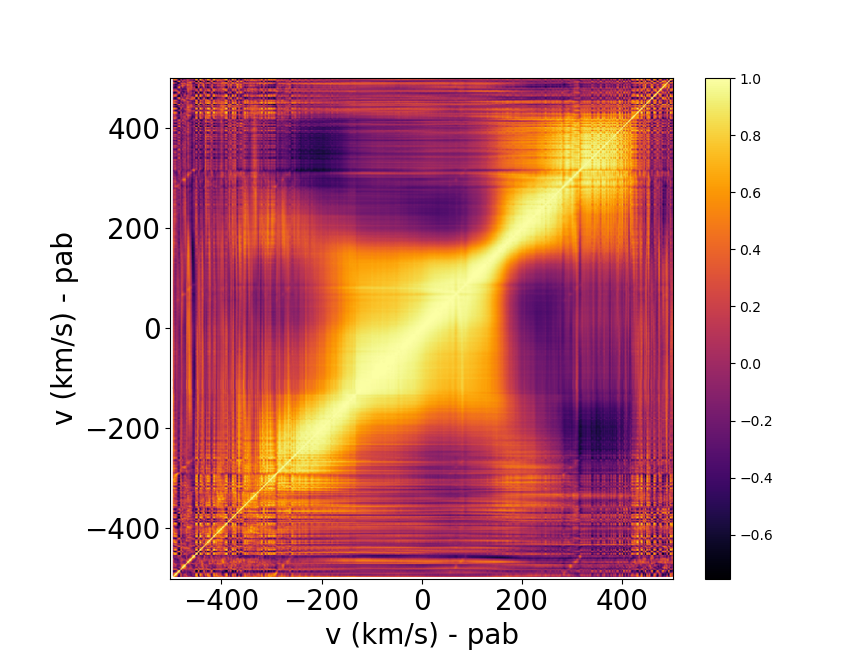}
  \includegraphics[width=0.3\textwidth]{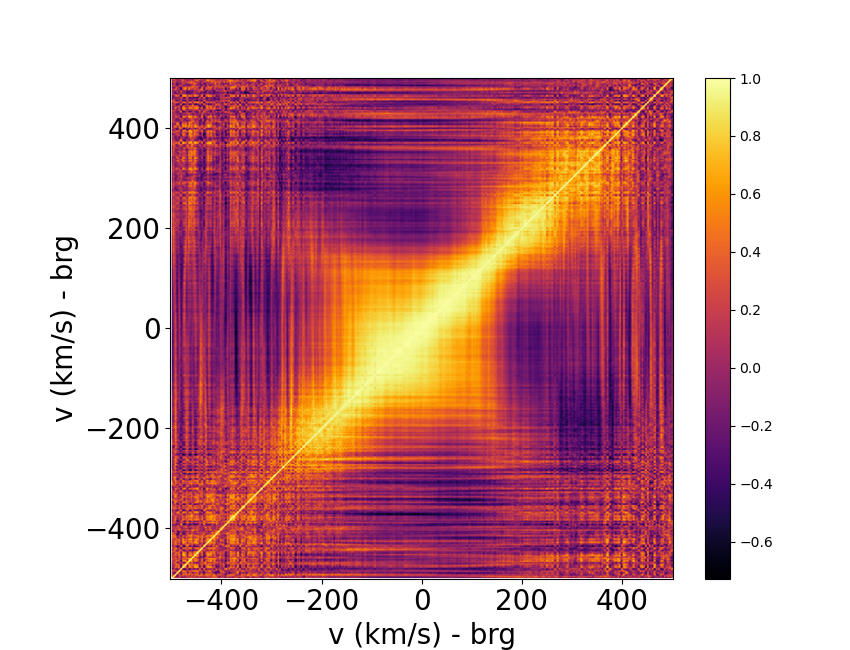}
  \includegraphics[width=0.3\textwidth]{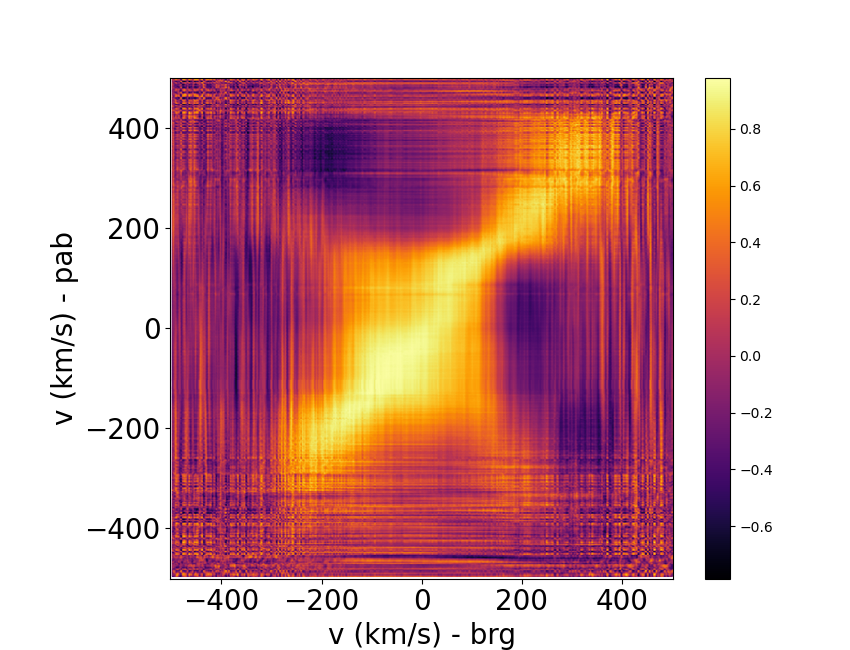}
  \caption{ Correlation matrices for the hydrogen near-infrared line profiles computed from the 34 CFHT/SPIRou spectra obtained during the campaign : \pab$\star$\pab\ ({\it left}), \brg$\star$\brg\ (\it{center}), \textup{and} \pab$\star$\brg\ (\it{right}).} 
  \label{fig:cmpabbrg}
\end{figure*}

\begin{figure*}
  \centering
    \includegraphics[width=0.3\textwidth]{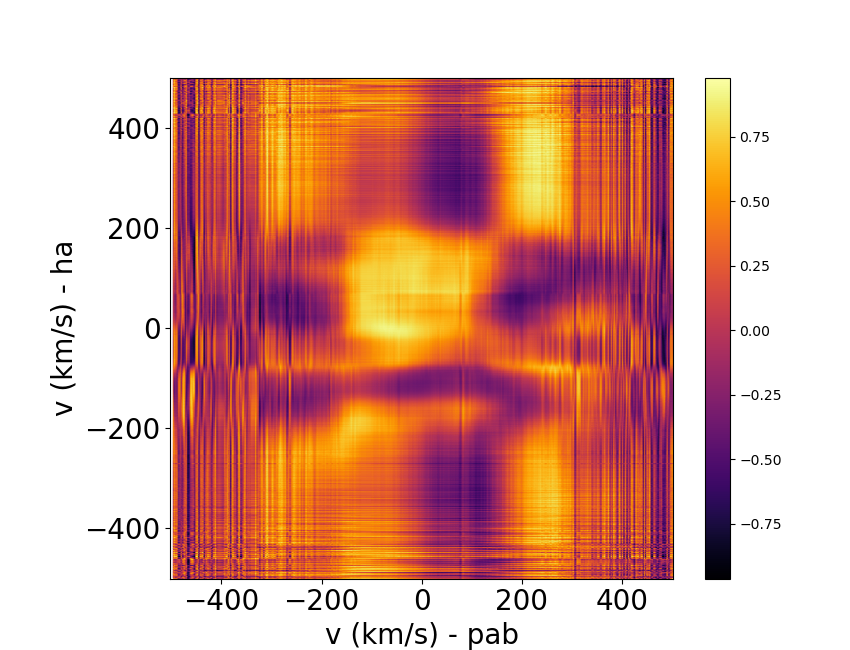}
  \includegraphics[width=0.3\textwidth]{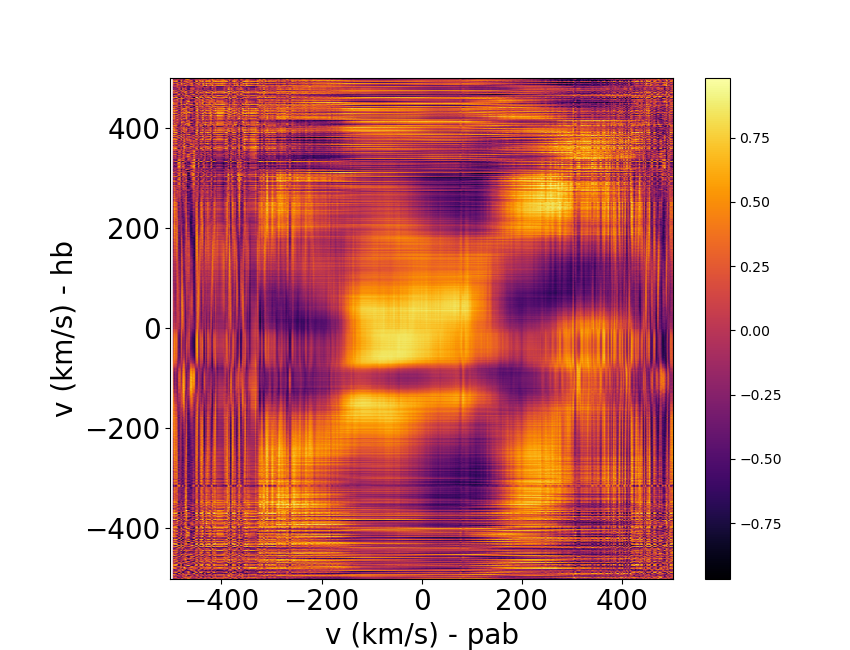}
  \includegraphics[width=0.3\textwidth]{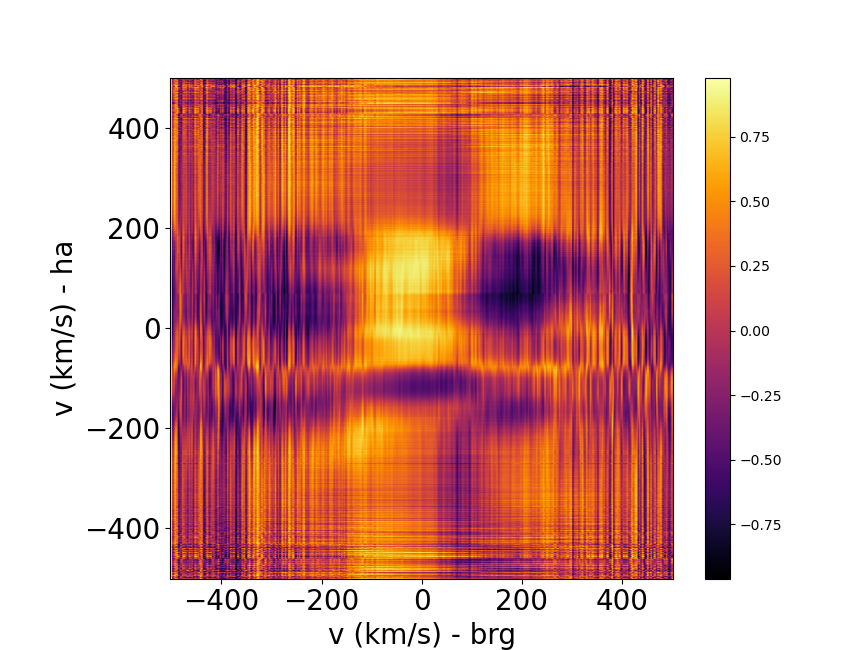}
  \includegraphics[width=0.3\textwidth]{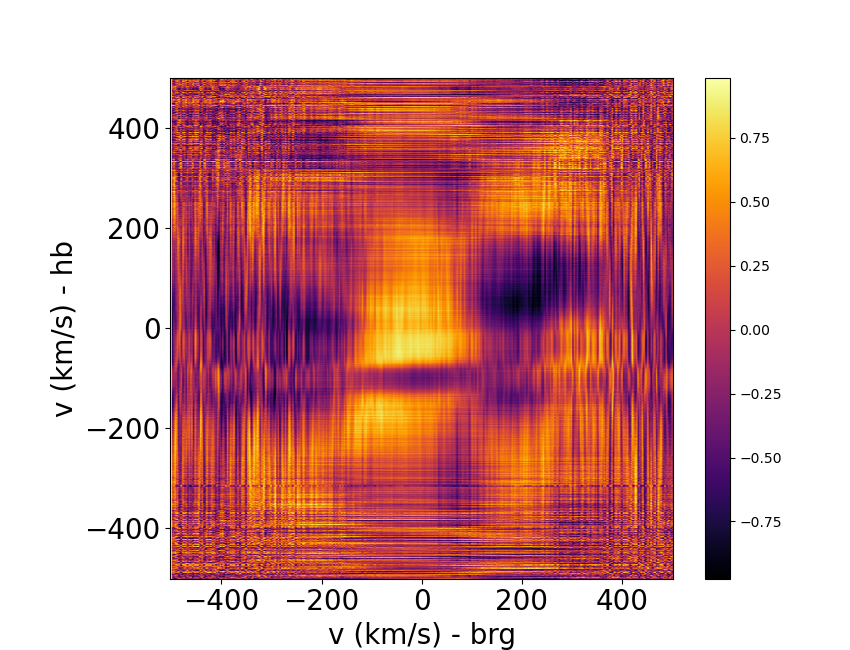}
  \caption{ Correlation matrices between optical and near-infrared hydrogen line profiles computed from ten OHP/SOPHIE and ten CFHT/SPIRou spectra obtained over the same nights during the October runs: \ha$\star$\pab\ ({\it left}), \hb$\star$\pab\ (\it{center left}), \ha$\star$\brg\ (\it{center right}), \textup{and} \hb$\star$\brg\ (\it{right}).} 
  \label{fig:cmhapab}
\end{figure*}

\begin{figure*}
  \centering
  \includegraphics[width=0.3\textwidth]{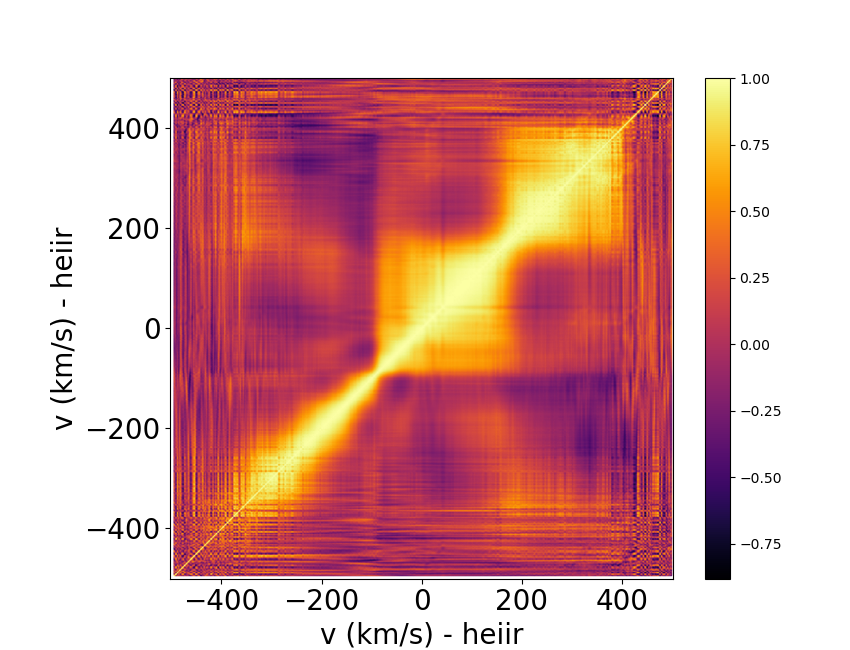}
  \includegraphics[width=0.3\textwidth]{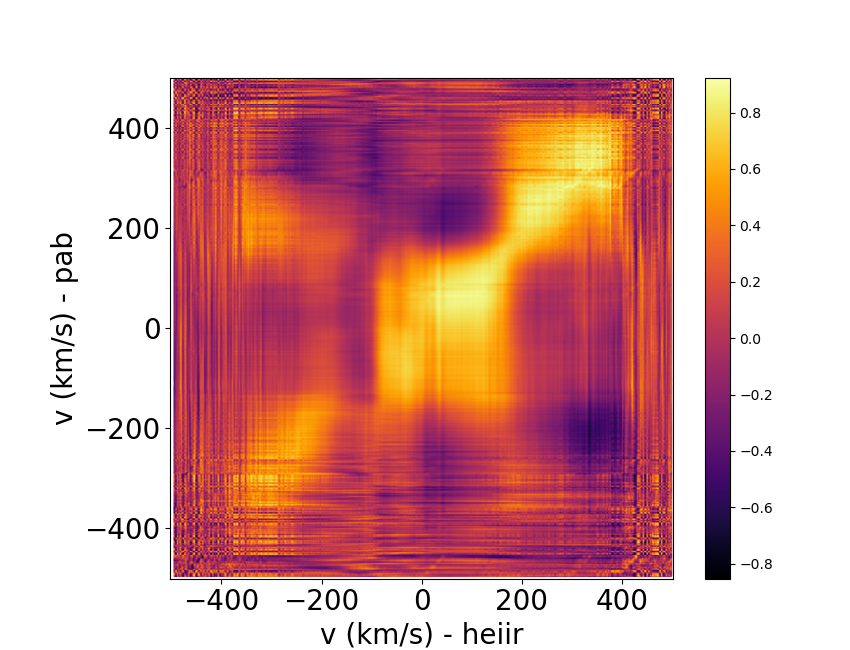}
  \includegraphics[width=0.3\textwidth]{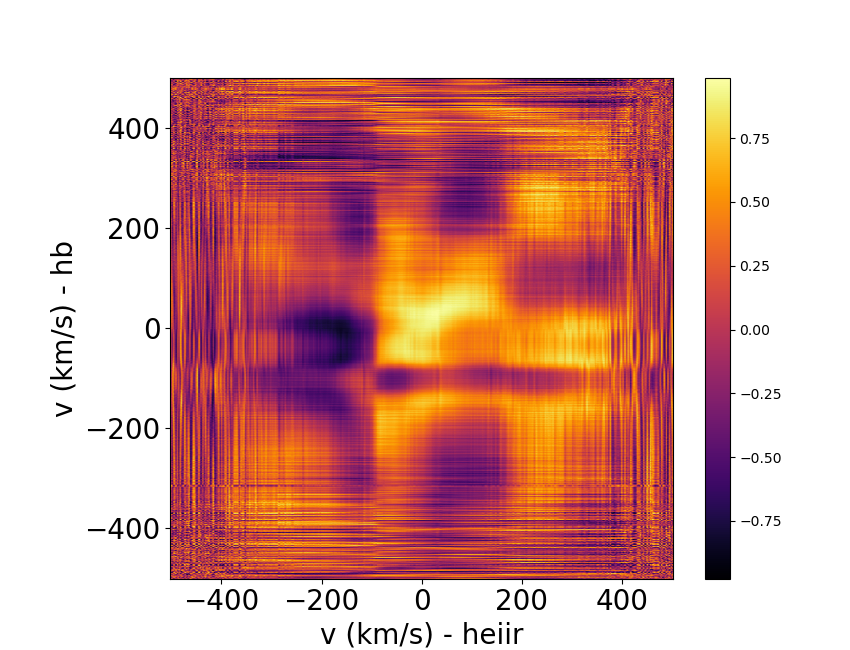}
  \caption{Correlation matrices for the \hei\ 10830 \AA\ and hydrogen lines:  \hei$\star$\hei\ ({\it left}), \pab$\star$\hei\ ({\it center}), and \hb$\star$\hei\ ({\it right}). The \brg$\star$\hei\ and \ha$\star$\hei\ matrices are not shown as they are similar to those of \pab$\star$\hei\ and \hb$\star$\hei, respectively.} 
  \label{fig:cmheipab}
\end{figure*}

This section provides correlation matrices between line profiles \citep{Johns95}. Correlation coefficients are computed across the line profiles for every pair of velocity channels common to the two profiles. Fig.~\ref{fig:cmhahb} to \ref{fig:cmheipab} present the correlation matrices of the optical and near-infrared line profiles studied here. 

\section{EWs and J-band measurements from the ExTrA spectra}
\label{app:extraewJ}

We provide the \hei\ 10830~\AA, \pab, \pac, and \pad\ line EWs and the J-band photometry measured from the ExTrA spectra  in Table~\ref{tab:extraewJ}. For each night, the table lists the mean observation time, the median EW measurement and its standard deviation for each spectral line, and the J-band magnitude and its error.

\begin{table}
  \centering
  \setlength{\tabcolsep}{4pt}
  \scriptsize
  \renewcommand{\arraystretch}{0.9}
  \caption{EW measurements and J-band photometry from the ExTrA spectra.}\label{table:EW_ExTrA}
\begin{tabular}{ccccccccccc}
\hline
Julian date &  EW(HeI) & EW(Pa$\beta$) & EW(Pa$\gamma$) & EW(Pa$\delta$) & J &errJ \\
(2,450,000+) & (\AA) & (\AA) & (\AA) & (\AA) & (mag) & (mag)\\
\hline
9501.89138 &  2.95 $\pm$ 0.80 &  10.45 $\pm$ 0.28 &  8.23 $\pm$ 0.88 &  2.37 $\pm$ 0.80       & 9.487 & 0.028 \\ 
9504.85514 &  2.43 $\pm$ 0.50 &  9.33 $\pm$ 0.61 &  4.93 $\pm$ 0.70 &  1.48 $\pm$ 0.67       & 9.499 & 0.027 \\ 
9506.86219 &  2.43 $\pm$ 0.58 &  10.91 $\pm$ 0.72 &  7.28 $\pm$ 0.60 &  2.80 $\pm$ 0.64       & 9.451 & 0.027 \\ 
9507.81204 &  7.87 $\pm$ 0.70 &  14.65 $\pm$ 1.00 &  9.57 $\pm$ 0.62 &  3.92 $\pm$ 0.73       & 9.411 & 0.027 \\ 
9509.84834 &  10.06 $\pm$ 0.61 &  16.56 $\pm$ 0.82 &  9.05 $\pm$ 0.60 &  4.23 $\pm$ 0.69          & 9.316 & 0.027 \\ 
9510.88153$^\dagger$  &  5.91 $\pm$ 0.61 &  12.95 $\pm$ 0.93 &  7.62 $\pm$ 0.58 &  2.42 $\pm$ 0.59          & --- & --- \\ 
9512.79907 &  1.87 $\pm$ 0.30 &  7.09 $\pm$ 1.20 &  7.14 $\pm$ 0.27 &  1.14 $\pm$ 0.02       & 9.461 & 0.027 \\ 
9513.79425 &  5.50 $\pm$ 0.69 &  10.50 $\pm$ 1.07 &  8.73 $\pm$ 1.39 &  2.90 $\pm$ 1.29       & 9.407 & 0.027 \\ 
9514.84366 &  2.16 $\pm$ 0.67 &  10.56 $\pm$ 0.83 &  7.42 $\pm$ 0.71 &  2.56 $\pm$ 0.68       & 9.444 & 0.027 \\ 
9517.84528 &  4.26 $\pm$ 0.74 &  11.42 $\pm$ 0.78 &  7.85 $\pm$ 0.95 &  2.74 $\pm$ 0.83       & 9.495 & 0.037 \\ 
9518.84178 &  1.08 $\pm$ 0.55 &  7.88 $\pm$ 0.73 &  7.77 $\pm$ 0.50 &  2.08 $\pm$ 0.82       & 9.482 & 0.027 \\ 
9519.86259 &  4.16 $\pm$ 0.75 &  9.48 $\pm$ 0.71 &  7.76 $\pm$ 0.56 &  3.05 $\pm$ 1.01       & 9.464 & 0.027 \\ 
9520.85926 &  4.42 $\pm$ 0.66 &  13.95 $\pm$ 0.57 &  8.53 $\pm$ 0.78 &  2.29 $\pm$ 0.35       & 9.444 & 0.027 \\ 
9521.87770 &  5.36 $\pm$ 0.48 &  12.38 $\pm$ 0.25 &  9.63 $\pm$ 0.94 &  1.63 $\pm$ 0.23       & 9.424 & 0.027 \\ 
9522.79321 &  3.04 $\pm$ 0.31 &  14.32 $\pm$ 0.41 &  7.54 $\pm$ 0.56 &  2.51 $\pm$ 0.48       & 9.448 & 0.027 \\ 
9524.87753 &  1.01 $\pm$ 0.84 &  12.81 $\pm$ 1.07 &  8.01 $\pm$ 0.69 &  1.40 $\pm$ 0.27       & 9.467 & 0.027 \\ 
9525.86383 &  2.47 $\pm$ 0.42 &  11.31 $\pm$ 0.86 &  6.61 $\pm$ 0.49 &  1.13 $\pm$ 0.76       & 9.473 & 0.027 \\ 
9526.86452 &  2.24 $\pm$ 0.47 &  8.44 $\pm$ 0.63 &  5.00 $\pm$ 0.70 &  1.17 $\pm$ 1.18       & 9.476 & 0.027 \\ 
9527.86505 &  2.71 $\pm$ 0.64 &  10.61 $\pm$ 0.76 &  5.91 $\pm$ 0.46 &  1.08 $\pm$ 0.74       & 9.442 & 0.027 \\ 
9528.82389 &  1.59 $\pm$ 0.52 &  8.35 $\pm$ 1.12 &  4.64 $\pm$ 0.72 &  0.02 $\pm$ 0.51       & 9.475 & 0.027 \\ 
9529.85920 &  0.03 $\pm$ 0.65 &  6.70 $\pm$ 0.70 &  3.97 $\pm$ 0.85 &  0.33 $\pm$ 0.59       & 9.463 & 0.027 \\ 
9530.85025 &  0.30 $\pm$ 0.81 &  9.43 $\pm$ 0.96 &  5.83 $\pm$ 1.17 &  1.13 $\pm$ 0.75       & 9.477 & 0.029 \\ 
9531.81805 &  2.70 $\pm$ 0.46 &  8.98 $\pm$ 0.62 &  6.50 $\pm$ 0.49 &  2.82 $\pm$ 0.46       & 9.460 & 0.027 \\ 
9532.82519 &  3.27 $\pm$ 0.61 &  10.56 $\pm$ 0.82 &  5.81 $\pm$ 0.57 &  2.67 $\pm$ 0.65       & 9.429 & 0.035 \\ 
9533.85615 &  4.04 $\pm$ 0.65 &  6.78 $\pm$ 0.60 &  4.30 $\pm$ 0.61 &  1.31 $\pm$ 0.63       & 9.428 & 0.027 \\ 
9534.84045 &  4.57 $\pm$ 0.86 &  8.43 $\pm$ 0.95 &  5.70 $\pm$ 0.60 &  2.12 $\pm$ 0.74       & 9.425 & 0.090 \\ 
9535.83697 &  1.85 $\pm$ 0.52 &  8.08 $\pm$ 0.71 &  5.85 $\pm$ 0.68 &  1.84 $\pm$ 0.62       & 9.456 & 0.053 \\ 
9536.86516 &  0.15 $\pm$ 0.64 &  6.93 $\pm$ 2.51 &  6.94 $\pm$ 0.30 &  2.34 $\pm$ 0.03       & 9.358 & 0.027 \\ 
9537.78315 &  3.45 $\pm$ 1.60 &  6.97 $\pm$ 0.98 &  5.43 $\pm$ 1.00 &  2.06 $\pm$ 0.69       & 9.545 & 0.027 \\ 
9538.80149 &  2.83 $\pm$ 1.64 &  5.43 $\pm$ 1.42 &  4.09 $\pm$ 1.20 &  1.59 $\pm$ 0.86       & 9.531 & 0.027 \\ 
9539.81823 &  3.43 $\pm$ 0.68 &  7.42 $\pm$ 1.27 &  5.50 $\pm$ 0.60 &  1.48 $\pm$ 0.65       & 9.396 & 0.028 \\ 
9540.84522 &  3.61 $\pm$ 0.72 &  10.38 $\pm$ 2.17 &  6.80 $\pm$ 0.84 &  2.58 $\pm$ 0.70       & 9.313 & 0.037 \\ 
9543.79258 &  2.94 $\pm$ 0.46 &  10.10 $\pm$ 0.79 &  6.48 $\pm$ 0.40 &  1.94 $\pm$ 0.61       & 9.419 & 0.027 \\ 
9544.73410 &  2.34 $\pm$ 0.48 &  9.12 $\pm$ 0.59 &  5.92 $\pm$ 0.42 &  1.68 $\pm$ 0.52       & 9.376 & 0.027 \\ 
9547.68435 &  0.08 $\pm$ 0.37 &  4.16 $\pm$ 0.69 &  4.52 $\pm$ 0.58 &  0.46 $\pm$ 0.36       & 9.443 & 0.027 \\ 
9548.72798 & -1.19 $\pm$ 0.62 &  2.79 $\pm$ 0.90 &  4.06 $\pm$ 0.62 &  0.30 $\pm$ 0.68       & 9.568 & 0.027 \\ 
9549.73881 &  1.79 $\pm$ 0.63 &  6.06 $\pm$ 1.24 &  5.67 $\pm$ 0.82 &  1.21 $\pm$ 0.69       & 9.537 & 0.027 \\ 
9550.69491 &  0.74 $\pm$ 0.41 &  3.22 $\pm$ 0.74 &  3.84 $\pm$ 0.59 &  1.16 $\pm$ 0.44       & 9.492 & 0.027 \\ 
9551.70856 &  2.48 $\pm$ 0.64 &  5.54 $\pm$ 0.70 &  4.66 $\pm$ 0.48 &  1.22 $\pm$ 0.54       & 9.474 & 0.028 \\ 
9553.69774 &  1.13 $\pm$ 0.64 &  6.05 $\pm$ 0.62 &  4.50 $\pm$ 0.63 &  0.87 $\pm$ 0.61       & 9.443 & 0.028 \\ 
9555.72248 &  4.05 $\pm$ 0.75 &  12.83 $\pm$ 1.35 &  7.78 $\pm$ 0.94 &  2.04 $\pm$ 0.72       & 9.266 & 0.028 \\ 
9556.67400 &  1.85 $\pm$ 0.57 &  11.64 $\pm$ 1.85 &  6.23 $\pm$ 1.62 &  1.48 $\pm$ 0.61       & 9.333 & 0.027 \\ 
9560.71477 &  0.99 $\pm$ 0.56 &  6.39 $\pm$ 0.68 &  5.64 $\pm$ 0.63 &  1.31 $\pm$ 0.70       & 9.507 & 0.028 \\ 
9561.73574 &  2.54 $\pm$ 0.44 &  7.87 $\pm$ 0.52 &  7.21 $\pm$ 0.44 &  2.16 $\pm$ 0.57       & 9.481 & 0.028 \\ 
9564.76316 &  6.31 $\pm$ 0.66 &  11.98 $\pm$ 0.74 &  6.73 $\pm$ 0.67 &  1.83 $\pm$ 0.66       & 9.394 & 0.029 \\ 
9566.66048 & -1.91 $\pm$ 0.85 &  3.89 $\pm$ 0.75 &  3.44 $\pm$ 0.59 &  0.30 $\pm$ 0.51       & 9.505 & 0.028 \\ 
9567.68335 &  2.96 $\pm$ 0.69 &  8.80 $\pm$ 0.92 &  5.48 $\pm$ 0.39 &  1.18 $\pm$ 0.64       & 9.470 & 0.027 \\ 
9569.70386 &  2.77 $\pm$ 0.56 &  7.92 $\pm$ 0.78 &  5.88 $\pm$ 0.90 &  1.35 $\pm$ 0.75       & 9.397 & 0.029 \\ 
9570.68277 &  4.73 $\pm$ 0.48 &  10.35 $\pm$ 1.11 &  7.26 $\pm$ 0.55 &  2.46 $\pm$ 0.79       & 9.399 & 0.027 \\ 
9573.66353 &  4.33 $\pm$ 0.48 &  10.45 $\pm$ 0.41 &  9.01 $\pm$ 0.66 &  2.64 $\pm$ 0.41       & 9.432 & 0.027 \\ 
9577.61918 &  2.11 $\pm$ 0.36 &  8.54 $\pm$ 0.65 &  5.62 $\pm$ 0.51 &  1.43 $\pm$ 0.57       & 9.401 & 0.027 \\ 
9578.66562 &  7.11 $\pm$ 0.65 &  12.34 $\pm$ 0.76 &  7.62 $\pm$ 0.91 &  3.88 $\pm$ 0.55       & 9.368 & 0.027 \\ 
9580.62157 &  4.88 $\pm$ 0.50 &  10.48 $\pm$ 0.72 &  7.06 $\pm$ 0.53 &  2.42 $\pm$ 0.56       & 9.410 & 0.027 \\ 
9581.66843 &  6.99 $\pm$ 0.42 &  12.16 $\pm$ 0.54 &  7.38 $\pm$ 0.64 &  2.40 $\pm$ 0.46       & 9.331 & 0.027 \\ 
9582.60702 &  3.56 $\pm$ 0.56 &  10.93 $\pm$ 0.72 &  6.81 $\pm$ 0.86 &  1.99 $\pm$ 0.52       & 9.390 & 0.027 \\ 
9586.69987 &  6.29 $\pm$ 0.58 &  13.34 $\pm$ 1.26 &  9.30 $\pm$ 0.66 &  3.94 $\pm$ 0.51       & 9.406 & 0.029 \\ 
9588.70981 &  2.08 $\pm$ 0.26 &  6.93 $\pm$ 0.69 &  6.47 $\pm$ 0.55 &  1.77 $\pm$ 0.67       & 9.426 & 0.027 \\ 
9589.60234 &  2.20 $\pm$ 0.68 &  9.23 $\pm$ 1.17 &  7.62 $\pm$ 0.58 &  2.38 $\pm$ 0.52       & 9.429 & 0.027 \\ 
9596.62015 &  2.49 $\pm$ 0.58 &  9.68 $\pm$ 1.33 &  6.58 $\pm$ 0.47 &  2.26 $\pm$ 0.62       & 9.458 & 0.028 \\ 
9597.67444 &  4.58 $\pm$ 0.65 &  11.33 $\pm$ 0.95 &  9.61 $\pm$ 0.56 &  2.91 $\pm$ 0.64       & 9.450 & 0.028 \\ 
9598.59529 &  4.30 $\pm$ 0.54 &  10.94 $\pm$ 0.86 &  7.80 $\pm$ 0.48 &  2.43 $\pm$ 0.63       & 9.397 & 0.028 \\ 
9600.55490 &  7.14 $\pm$ 0.48 &  16.46 $\pm$ 0.86 &  10.65 $\pm$ 0.68 &  4.19 $\pm$ 0.58          & 9.424 & 0.028 \\ 
9601.56718 &  2.41 $\pm$ 0.72 &  8.23 $\pm$ 1.12 &  6.02 $\pm$ 0.59 &  1.59 $\pm$ 0.60       & 9.456 & 0.028 \\ 
9602.53523 &  1.35 $\pm$ 0.63 &  9.00 $\pm$ 0.75 &  6.44 $\pm$ 0.80 &  2.51 $\pm$ 0.61       & 9.486 & 0.027 \\ 
9605.54455 &  10.34 $\pm$ 0.74 &  17.45 $\pm$ 1.23 &  11.39 $\pm$ 0.82 &  4.42 $\pm$ 0.90         & 9.372 & 0.027 \\ 
9606.55663 &  4.71 $\pm$ 0.58 &  12.46 $\pm$ 1.33 &  7.92 $\pm$ 0.82 &  2.47 $\pm$ 0.82       & 9.409 & 0.028 \\ 
9607.55053 &  2.61 $\pm$ 0.51 &  11.21 $\pm$ 1.02 &  7.28 $\pm$ 0.60 &  2.37 $\pm$ 0.57       & 9.443 & 0.027 \\ 
9608.52952 & -0.27 $\pm$ 0.75 &  7.99 $\pm$ 1.16 &  5.62 $\pm$ 1.00 &  1.60 $\pm$ 0.60       & 9.484 & 0.027 \\ 
9615.52033 &  5.17 $\pm$ 0.56 &  13.40 $\pm$ 1.17 &  8.94 $\pm$ 0.97 &  3.34 $\pm$ 0.62       & 9.409 & 0.027 \\ 
9616.55954 &  6.72 $\pm$ 0.64 &  13.72 $\pm$ 1.09 &  9.56 $\pm$ 0.88 &  3.19 $\pm$ 0.62       & 9.360 & 0.027 \\ 
9617.55930 &  6.54 $\pm$ 0.92 &  15.62 $\pm$ 0.69 &  9.79 $\pm$ 0.72 &  3.73 $\pm$ 0.72       & 9.354 & 0.030 \\ 
9618.55535 &  8.27 $\pm$ 1.11 &  18.41 $\pm$ 1.05 &  11.10 $\pm$ 0.65 &  4.79 $\pm$ 0.72          & 9.332 & 0.028 \\ 
9619.56342 &  7.84 $\pm$ 0.76 &  14.54 $\pm$ 0.90 &  9.04 $\pm$ 0.72 &  3.59 $\pm$ 0.78       & 9.417 & 0.028 \\ 
9623.52755 &  11.79 $\pm$ 0.81 &  17.61 $\pm$ 1.58 &  10.38 $\pm$ 0.73 &  4.42 $\pm$ 0.75         & 9.320 & 0.027 \\ 
9624.55795 &  9.62 $\pm$ 0.78 &  14.33 $\pm$ 1.01 &  8.17 $\pm$ 0.75 &  3.10 $\pm$ 0.65       & 9.365 & 0.028 \\ 
9625.53847 &  6.42 $\pm$ 0.92 &  10.27 $\pm$ 1.41 &  6.52 $\pm$ 0.85 &  1.64 $\pm$ 0.70       & 9.438 & 0.028 \\ 
9626.54670 & -0.82 $\pm$ 0.70 &  5.39 $\pm$ 1.01 &  4.91 $\pm$ 0.92 &  1.34 $\pm$ 0.86       & 9.527 & 0.030 \\ 
9629.53317 &  7.30 $\pm$ 0.63 &  12.52 $\pm$ 0.87 &  7.25 $\pm$ 0.65 &  2.74 $\pm$ 0.60       & 9.336 & 0.028 \\ 
9630.53061 &  11.17 $\pm$ 0.86 &  18.58 $\pm$ 0.89 &  10.69 $\pm$ 0.81 &  4.68 $\pm$ 0.87         & 9.342 & 0.027 \\ 
9631.50698 &  8.20 $\pm$ 0.41 &  15.59 $\pm$ 1.21 &  9.15 $\pm$ 0.94 &  3.27 $\pm$ 0.30       & 9.409 & 0.027 \\ 
9632.52380 &  1.15 $\pm$ 0.82 &  9.75 $\pm$ 1.99 &  5.78 $\pm$ 0.74 &  1.04 $\pm$ 0.73       & 9.627 & 0.028 \\ 
9633.54300 &  1.99 $\pm$ 0.41 &  10.77 $\pm$ 1.32 &  6.97 $\pm$ 0.60 &  1.70 $\pm$ 0.31       & 9.445 & 0.027 \\ 
9634.53461 &  2.68 $\pm$ 0.59 &  8.75 $\pm$ 1.34 &  5.18 $\pm$ 0.56 &  1.25 $\pm$ 0.63       & 9.413 & 0.027 \\ 
9643.52095 &  1.44 $\pm$ 0.58 &  14.07 $\pm$ 1.00 &  10.88 $\pm$ 0.90 &  3.03 $\pm$ 0.62          & 9.393 & 0.028 \\ 
9644.52110 &  1.53 $\pm$ 0.68 &  10.47 $\pm$ 1.09 &  8.49 $\pm$ 0.89 &  1.87 $\pm$ 0.83       & 9.467 & 0.029 \\ 
9645.51054 &  2.58 $\pm$ 0.50 &  10.74 $\pm$ 2.45 &  9.04 $\pm$ 0.41 &  2.61 $\pm$ 0.71       & 9.416 & 0.027 \\ 
9646.52604 &  6.67 $\pm$ 1.02 &  14.56 $\pm$ 1.53 &  9.61 $\pm$ 0.92 &  2.84 $\pm$ 0.83       & 9.375 & 0.028 \\ 
9647.50937 &  2.58 $\pm$ 0.66 &  10.79 $\pm$ 1.33 &  7.04 $\pm$ 1.01 &  1.48 $\pm$ 0.64       & 9.392 & 0.028 \\ 
\hline
\end{tabular}
\noindent $^\dagger$ No J-band photometry could be obtained for that night due to poor seeing. 
\label{tab:extraewJ}
\end{table}

\section{nIR veiling and line EWs }
\label{app:nirveil}

\begin{figure}
  \centering
  \includegraphics[width=0.49\textwidth]{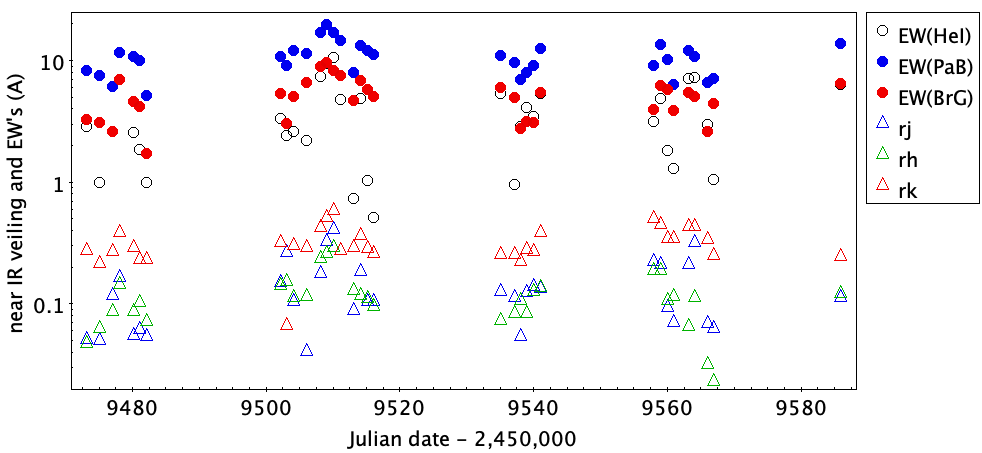}
   \includegraphics[width=0.48\textwidth]{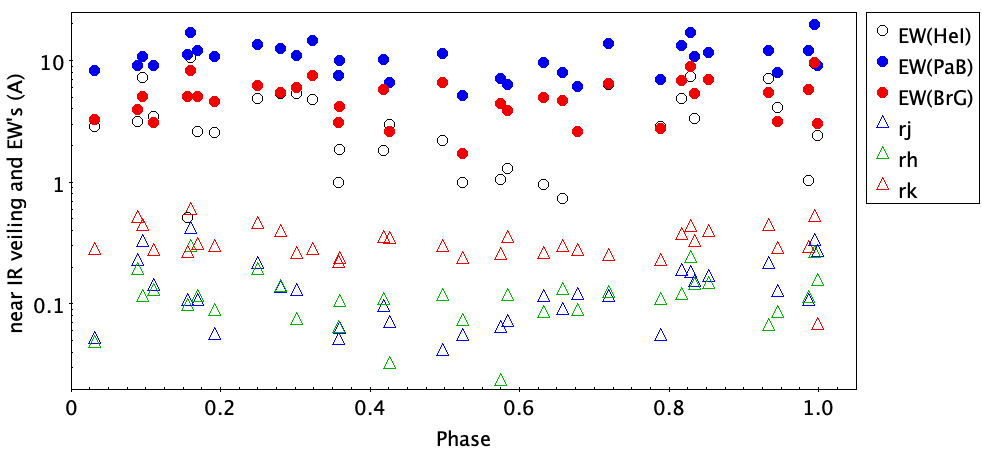}

  \caption{ Veiling measured on SPIRou spectra in the JHK bands ({\it triangles}) and the EW of the \hei, \pab, and \brg\ lines ({\it dots}) plotted as a function of Julian date {\it (top)} and rotational phase ({\it bottom}). } \label{fig:spirouveilew}
\end{figure}

We present in Figure~\ref{fig:spirouveilew}  the evolution of near-infrared veiling in the JHK bands and of the \hei, \pab, and \brg\ line EWs over the course of the campaign. 

\section{Line profile variability on successive SPIRou runs}
\label{app:spirouruns}

\begin{figure*}
   \centering
   \includegraphics[width=0.25\hsize]{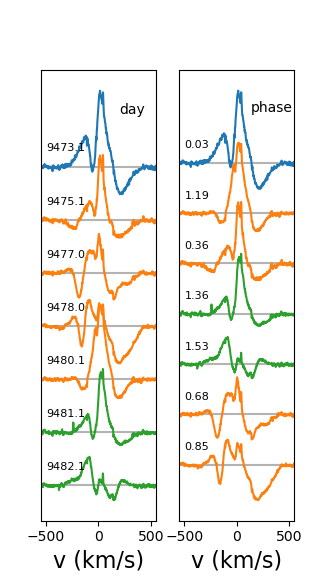}
   \includegraphics[width=0.25\hsize]{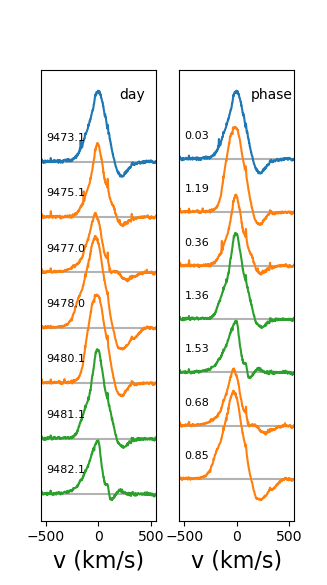}
   \includegraphics[width=0.25\hsize]{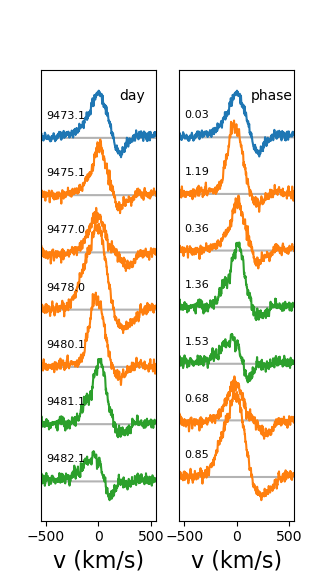}
   \includegraphics[width=0.25\hsize]{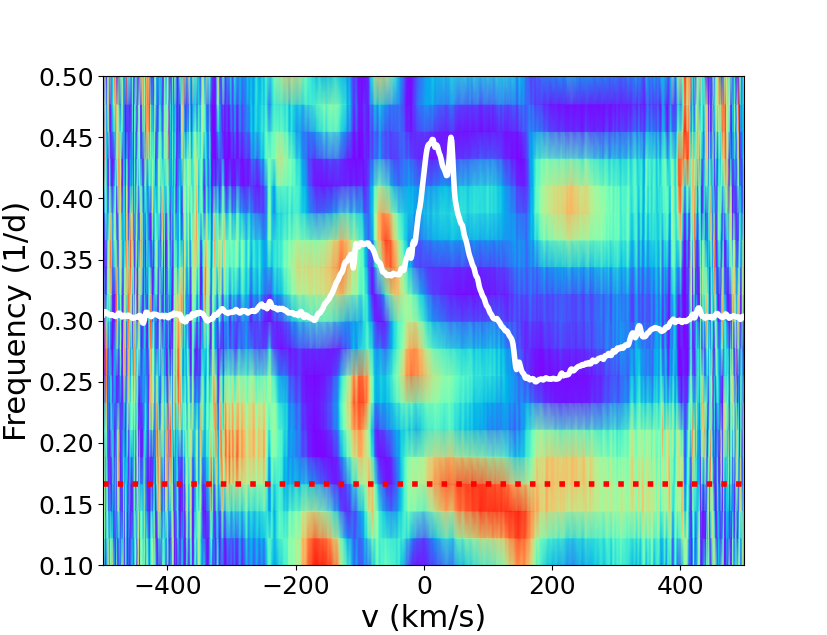}
   \includegraphics[width=0.25\hsize]{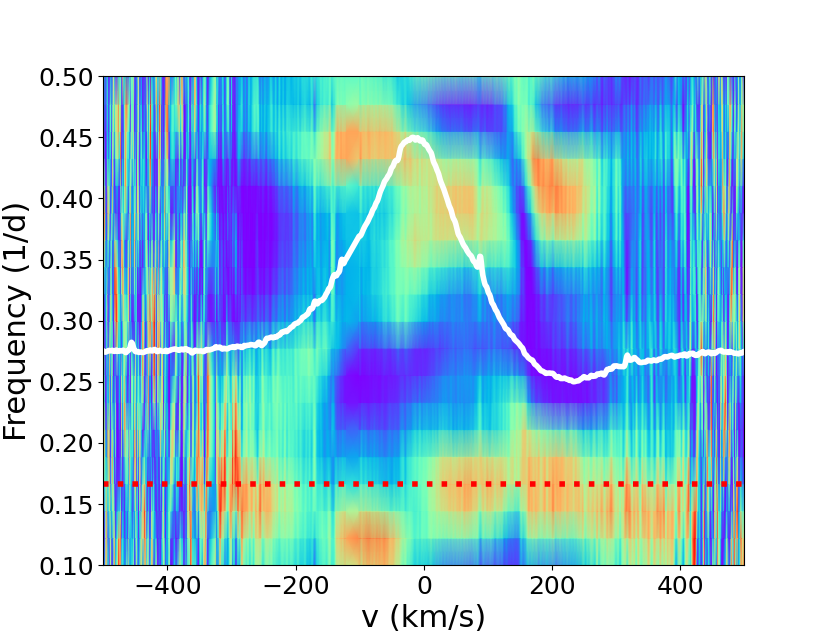}
   \includegraphics[width=0.25\hsize]{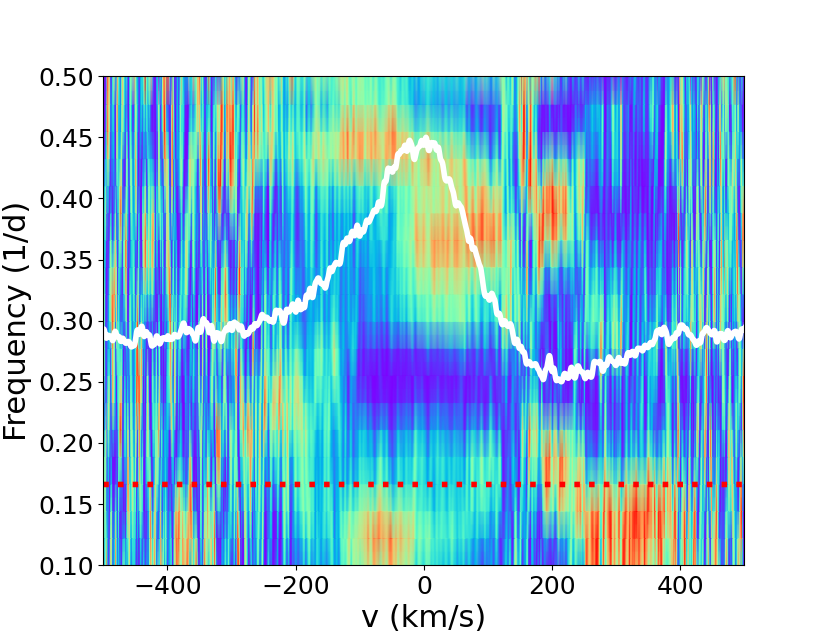}
   \caption{Near-infrared \hei\  ({\it left}), \pab\ ({\it center}), and \brg\ ({\it right}) line profiles obtained over nine days during the September 2021 SPIRou run.  {\it Top:} Line profiles are plotted as a function of Julian date ({\it left subpanels}) and rotational phase ({\it right subpanels}).  The colors represent successive rotational cycles. {\it Bottom:} 2D periodograms across the line profiles. The dotted horizontal red line drawn at a frequency of 0.166 day$^{-1}$ indicates the stellar rotational period. The white curve displays the mean line profile. The color code reflects the periodogram power from zero ({\it blue}) to 1 ({\it red}).}
              \label{fig:spirou_sep}%
    \end{figure*}

\begin{figure*}
   \centering
   \includegraphics[width=0.25\hsize]{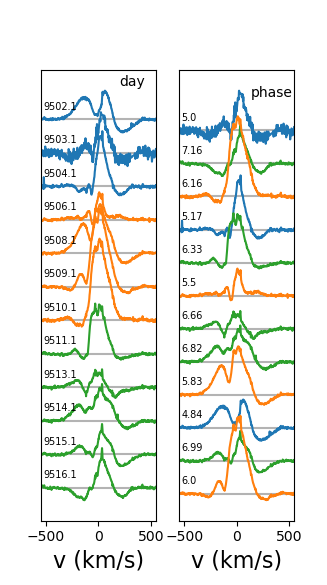}
   \includegraphics[width=0.25\hsize]{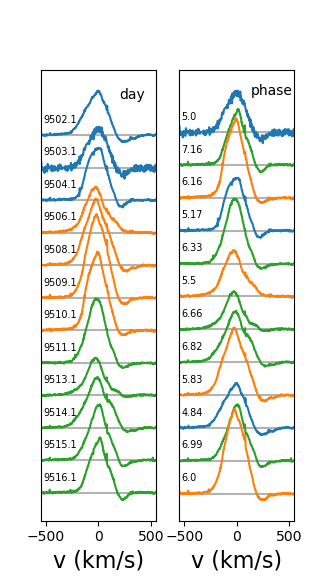}
   \includegraphics[width=0.25\hsize]{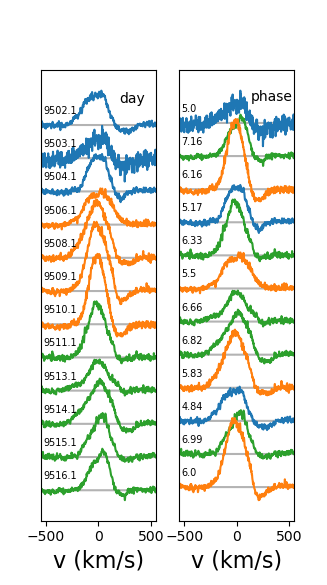}
  \includegraphics[width=0.25\hsize]{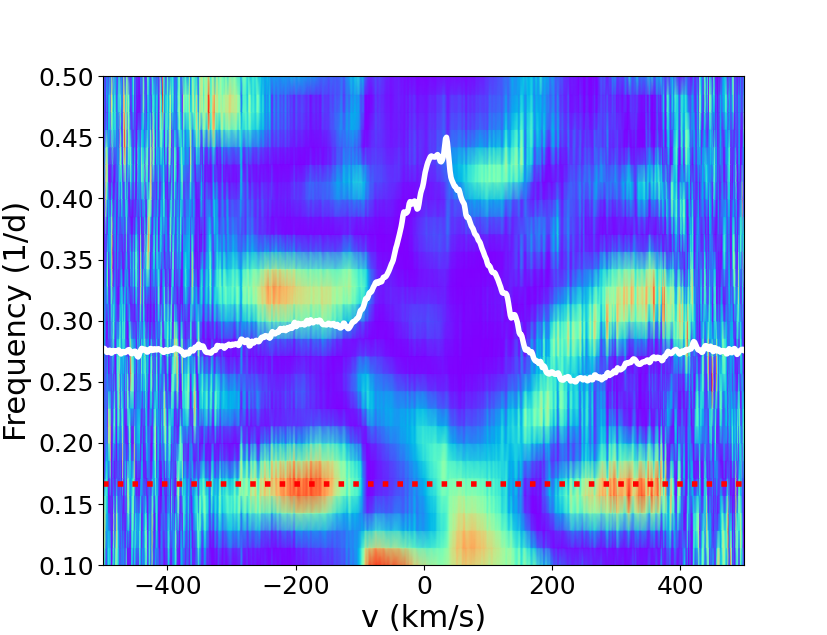}
   \includegraphics[width=0.25\hsize]{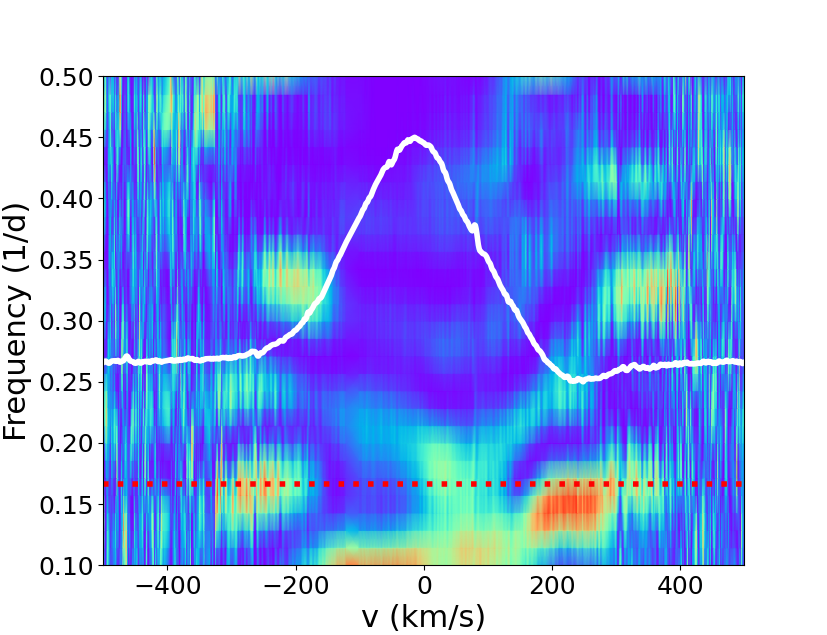}
   \includegraphics[width=0.25\hsize]{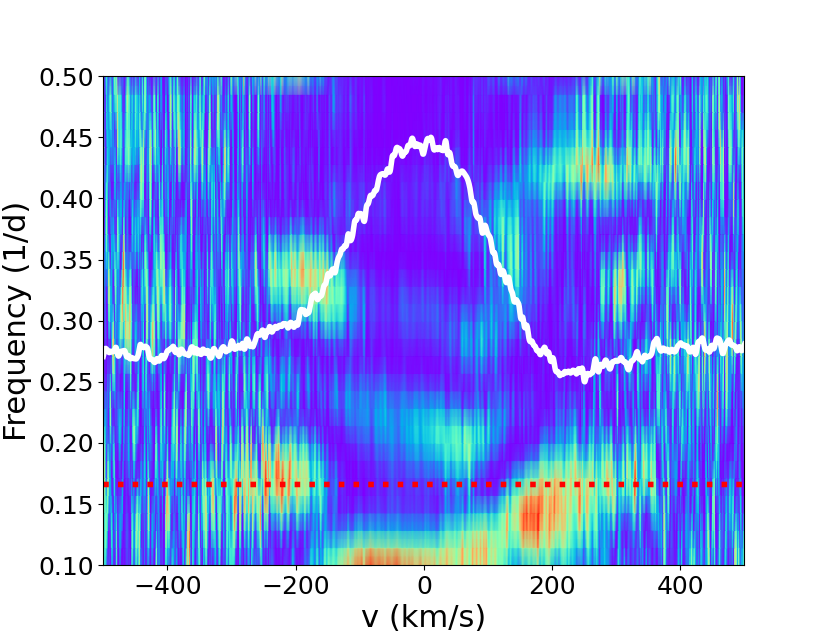}
   \caption{Near-infrared \hei\ ({\it left}), \pab\ ({\it center}), and \brg\ ({\it right}) line profiles obtained over 14 days during the October 2021 SPIRou run.  {\it Top:} Line profiles are plotted as a function of Julian date ({\it left subpanels}) and rotational phase ({\it right subpanels}).  The colors represent successive rotational cycles. {\it Bottom:} 2D periodograms across the line profiles. The dotted horizontal red line drawn at a frequency of 0.166 day$^{-1}$ indicates the stellar rotational period. The white curve displays the mean line profile. The color code reflects the periodogram power from zero ({\it blue}) to 0.9 ({\it red}).}
              \label{fig:spirou_oct}%
    \end{figure*}

\begin{figure*}
   \centering
   \includegraphics[width=0.25\hsize]{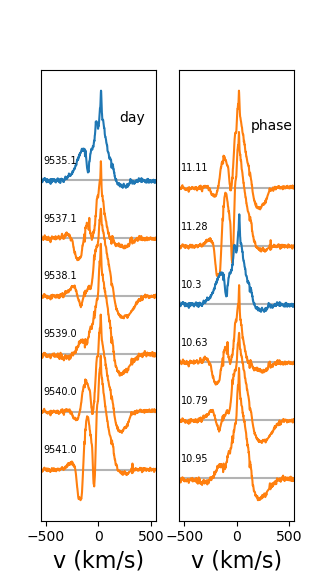}
   \includegraphics[width=0.25\hsize]{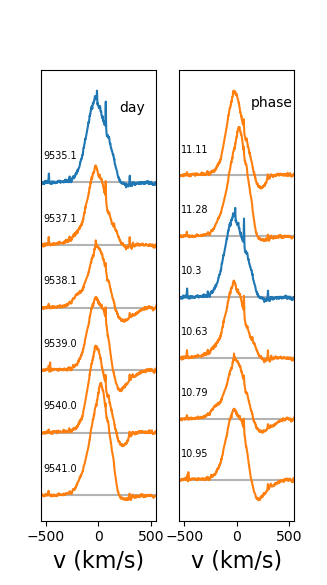}
   \includegraphics[width=0.25\hsize]{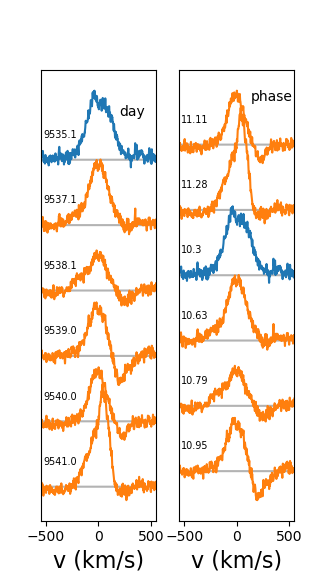}
   \includegraphics[width=0.25\hsize]{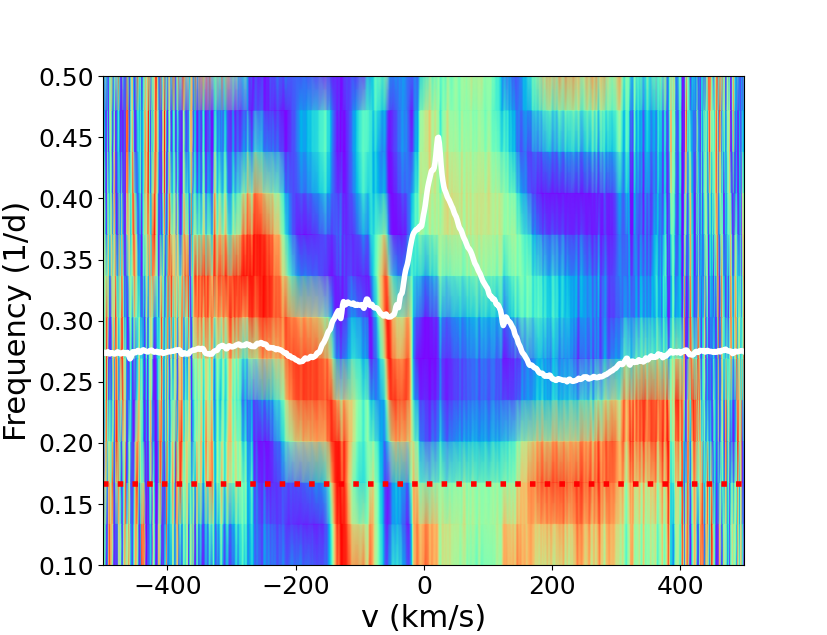}
   \includegraphics[width=0.25\hsize]{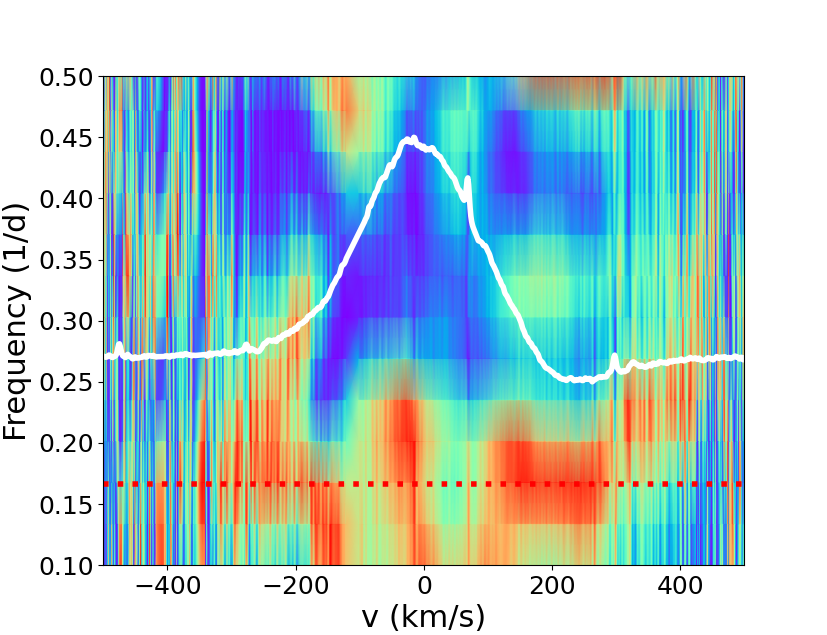}
   \includegraphics[width=0.25\hsize]{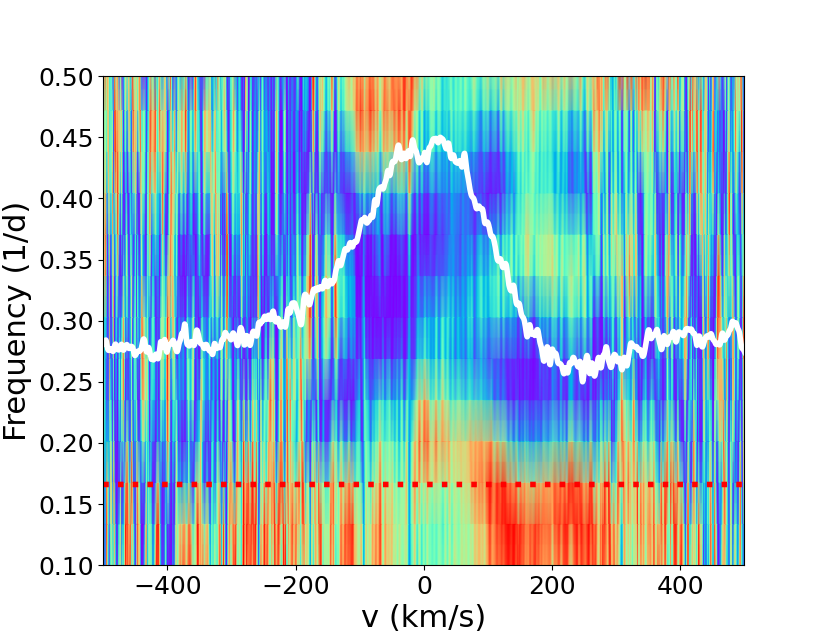}
   \caption{Near-infrared \hei\ ({\it left}), \pab\ ({\it center}), and \brg\ ({\it right}) line profiles obtained over six days during the November 2021 SPIRou run.  {\it Top:} Line profiles are plotted as a function of Julian date ({\it left subpanels}) and rotational phase ({\it right subpanels}).  The colors represent successive rotational cycles. {\it Bottom:} 2D periodograms across the line profiles. The dotted horizontal red line drawn at a frequency of 0.166 day$^{-1}$ indicates the stellar rotational period. The white curve displays the mean line profile. The color code reflects the periodogram power from zero ({\it blue}) to 1 ({\it red}).}
              \label{fig:spirou_nov}%
    \end{figure*}

\begin{figure*}
   \centering
   \includegraphics[width=0.25\hsize]{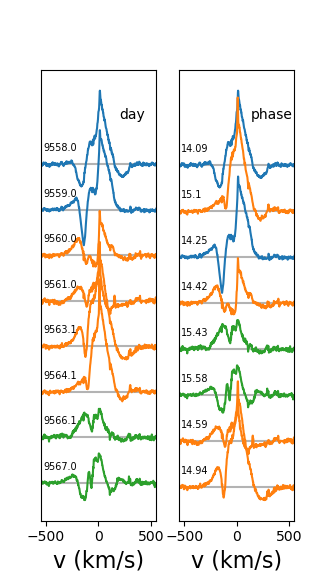}
   \includegraphics[width=0.25\hsize]{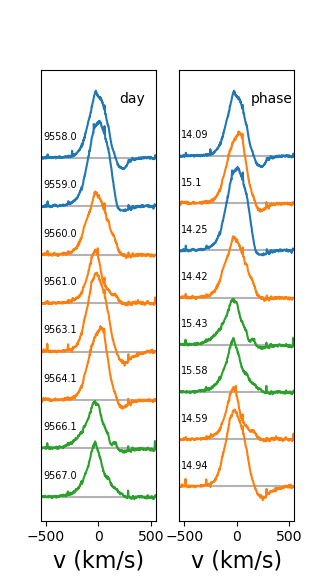}
   \includegraphics[width=0.25\hsize]{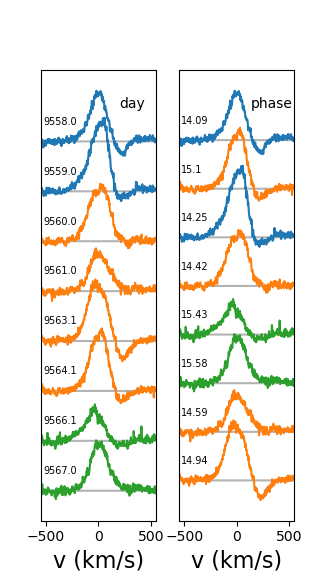}
 \includegraphics[width=0.25\hsize]{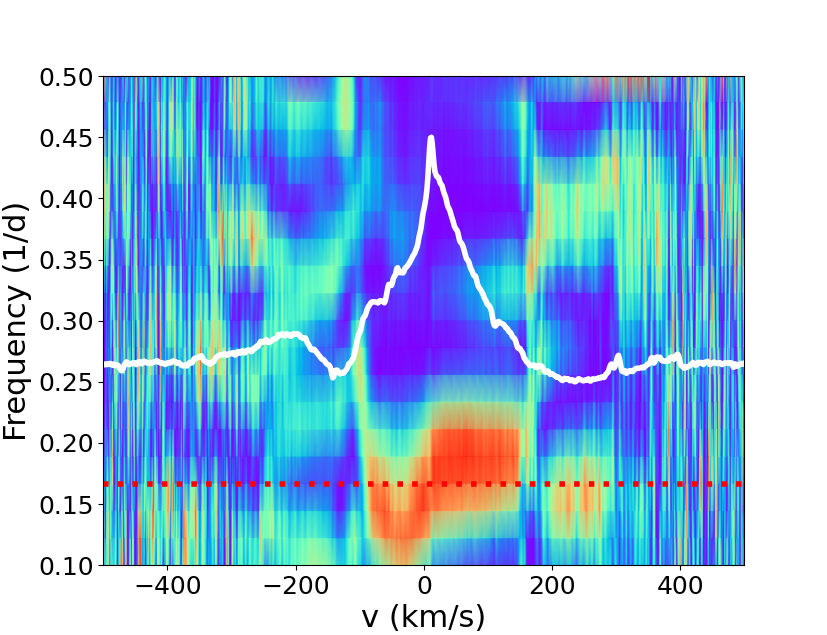}
   \includegraphics[width=0.25\hsize]{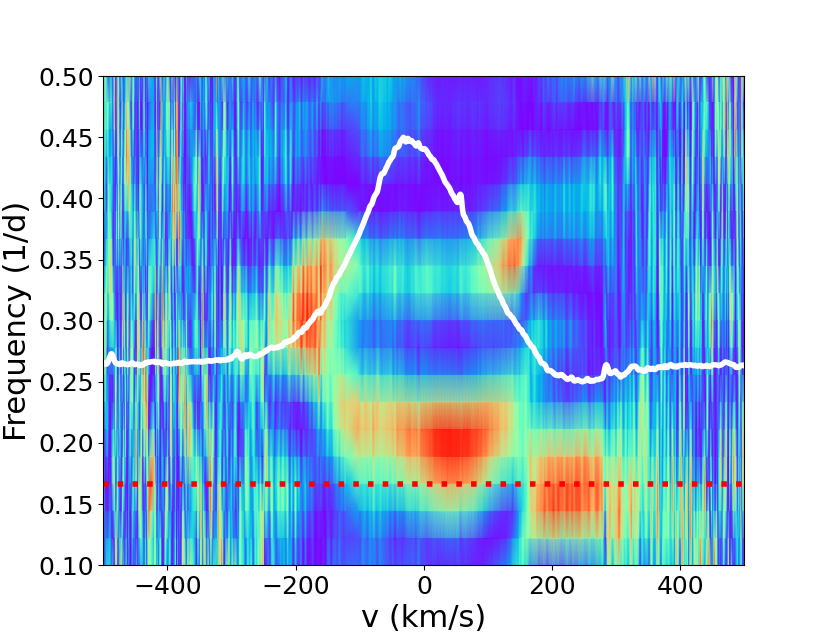}
   \includegraphics[width=0.25\hsize]{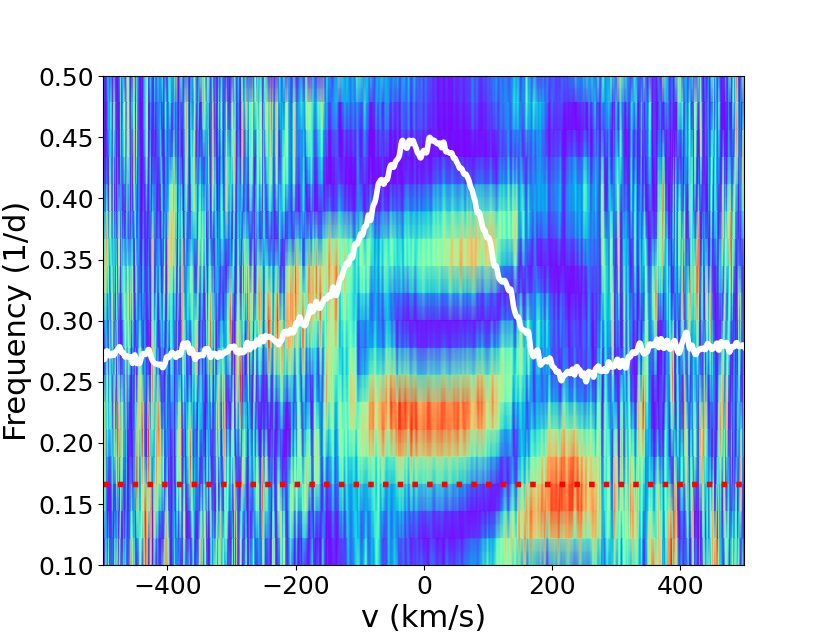}
   \caption{Near-infrared \hei\ ({\it left}), \pab\ ({\it center}), and \brg\ ({\it right}) line profiles obtained over nine days during the December 2021 SPIRou run.  {\it Top:} Line profiles are plotted as a function of Julian date ({\it left subpanels}) and rotational phase ({\it right subpanels}).  The colors represent successive rotational cycles. {\it Bottom:} 2D periodograms across the line profiles. The dotted horizontal red line drawn at a frequency of 0.166 day$^{-1}$ indicates the stellar rotational period. The white curve displays the mean line profile. The color code reflects the periodogram power from zero ({\it blue}) to 1 ({\it red}).}
              \label{fig:spirou_dec}%
    \end{figure*}

Figures~\ref{fig:spirou_sep} to \ref{fig:spirou_dec} show the line profile variability of the \hei\ 10830 \AA, \pab, and \brg\ line profiles for the successive SPIRou runs in September, October, November, and December 2021. The figures also include 2D periodograms for each line. We note, however, that only the October SPIRou run is long enough, extending over 14 days, to yield significant results when searching for a modulation of the line profiles around the rotational period of 6.04 days. The September, November, and December runs lasted for only 9, 6, and 9 days, respectively, which is too short to reliably investigate periods longer than 5 days.

\end{appendix}

\end{document}